\documentclass[twocolumn]{aastex631}
\usepackage{CJK} 

\usepackage{amsmath}
\usepackage{amssymb}
\usepackage{epsfig}
\usepackage{apjfonts}
\usepackage{natbib}
\usepackage{hyperref}
\usepackage{epstopdf}
\usepackage{verbatim}
\usepackage{enumitem}
\usepackage{color}
\setlist[enumerate]{itemsep=-1mm}
\usepackage{soul, xcolor}
\setstcolor{blue}
\usepackage{comment}

\usepackage{textcase}

\usepackage{mathtools}

\usepackage{bigints}

\submitjournal{AJ (Accepted: January 24, 2026)}

\begin{document}


\begin{CJK*}{UTF8}{gbsn}

\title{Uniform Forward-Modeling Analysis of Ultracool Dwarfs. IV. Benchmarking the Sonora Diamondback and Saumon \& Marley (2008) Atmospheric Models Across Late-M, L, and T types with Low-Resolution 0.8--2.5~$\mu$m Spectroscopy}

\author[0009-0002-9013-8005]{Emily Mader}
\affiliation{Department of Astronomy \& Astrophysics, University of California, Santa Cruz, CA 95064, USA}
\affiliation{School of Information, University of California, Berkeley, CA 94720, USA}

\author[0000-0002-3726-4881]{Zhoujian Zhang (张周健)} \thanks{NASA Sagan Fellow}
\affiliation{Department of Physics \& Astronomy, University of Rochester, Rochester, NY 14627, USA}
\affiliation{Department of Astronomy \& Astrophysics, University of California, Santa Cruz, CA 95064, USA}

\author[0000-0002-9843-4354]{Jonathan J. Fortney}
\affiliation{Department of Astronomy \& Astrophysics, University of California, Santa Cruz, CA 95064, USA}

\author[0000-0002-4404-0456]{Caroline V. Morley}
\affiliation{Department of Astronomy, The University of Texas at Austin, Austin, TX 78712, USA}

\author[0009-0004-0725-0034]{Malik Bossett}
\affiliation{Department of Astronomy \& Astrophysics, University of California, Santa Cruz, CA 95064, USA}

\author[0000-0002-5251-2943]{Mark S. Marley}
\affiliation{Lunar and Planetary Laboratory, University of Arizona, Tucson, AZ 85721, USA}

\author[0000-0003-1622-1302]{Sagnick Mukherjee}
\affiliation{Department of Astronomy \& Astrophysics, University of California, Santa Cruz, CA 95064, USA}

\author[0000-0003-2649-2288]{Brendan P. Bowler}
\affiliation{Department of Physics, University of California, Santa Barbara, Santa Barbara, CA 93106, USA}

\author[0000-0003-2232-7664]{Michael C. Liu}
\affiliation{Institute for Astronomy, University of Hawai`i at M\={a}noa. 2680 Woodlawn Drive, Honolulu, HI 96822, USA}

\correspondingauthor{Zhoujian Zhang}
\email{zjzhang@rochester.edu}

\begin{abstract}
We present a systematic assessment of two major cloudy atmospheric model grids --- \texttt{SM08} \citep{2008ApJ...689.1327S} and \texttt{Sonora~Diamondback} \citep{2024ApJ...975...59M} --- when applied to low-resolution near-infrared (0.8--2.5~$\mu$m) spectroscopy. Our analysis focuses on a uniform sample of 142 age-benchmark brown dwarfs and planetary-mass objects spanning late-M, L, and T spectral types, with independently determined ages from 10~Myr to 10~Gyr. We perform forward-model spectral fitting for all benchmarks' IRTF/SpeX spectra ($R\sim$80--250) using both \texttt{SM08} and \texttt{Sonora Diamondback} atmospheric models to infer effective temperatures, surface gravities, metallicities, radii, and cloud sedimentation efficiencies. The two model grids yield broadly consistent results. Among L4--L9 dwarfs, we identify a statistically significant, population-level age dependence of the cloud parameter $f_{\rm sed}$, with young benchmarks ($<300$~Myr) exhibiting systematically lower $f_{\rm sed}$ values than older counterparts. This trend is absent across L0--T5 and T0--T5, demonstrating that cloud properties vary with age and surface gravity and offering explanations for the observed gravity-dependent photometric properties at the late-L end of the L/T transition. By comparing spectroscopically inferred parameters with predictions from evolution models, we quantify systematic errors in the fitted atmospheric parameters and establish empirical calibrations to anchor future studies using these atmospheric models. Stacked residuals across the sample reveal wavelength-dependent data-model mismatches associated with key atomic and molecular absorption bands, highlighting the need for improved opacities and rainout chemistry. In particular, persistent residuals in FeH bands likely contribute to the difficulty of robustly constraining $\log{(g)}$ and mass from spectral fitting for late-M and L dwarfs. Finally, we show that including an interstellar-medium-like extinction term significantly improves the spectral fits, confirming and broadening previous findings and suggesting missing opacity sources in current cloudy models.
\end{abstract}

\section{Introduction} 
\label{sec:intro}

Brown dwarfs bridge the mass gap between stars and exoplanets, providing unique laboratories for studying physical and chemical processes in ultracool ($\lesssim 2400$~K), molecule-rich, self-luminous atmospheres. Their characterization has traditionally relied on forward models of atmospheres and thermal evolution, which are precomputed across grids of fundamental parameters, such as effective temperature ($T_{\rm eff}$), surface gravity ($\log{(g)}$), and metallicity ([M/H]), along with self-consistent assumptions \citep[e.g.,][]{1999ApJ...512..843B, 2015ARA&A..53..279M}. These forward models are essential for contextualizing the atmospheric properties of brown dwarfs and exoplanets, planning observational programs, and guiding the design of next-generation instruments.

Forward-modeling analysis --- fitting observed spectra with a grid of synthetic spectra generated from atmospheric models --- has been a major approach for constraining the fundamental properties of brown dwarfs and self-luminous exoplanets over the past 30 years \citep[e.g.,][]{1996ApJ...465L.123A, 1996Sci...272.1919M, 2000ApJ...541..374S, 2006ApJ...647..552S, 2001ApJ...556..373G, 2007ApJ...667..537L, 2008ApJ...678.1372C, 2009ApJ...702..154S, 2010ApJS..186...63R, 2014A&A...562A.127B, 2021ApJ...920..146L, 2021ApJ...916...53Z, 2021ApJ...921...95Z, 2025AJ....169....9Z, 2023ApJ...951L..48B, 2023A&A...670A..90P, 2023ApJ...959...63S, 2024ApJ...961..121H, 2024ApJ...963...73M, 2024ApJ...966L..11P, 2024ApJ...961..210P, 2025arXiv250500978T}. However, forward-modeling results must be interpreted within the context of the underlying model assumptions, which may not always accurately represent the atmospheric properties of certain targets. These limitations can lead to wavelength-dependent mismatches between observed and modeled spectra, identifying model assumptions that require refinement. For instance, data-model discrepancies in the $Y$ band often arise in the analysis of T/Y brown dwarfs, implying uncertainties in alkali condensation models and opacities \citep[e.g.,][]{2017ApJ...850..150B, 2020A&A...637A..38P,  2021ApJ...916...53Z, 2021ApJ...921...95Z, 2025AJ....169....9Z}. For late-M and L dwarfs, potential limitations in the FeH line list and opacities can lead to data-model offsets in, for example, the $H$ band \citep[e.g.,][]{2024ApJ...961..121H}. Across the L--T--Y spectral sequence, discrepancies in the near-infrared wavelengths often indicate inadequacies in cloud model assumptions \citep[e.g.,][]{2016ApJ...830...96H, 2020A&A...633A.124P, 2024ApJ...966L..11P, 2024ApJ...961..121H} or additional processes unrelated to cloud formation \citep[e.g., non-adiabatic thermo-compositional instabilities;][]{2015ApJ...804L..17T, 2016ApJ...817L..19T, 2021ApJ...918...11L, 2025AJ....169....9Z}. Moreover, for objects near the L/T transition, analyzing the shape of data-model offsets near the 10~$\mu$m silicate absorption feature effectively assesses the composition and particle size of clouds as assumed by atmospheric models \citep[e.g.,][]{2006ApJ...648..614C, 2021ApJ...920..146L, 2023MNRAS.523.4739S, 2025Natur.643..938H, 2025A&A...703A..79M}. 

The presence of data-model mismatch suggests that the physical properties derived from spectral fits may be systematically over- or under-estimated, emphasizing the importance of quantifying these systematic errors for reliable interpretation of observations. This challenge can be addressed by studying a rare and valuable subset of substellar objects known as ``age benchmarks'' --- brown dwarfs or planetary-mass objects that either orbit stars as companions or are kinematic members of young moving groups (YMGs) or star clusters \citep[e.g.,][]{2006MNRAS.368.1281P, 2008ApJ...689..436L, 2014ApJ...792..119D, 2018ApJ...856...23G, 2018ApJ...858...41Z, 2020ApJ...891..171Z, 2021ApJ...911....7Z, 2022ApJ...935...15Z, 2024arXiv241204597B, 2024ApJ...963...67C, 2024ApJ...961..210P, 2024AJ....167..253R}. Benchmark objects are particularly valuable because their ages can be independently determined from their parent stars or other YMG members, assuming coevality within binaries or YMGs \citep[though see][for a counter example]{2024RNAAS...8..114Z}. The independent age determination, which is infeasible for most non-benchmark brown dwarfs, allows for precise characterization of fundamental properties through thermal evolution models when combined with these objects' bolometric luminosities. Thermal evolution models incorporate atmospheric models as boundary conditions but primarily describe the time evolution of bulk properties, thereby making them less vulnerable to wavelength-dependent systematics, such as uncertainties of molecular opacities and line lists. Consequently, evolution-based parameters of benchmark objects serve as empirical anchors for validating and testing the parameters inferred from spectral fits via atmospheric models. Comparing benchmarks' properties derived from evolution models with those from atmospheric models thus offers a promising avenue for directly quantifying the systematic errors inherent in forward-modeling analyses.

\begin{figure} 
\includegraphics[width=\linewidth]{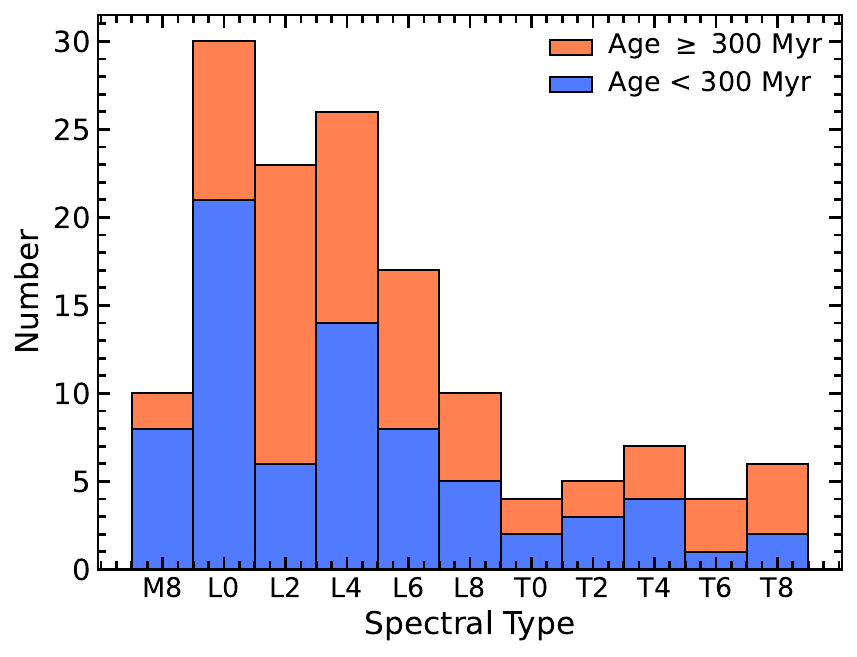}
\caption{Spectral type distribution of our sample of benchmark brown dwarfs and planetary-mass objects, with ages younger (blue) or older (orange) than 300~Myr.}
\label{fig:spt_hist}
\end{figure}

\begin{figure*}[t]
\begin{center}
\includegraphics[width=6.in]{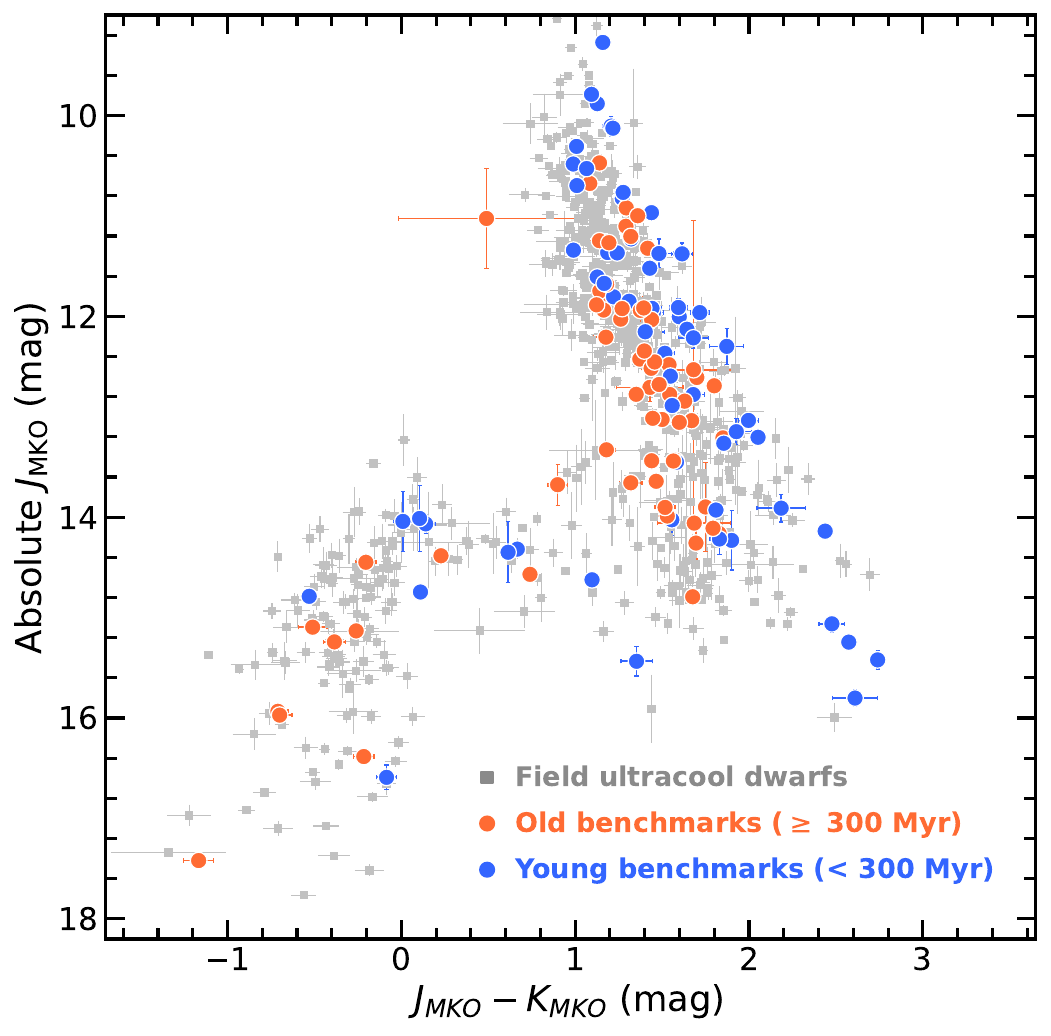}
\caption{$J$-band absolute magnitudes and $J-K$ colors of our sample. Young benchmarks with ages below 300~Myr are shown as blue circles, while those older than 300~Myr are shown in orange. We overlay previously known late-M, L, T, and Y dwarfs from the UltracoolSheet \citep{ultracoolsheet}; these objects have precise photometry (S/N$>5$) and are neither candidate nor resolved binaries.}
\label{fig:color_mag}
\end{center}
\end{figure*}

To evaluate the performance of state-of-the-art atmospheric models, we are conducting the ``Uniform Forward-Modeling Analysis of Ultracool Dwarfs`` program by analyzing uniform spectroscopic samples of brown dwarfs and planetary-mass objects across a wide range of spectral types, spectral resolutions, and model assumptions. In paper~I, \cite{2021ApJ...916...53Z} examined low-resolution ($R \sim 80-250$) near-infrared (1.0--2.5~$\mu$m) spectra of three benchmark late-T dwarfs using the \texttt{Sonora Bobcat} models \citep{2021ApJ...920...85M}, which assume cloud-free atmospheres in chemical equilibrium. Their analysis incorporated the uncertainties from model interpolation and correlated residuals into the forward-modeling framework \citep{2015ApJ...812..128C, 2017ApJ...836..200G}. For two benchmarks (GJ~570D and HD~3651B; T7.5) with $T_{\rm eff}$ of 780--810~K and field ages of $\gtrsim 1.5$~Gyr, atmospheric and evolution models yielded significantly different surface gravities and metallicities. Specifically, the $\log{(g)}$ and [M/H] values as determined via spectral fits are lower than evolution-based values by about $1.2$~dex (a factor of 16) and $0.35$~dex (a factor of 2), respectively. \cite{2021ApJ...916...53Z} attributed these mismatches to the uncertainties in potassium opacities adopted by the \texttt{Sonora Bobcat} models. For the third benchmark (Ross~458C; T8.5), which has a relatively cooler $T_{\rm eff}$ ($\sim 680$~K) and younger age (150--800~Myr), the $\log{(g)}$ and [M/H] inferred from spectral fits are consistent with those from thermal evolution models. However, the spectral fits yielded a higher $T_{\rm eff}$ by $\sim 120$~K and a smaller radius by a factor of 1.6. \cite{2021ApJ...916...53Z} proposed that incorporating additional physical processes, such as cloud formation, disequilibrium chemistry, and modifications to the atmospheric adiabatic index, into model assumptions may improve the consistency. 

In paper II, \cite{2021ApJ...921...95Z} expanded the forward-modeling analysis to a larger sample of 55 late-T dwarfs using the \texttt{Sonora Bobcat} models. Although most of these objects are not benchmarks, this ensemble analysis allowed for a population-level investigation of potential systematic errors in inferred parameters. \cite{2021ApJ...921...95Z} found that systematic errors in $\log{(g)}$, [M/H], and $T_{\rm eff}$ derived from spectral fits are consistent with those identified for late-T benchmarks in \cite{2021ApJ...916...53Z}. Specifically, [M/H] values inferred from spectral fits are likely underestimated by 0.3--0.4~dex. Inferred $\log{(g)}$ values are systematically low, corresponding to unrealistically young ages of $10-400$~Myr and small masses of $1-8$~M$_{\rm Jup}$. For brown dwarfs with spectral types later than T8, $T_{\rm eff}$ values from spectral fits are likely overestimated by 50--200~K. Additionally, they quantified the degeneracy between surface gravity and metallicity in the forward-modeling analysis of late-T dwarfs as $\Delta\log{(g)} = 3.42 \times \Delta$[M/H]. Furthermore, by stacking the residuals between observed and fitted spectra of individual late-T dwarfs, \cite{2021ApJ...921...95Z} revealed systematic wavelength-dependent structures. The fitted model spectra exhibit brighter fluxes in the $J$ band and fainter fluxes in the $Y$ and $H$ bands. These mismatches highlight the importance of incorporating more sophisticated model assumptions --- cloud scattering, disequilibrium chemistry, and modifications of the atmospheric adiabatic index --- for accurately characterizing late-T dwarf atmospheres. 

In paper III, \cite{2024ApJ...961..121H} extended the analysis from the late-T regime to earlier spectral types. They analyzed low-resolution ($R\sim150$) near-infrared (0.9--2.4~$\mu$m) spectra of 90 late-M and L benchmarks from several nearby YMGs (10--200~Myr) using the \texttt{BT-Settl} models \citep{2012RSPTA.370.2765A, 2015A&A...577A..42B}, which incorporate cloud formation by coupling the mixing and diffusion processes as predicted by hydrodynamic simulations. Their analysis revealed that for a subset of M8--L6 dwarfs with relatively red $J-K$ colors, $T_{\rm eff}$ values inferred from spectral fits are clustered near 1800~K, and the $\log{(g)}$ posteriors peak at the upper boundary of the model grid at 5.5~dex. We note that $T_{\rm eff} = 1800$~K corresponds to a critical effective temperature below which the synthetic BT-Settl spectra deviate from the monotonic trends established at higher temperatures. Additionally, the analysis of \cite{2024ApJ...961..121H} revealed that the fitted model spectra exhibit systematically brighter fluxes in the $J$ and $H$ bands near the M/L transition, implying that the treatment of cloud condensation and sedimentation in these models requires further refinement. Interestingly, \cite{2024ApJ...961..121H} found that the data-model agreement near the M/L transition could be significantly improved by incorporating an additional interstellar-medium-like (ISM-like) extinction term into the spectral fits \citep[see also][]{2020A&A...633A.124P, 2024ApJ...966L..11P}, and that the amount of such extinction varies with spectral types.

In this work, we extend the forward-modeling analysis to an even larger sample of 142 benchmark brown dwarfs and planetary-mass objects spanning the late-M, L, and T spectral sequence, with ages ranging from 10~Myr to 10~Gyr (Section~\ref{sec:sample}). We focus on analyzing a uniform dataset of low-resolution ($R\sim$80--250), near-infrared (0.8--2.5~$\mu$m) spectra obtained with the SpeX spectrograph \citep{2003PASP..115..362R} on the NASA Infrared Telescope Facility (IRTF).\footnote{The goal of this work is to benchmark the performance of cloudy atmospheric models specifically for low-resolution ($R\sim$80--250), 0.8--2.5~$\mu$m spectra and to establish empirical calibrations that are directly applicable for future studies using data with similar properties. Beyond the dataset used in this work, there are other valuable spectral libraries of brown dwarfs and planetary-mass objects that span different/broader wavelength ranges and/or higher spectral resolutions \citep[e.g.,][]{2003ApJ...596..561M, 2007ApJ...658.1217M, 2007A&A...473..245R, 2014A&A...562A.127B, 2019AJ....157..101M, 2020MNRAS.491.5925M, 2022MNRAS.513.5701S, 2025arXiv251008691K, 2025A&A...701A.208P}. These datasets can offer complementary insights to this particular paper and will be investigated in future work.} Our analysis employs two model grids: the \cite{2008ApJ...689.1327S} models (\texttt{SM08}) and the \texttt{Sonora Diamondback} models \citep{2024ApJ...975...59M}.\footnote{The \texttt{Sonora Diamondback} models are publicly available on Zenodo: \url{https://doi.org/10.5281/zenodo.12735103}. The \texttt{SM08} atmospheric models were obtained from D. Saumon via private communication.} Both grids assume chemical equilibrium, incorporate cloud properties following the \cite{2001ApJ...556..872A} prescription, and self-consistently couple atmospheric models with thermal evolution models. The \texttt{Sonora Diamondback} models, however, include updated atomic and molecular opacities that have become available since the development of \texttt{SM08} and span a significantly broader metallicity range. The primary goal of this work is to systematically assess the performance of these two major cloudy atmospheric model grids when applied to the low-resolution 0.8-2.5~$\mu$m spectroscopy.

We first use the thermal evolution models of \texttt{Sonora Diamondback} to contextualize the fundamental properties of our benchmark targets (Section~\ref{sec:evo}) and then use both sets of atmospheric models to fit these objects' low-resolution near-infrared spectra (Section~\ref{sec:FM}). These analyses allow us to evaluate state-of-the-art cloudy models for brown dwarfs and planetary-mass companions across late-M, L, and T types, providing context for future studies that utilize these models. The implications of our work are discussed in Section~\ref{sec:discussion}, with a summary presented in Section~\ref{sec:summary}.

\section{Sample}
\label{sec:sample}
We used the UltracoolSheet \citep{ultracoolsheet} to identify benchmark brown dwarfs and planetary-mass objects with late-M, L, and T types, and compiled their published low-resolution ($R \approx$ 80--250) near-infrared (0.8--2.5 $\mu$m) spectra from the literature \citep{2006ApJ...637.1067B, 2006ApJ...639.1095B, 2007ApJ...658..617B, 2008ApJ...681..579B, 2010ApJ...710.1142B, 2011AJ....141...70B, 2013ApJ...772..129B, 2006AJ....131.2722C, 2006ApJ...639.1120K, 2010ApJS..190..100K, 2011ApJS..197...19K, 2006AJ....132..891R, 2007AJ....134.1162L, 2007ApJ...654..570L, 2007AJ....133.2320S, 2010AJ....139.1045S, 2011AJ....142...77D, 2014ApJ...792..119D, 2017MNRAS.467.1126D, 2010AJ....139..176F, 2011AJ....141...71F, 2011ApJ...732...56G, 2012ApJ...753..142B, 2012AJ....144...94G, 2013ApJ...777...84B, 2015ApJ...814..118B, 2013ApJ...772...79A, 2014ApJ...794..143B, 2014ApJ...785L..14G, 2015ApJS..219...33G, 2014AJ....147...34S, 2016ApJS..225...10F, 2017ASInC..14....7B, 2020ApJ...891..171Z, 2021ApJ...921...95Z, 2022ApJ...935...15Z, 2023ApJ...959...63S}. We restricted the sample to spectra obtained with IRTF/SpeX. This selection yielded 142 benchmarks, including 63 known companions to stars and 79 free-floating YMG members, with properties summarized in Table~\ref{tab:sample_dat}. Their spectral types span M7--T8 (Figure~\ref{fig:spt_hist}) with near-infrared photometry shown in Figure \ref{fig:color_mag}. 

\begin{figure}
\includegraphics[width=\linewidth]{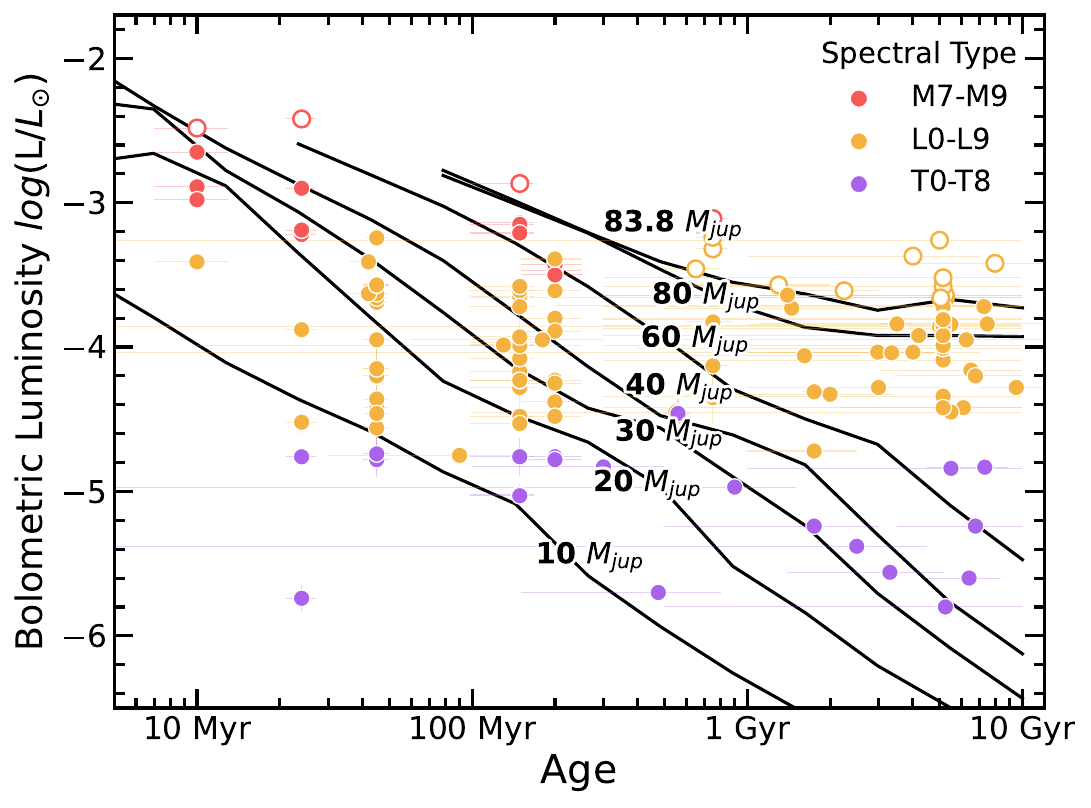}
\caption{Ages and bolometric luminosities of our benchmark targets (Table~\ref{tab:evo}), color-coded by spectral types. We use open circles to mark the benchmarks excluded from the evolution model analysis, including binaries and several objects whose ages and/or $L_{\rm bol}$ fall outside the convex hull of the evolution models (see Section~\ref{sec:evo}). The \texttt{Sonora Diamondback hybrid-grav} evolution models (black lines) with [M/H]$=0$~dex are overlaid. }
\label{fig:M_tracks}
\end{figure}

The ages of these benchmarks range from 10~Myr to 10~Gyr (Figure~\ref{fig:M_tracks}), determined from their parent stars or host YMGs, as summarized by \cite{2014ApJ...792..119D}, \cite{2020ApJ...891..171Z, 2021ApJ...911....7Z, 2022ApJ...935...15Z}, \cite{2023ApJ...959...63S}, and \cite{ultracoolsheet}. Most targets have bolometric luminosities ($L_{\rm bol}$) derived by \cite{2023ApJ...959...63S}, who computed $L_{\rm bol}$ by integrating observed IRTF/SpeX spectra in combination with extrapolated fluxes from best-fitting synthetic spectra of atmospheric models. For objects not included in that work, we adopt the $L_{\rm bol}$ values from the literature: HD~203030B \citep{2017AJ....154..262M}, ULAS~J141623.94+134836.3 \citep{2020ApJ...905...46G}, COCONUTS-1B \citep{2020ApJ...891..171Z}, COCONUTS-3B \citep{2022ApJ...935...15Z}, and 2MASSW~J1207334$-$393254 (Zhang et al. in prep). For the remaining two objects, 2MASS~J23520507$-$1100435 and WISE~J003110.04+574936.3, we estimate $L_{\rm bol}$ from their $H_{\rm MKO}$-band absolute magnitudes using the empirical relation of \cite{2023ApJ...959...63S}. 

Our sample also includes fourteen objects whose binarity has been suggested or confirmed by adaptive optics (AO) observations (Table~\ref{tab:sample_dat}). We do not compute $L_{\rm bol}$ values for these binaries and exclude them from subsequent analyses in Sections~\ref{sec:evo} and \ref{sec:discussion}. Nevertheless, we fit their spectra using atmospheric models by treating each system as a single object (Section~\ref{sec:FM}).

The IRTF/SpeX spectra of our benchmark targets have an averaged signal-to-noise ratio (S/N) of $\approx$ 124 per pixel in the $J$ band (1.15--1.3~$\mu$m). We flux-calibrated the SpeX spectra using these objects' $J_{\rm MKO}$ magnitudes along with the corresponding filter response \citep{2006MNRAS.367..454H} and zero-point flux \citep{2007MNRAS.379.1599L}. For one benchmark, GJ~1048B (L1), we used the $H_{\rm MKO}$ magnitude for flux calibration due to a large uncertainty (0.5~mag) in its $J$-band magnitude.

\begin{figure*}[t]
\begin{center}
\includegraphics[width=7in]{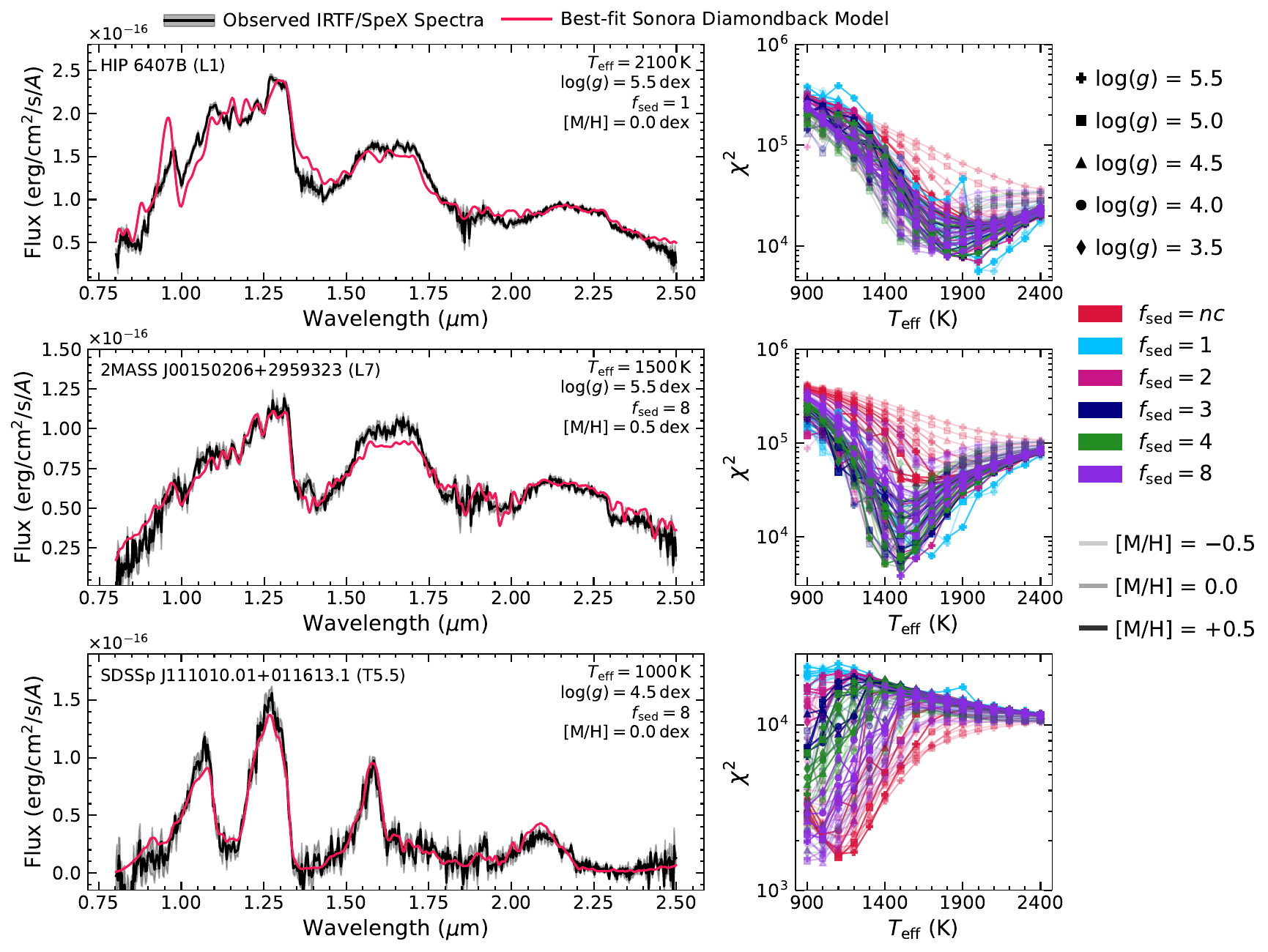} 
\caption{$\chi^{2}$-based spectral fitting results of three representative targets based on the \texttt{Sonora Diamondback} atmospheric models. In the left panels, the observed IRTF/SpeX spectra (black) are compared with the best-fit model spectra (red), and the corresponding best-fit parameters are labeled. Fluxes within the broad water bands (shaded in gray) are masked in spectral fits. The right panels present the  $\chi^{2}$ values as a function of effective temperatures, with variations in symbol, color, and transparency indicating different values of $\log{(g)}$, $f_{\rm sed}$, and [M/H], respectively.}
\label{fig:DB_specfit_Gk}
\end{center}
\end{figure*}

\begin{figure*}[t]
\begin{center}
\includegraphics[width=7.3 in]{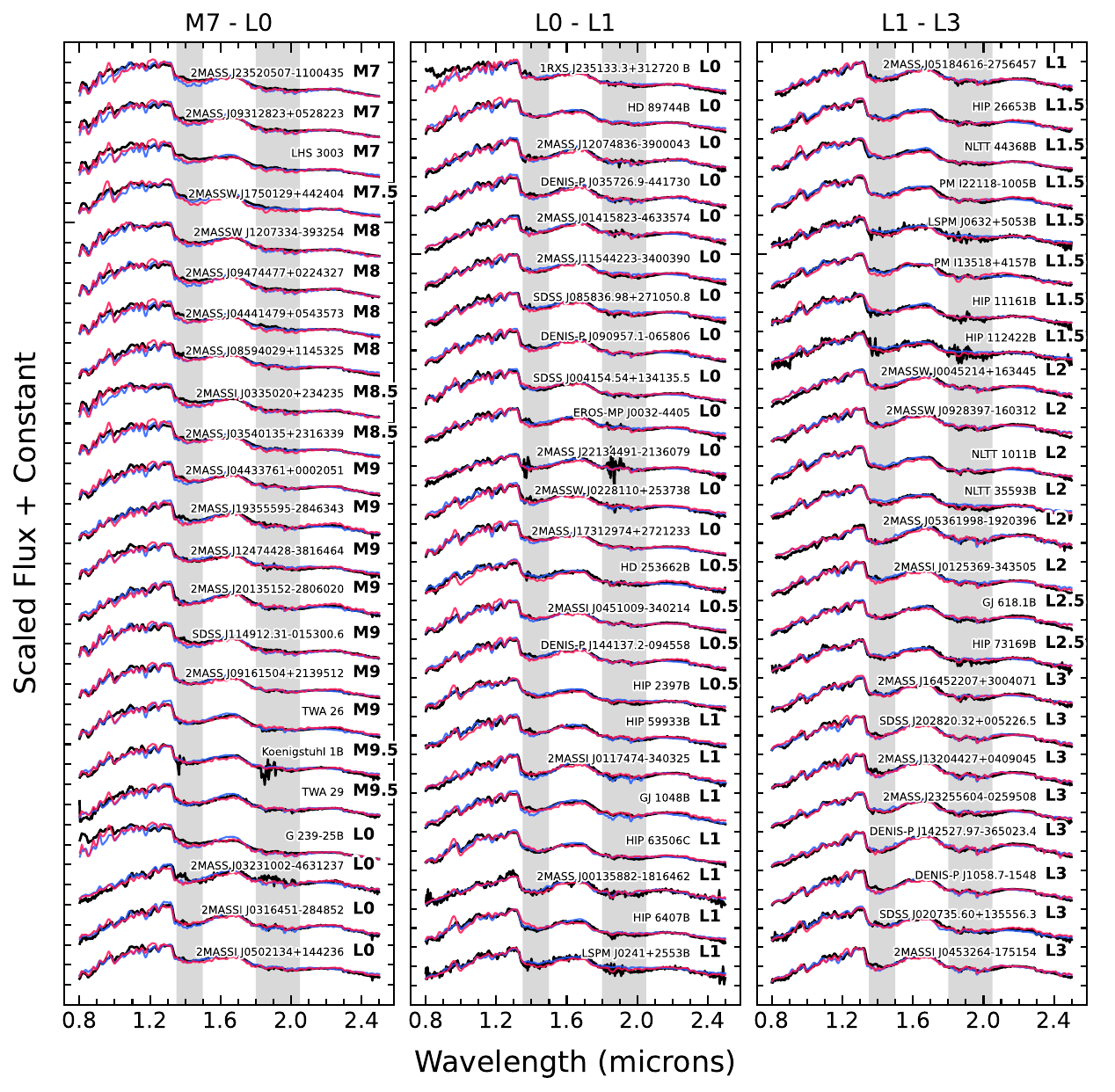}
\caption{Best-fit synthetic spectra for all targets using the \texttt{SM08} models (blue) and \texttt{Sonora Diamondback} (red) models, overlaid with the observed IRTF/SpeX spectra (black). Fluxes within the broad water bands (shaded in gray) are masked in spectral fits. Targets are ordered by spectral types, spanning M7 to L3.}
\label{fig:chi_fits_1}
\end{center}
\end{figure*}

\renewcommand{\thefigure}{\arabic{figure}}
\addtocounter{figure}{-1}
\begin{figure*}[t]
\begin{center}
\includegraphics[width=7.3 in]{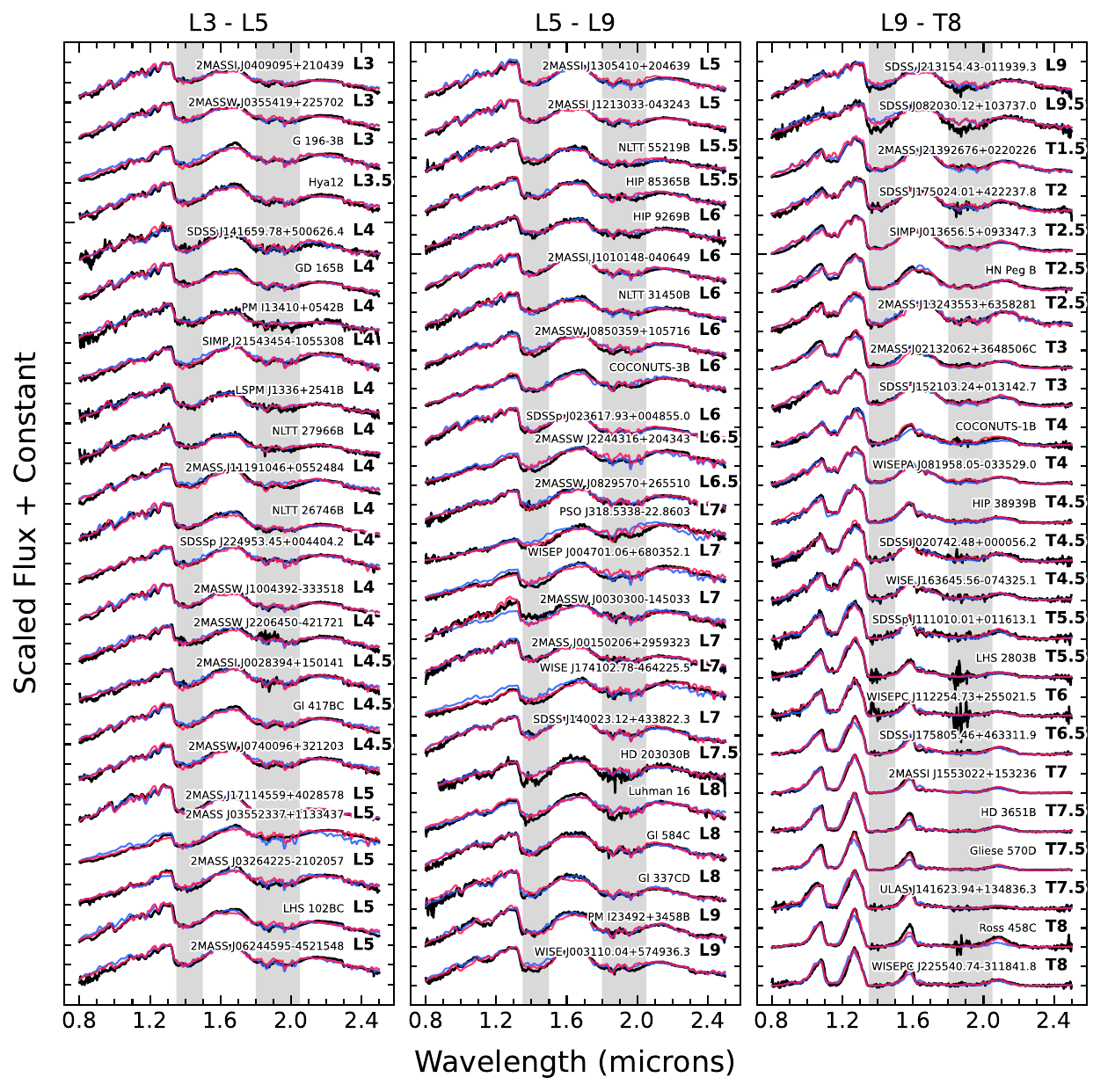}
\caption{Continued. Spectral fitting results for L3--T8 targets.}
\end{center}
\end{figure*}

\section{Contextualizing the Properties of Benchmarks via Evolution Models} 
\label{sec:evo}

Before fitting the observed spectra of our benchmark targets with atmospheric model grids, we first use thermal evolution models to contextualize their physical properties, including $T_{\rm eff}$, $\log{(g)}$, radius ($R$), and mass ($M$). These evolution-based properties are subsequently compared to the same set of parameters derived from spectral fits using atmospheric models (Section~\ref{sec:discussion}). As demonstrated in previous works within this program \citep{2021ApJ...916...53Z, 2021ApJ...921...95Z, 2024ApJ...961..121H}, such comparison helps quantify the systematic errors of the fitted atmospheric models.

The properties of benchmarks are determined using the \texttt{Sonora Diamondback hybrid-grav} evolution models \citep{2024ApJ...975...59M}, which incorporate a gravity-dependent cloud-clearing process across the L/T transition, as supported by observational evidence \citep[e.g.,][]{2006ApJ...651.1166M, 2008Sci...322.1348M, 2011ApJ...735L..39B, 2015ApJ...805...56D, 2015ApJ...810..158F, 2019MNRAS.483..480V, 2020ApJ...891..171Z}. These models span a wide parameter space, covering ages from 0.1~Myr to 13~Gyr, logarithmic bolometric luminosities ($L_{\rm bol}$) from $-9.1$~dex to $-1.5$~dex, effective temperatures from 100~K to 2800~K, logarithmic surface gravities from $2.15$~dex to $5.5$~dex, radii from $0.75$~R$_{\rm Jup}$ to 7.22~R$_{\rm Jup}$, and masses from $0.5$~M$_{\rm Jup}$ to $84$~M$_{\rm Jup}$. Additionally, these models are provided at three metallicities: [M/H]$=-0.5$~dex, 0~dex, and $+0.5$~dex.

To estimate the properties of benchmarks, we draw $10^{6}$ random samples of their ages, bolometric luminosities, and metallicities, assuming specific distributions as described below, and use these samples to interpolate the evolution models. For benchmarks with ages provided as values and error bars in Table~\ref{tab:evo}, we adopt normal distributions.\footnote{For ages with asymmetric error bars, we assume a distribution composed of two half-Gaussians with the same mean but standard deviations corresponding to the upper and lower error bars, respectively.} For benchmarks with ages presented as ranges in Table~\ref{tab:evo}, we assume uniform distributions. We also adopt normal distributions for $L_{\rm bol}$ and a uniform distribution for [M/H] spanning $-0.5$~dex to $+0.5$~dex. The evolution models are interpolated in linear scales for $\log{(L_{\rm bol}/L_{\odot})}$, $\log{(g)}$, and [M/H], and in logarithmic scales for age, $T_{\rm eff}$, $R$, and $M$. 

We exclude two subsets of benchmarks from this evolution model analysis: (1) fourteen benchmarks with suggested or confirmed binarity (see Section~\ref{sec:sample}), and (2) twenty-four benchmarks (seven of which are also part of the first subset) whose ages and/or bolometric luminosities have median values outside the convex hull of the evolution model grid. 

Table~\ref{tab:evo} summarizes the inferred evolution-based properties of our benchmarks. We also repeat the analysis by using the \texttt{hybrid} version of \texttt{Sonora Diamondback} evolution models. These models account for cloud-clearing at a critical $T_{\rm eff}$ of 1200~K while neglecting the gravity dependence of this process, following \cite{2008ApJ...689.1327S}. As shown in Appendix~\ref{app:hybrid_grav_vs_hybrid}, the differences in the inferred parameters between \texttt{hybrid-grav} and \texttt{hybrid} models are not significant, with differences of $10 \pm 105$~K in $T_{\rm eff}$, $0.004 \pm 0.148$~dex in $\log{(g)}$, $-0.02 \pm 0.10$~$R_{\rm Jup}$ in $R$, and $-0.7 \pm 7.0$~$M_{\rm Jup}$ in $M$.

\begin{figure*}[t]
\begin{center}
\includegraphics[width=6.5in]{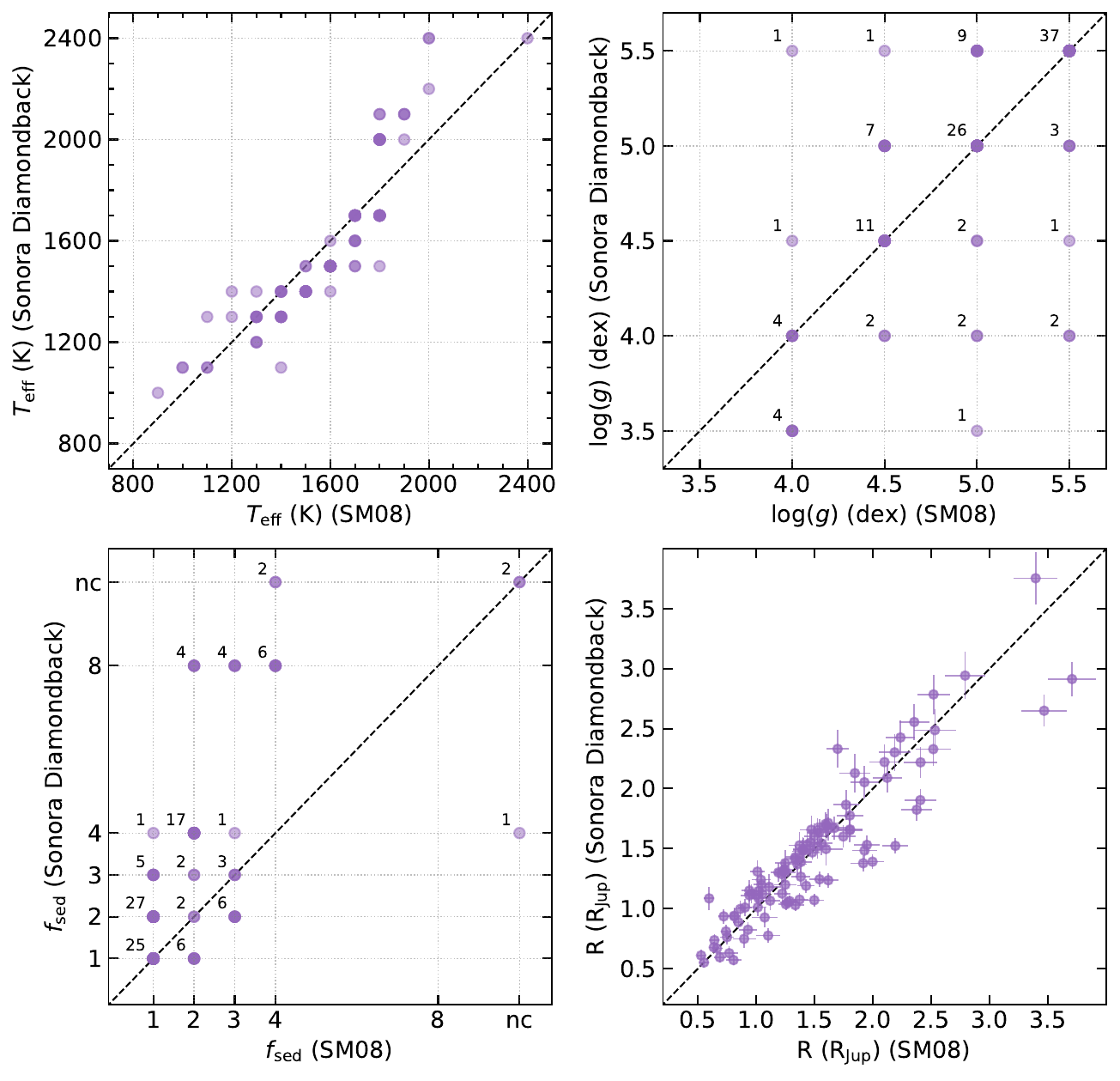} 
\caption{Comparison of spectroscopically inferred $T_{\rm eff}$, $\log{(g)}$, $f_{\rm sed}$, and $R$ values between \texttt{Sonora Diamondback} and \texttt{SM08} model grids. Uncertainties in $T_{\rm eff}$, $\log{(g)}$, and $f_{\rm sed}$ are adopted as half the model grid spacing and are not plotted for clarity. Model grids are indicated by grey dashed lines. In the $\log{(g)}$ and $f_{\rm sed}$ panels, data points are clustered at discrete grid values; the number of targets at each value is labeled.}
\label{fig:parameter_comparison}
\end{center}
\end{figure*}

\section{Forward-modeling Analysis} 
\label{sec:FM}

We fit the IRTF/SpeX spectra of our targets by using synthetic spectra from two grids of atmospheric models: \texttt{SM08} \citep{2008ApJ...689.1327S} and \texttt{Sonora Diamondback} \citep{2024ApJ...975...59M}. The \texttt{SM08} grid spans $T_{\rm eff}$ values from 800 K to 2400 K (in 100 K steps) and $\log{(g)}$ from 4.0 dex to 5.5 dex (in 0.5 dex steps), with metallicity ([M/H]) fixed at the solar value (0~dex). These models incorporate cloud effects based on the \cite{2001ApJ...556..872A} prescriptions, parameterized by the condensation sedimentation efficiency factor ($f_{\rm sed}$), which ranges from 1 to 4 (integers). A set of cloud-free models is also included, labeled as ``nc'' ($f_{\rm sed} \to \infty$). 

The \texttt{Sonora Diamondback} grid spans $T_{\rm eff}$ values from 900~K to 2400~K (in 100~K steps), $\log{(g)}$ from 3.5~dex to 5.5~dex (in 0.5~dex steps), and metallicities from $-0.5$~dex to $+0.5$~dex (in 0.5~dex steps). It includes $f_{\rm sed}$ values of 1, 2, 3, 4, and 8, as well as a cloud-free ``nc'' set. Compared to \texttt{SM08}, \texttt{Sonora Diamondback} models represent a significant update, offering a broader coverage in [M/H] and incorporating updated opacities for atomic and molecular species, including their isotopologues \citep[e.g.,][]{2020JQSRT.25507228T, 2021ApJS..254...34G, 2022JQSRT.27707949G}.

For each grid, we first convolve the individual synthetic model spectra with a Gaussian kernel of wavelength-dependent width to match the spectral resolution of the IRTF/SpeX prism \citep{2003PASP..115..362R}, following \cite{2021ApJ...916...53Z, 2021ApJ...921...95Z}. We adopt the empirical resolution curve provided by SpeX documentation\footnote{\url{https://irtfweb.ifa.hawaii.edu/~spex/uprism_R.jpg}}, which corresponds to the $0.3''$ slit, and scale it inversely with slit width to match the spectral resolution of each benchmark target. The degraded model spectra are then resampled onto the wavelength grid of each target.

Our spectral fitting is performed over 0.8--1.35~$\mu$m, 1.5--1.8~$\mu$m, and 2.05--2.5~$\mu$m, with fluxes in broad water bands (1.35--1.5~$\mu$m and 1.8--2.05~$\mu$m) masked to mitigate systematics from telluric correction during the data reduction. We evaluate the spectral fits by computing the $\chi^{2}$ statistic, following \cite{2008ApJ...678.1372C} and \cite{2024ApJ...961..210P}:
\begin{equation} \label{G_k}
    \chi^{2} = \sum^{N}_{i=1} \left( \frac{f_{\lambda_{i}} - \alpha F_{\lambda_{i}}}{\sigma_{\lambda_{i}}} \right) ^{2}
\end{equation}
Here, $N$ is the number of wavelength pixels; $f_{\lambda_{i}}$ and $F_{\lambda_{i}}$ are the observed and modeled spectral fluxes, respectively; and $\sigma_{\lambda_{i}}$ denotes the observed flux uncertainties. The scaling factor $\alpha$, which minimizes $\chi^{2}$, is given by:
\begin{equation} \label{C_k}
    \alpha = \frac{ \sum^{N}_{i=1} f_{\lambda_{i}} F_{\lambda_{i}} / \sigma_{\lambda_{i}}^{2} }{ \sum^{N}_{i=1} F_{\lambda_{i}}^{2} / \sigma_{\lambda_{i}}^{2} } \equiv \left(\frac{R}{d}\right)^{2}
\end{equation}
Here, $\alpha$ corresponds to $(R/d)^{2}$, where $R$ is the object's radius and $d$ is its distance. For benchmarks with known parallaxes, and thereby distances, our fitted $\alpha$ thus allows the inference of radii.   

We adopt uncertainties of half the grid spacing for $T_{\rm eff}$, $\log{(g)}$, [M/H], and $f_{\rm sed}$, following the practice established in previous $\chi^{2}$-based spectral fitting studies \citep[e.g.,][]{2008ApJ...678.1372C, 2010ApJS..186...63R, 2020ApJ...891..171Z, 2024ApJ...961..210P}. This approach is supported by \cite{2021ApJ...921...95Z}, who applied the \texttt{Starfish} forward-modeling framework \citep{2015ApJ...812..128C} to incorporate uncertainties from model interpolation and correlated residuals. Their analysis of late-T dwarfs demonstrated that parameter uncertainties typically range from 1/3 to 1/2 of the model grid spacing. Additionally, we propagate the uncertainty in $T_{\rm eff}$ to the derived $R$ by assuming constant bolometric luminosity, based on the best-fit $T_{\rm eff}$ and $R$ values, as well as the Stefan-Boltzmann law.

Table~\ref{tab:chi2} summarizes the spectroscopically inferred physical properties for all targets. Figure~\ref{fig:DB_specfit_Gk} presents the best-fit \texttt{Sonora Diamondback} model spectra and corresponding $\chi^{2}$ values for three representative targets. Figure~\ref{fig:chi_fits_1} summarizes the best-fit spectra of both \texttt{Sonora Diamondback} and \texttt{SM08} models for the entire sample.

We compare the spectroscopically inferred $T_{\rm eff}$, $\log{(g)}$, $f_{\rm sed}$, and $R$ between the \texttt{Sonora Diamondback} and \texttt{SM08} model grids in Figure~\ref{fig:parameter_comparison}. For this comparison, as well as for the subsequent discussion in Section~\ref{sec:discussion}, we focus on a subset of 111 out of 142 initial benchmark targets, after excluding (1) 14 suggested or confirmed binaries by AO imaging (see Section~\ref{sec:sample}); and (2) 18 objects (including one object overlapping with the binary subset) whose best-fit $T_{\rm eff}$ values lie at the boundaries of the \texttt{Sonora Diamondback} or \texttt{SM08} model grids --- either the lower edge (800~K for \texttt{SM08}, 900~K for \texttt{Sonora Diamondback}) or the upper edge (2400~K) --- suggesting that their actual effective temperatures likely fall outside the modeled parameter range (Table~\ref{tab:chi2}). The resulting 111 benchmarks have spectral types spanning M8--T6. 

Overall, the fitted $T_{\rm eff}$ and $R$ values are broadly consistent between the \texttt{Sonora Diamondback} and \texttt{SM08} models, with typical differences in $T_{\rm eff}$ on the order of one grid spacing ($100$~K) across the sample. The inferred $\log{(g)}$ values exhibit substantially larger scatter, although $93\%$ (106 out of 114) targets remain consistent within one grid step ($\pm 0.5$~dex). For $f_{\rm sed}$, $70\%$ (80 out of 114) targets show agreement within one grid spacing, with values inferred from the \texttt{Sonora Diamondback} models systematically larger than or equal to those obtained from \texttt{SM08}.  In Appendix~\ref{app:compare}, we further compare the relative quality of spectral fits between the \texttt{Sonora Diamondback} and \texttt{SM08} models.

In the following section, we discuss the implications of these results in more detail. In particular, in Section~\ref{subsec:calib}, we empirically calibrate the spectroscopically inferred parameters from both the \texttt{Sonora Diamondback} and \texttt{SM08} atmospheric models based on the same \texttt{Sonora Diamondback} evolutionary models. For this empirical calibration, we further restrict the sample to 100 out of the 111 benchmarks by excluding an additional 11 objects that lack evolution-based parameters, because their median ages and bolometric luminosities fall outside the convex hull of the \texttt{Sonora Diamondback} evolution models (see Section~\ref{sec:evo}).

\section{Discussion} 
\label{sec:discussion}

\begin{figure*}
\begin{center}
\includegraphics[width=6.in]{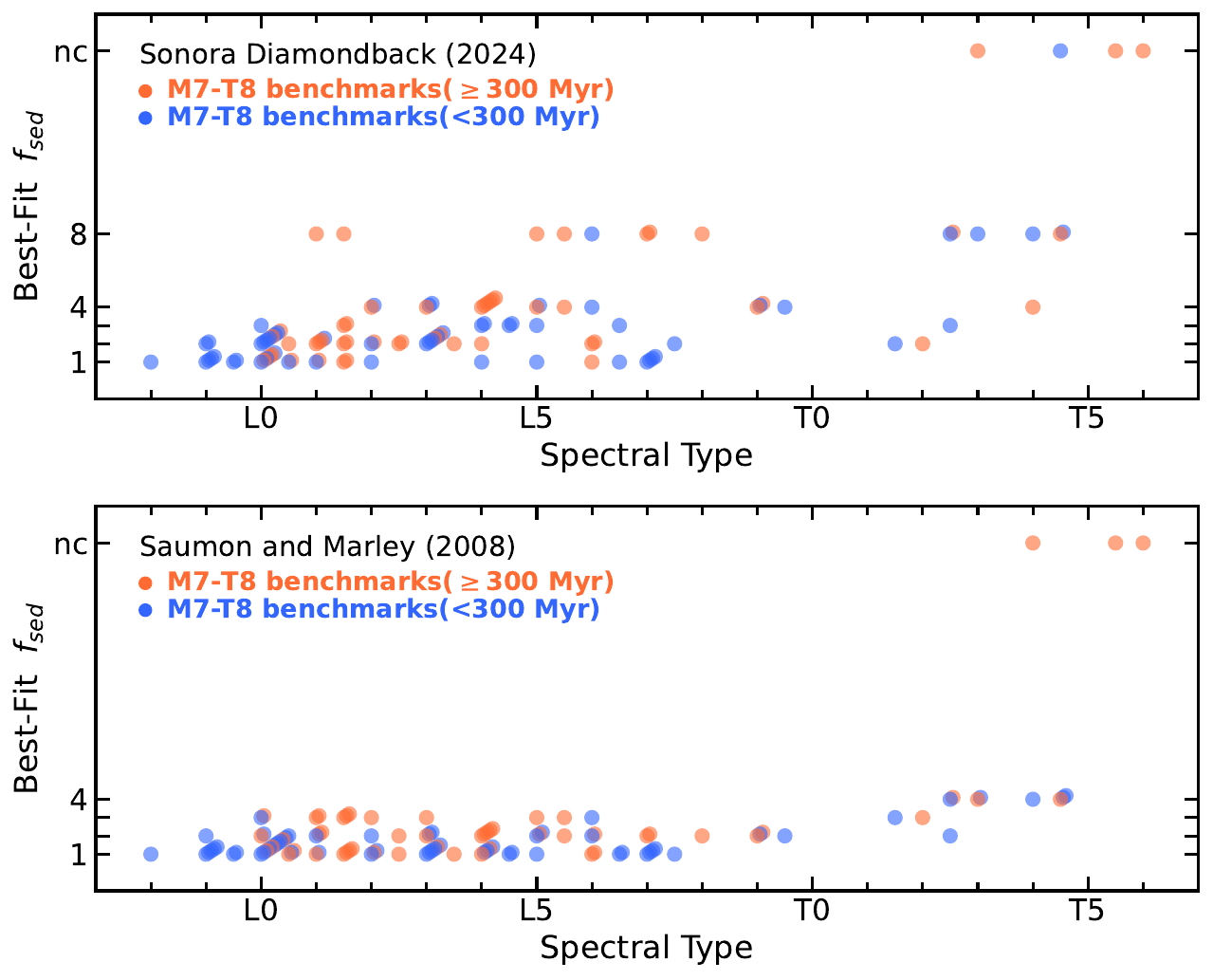}
\caption{Best-fit sedimentation efficiency $f_{sed}$ derived from the \texttt{Sonora Diamondback} (top) and \texttt{SM08} (bottom) model grids for our benchmark targets, with old ($\geq 300$~Myr) and young ($<300$~Myr) objects shown in red and blue, respectively. Although the ``nc'' models correspond to $f_{\rm sed} \rightarrow \infty$, we plot them at a value of $18$ for visualization. For targets with identical spectral types and best-fit $f_{\rm sed}$ values, data points are slightly offset toward the upper right in these panels.}
\label{fig:fsed_spt}
\end{center}
\end{figure*}

\begin{figure*}
\begin{center}
\includegraphics[width=5.in]{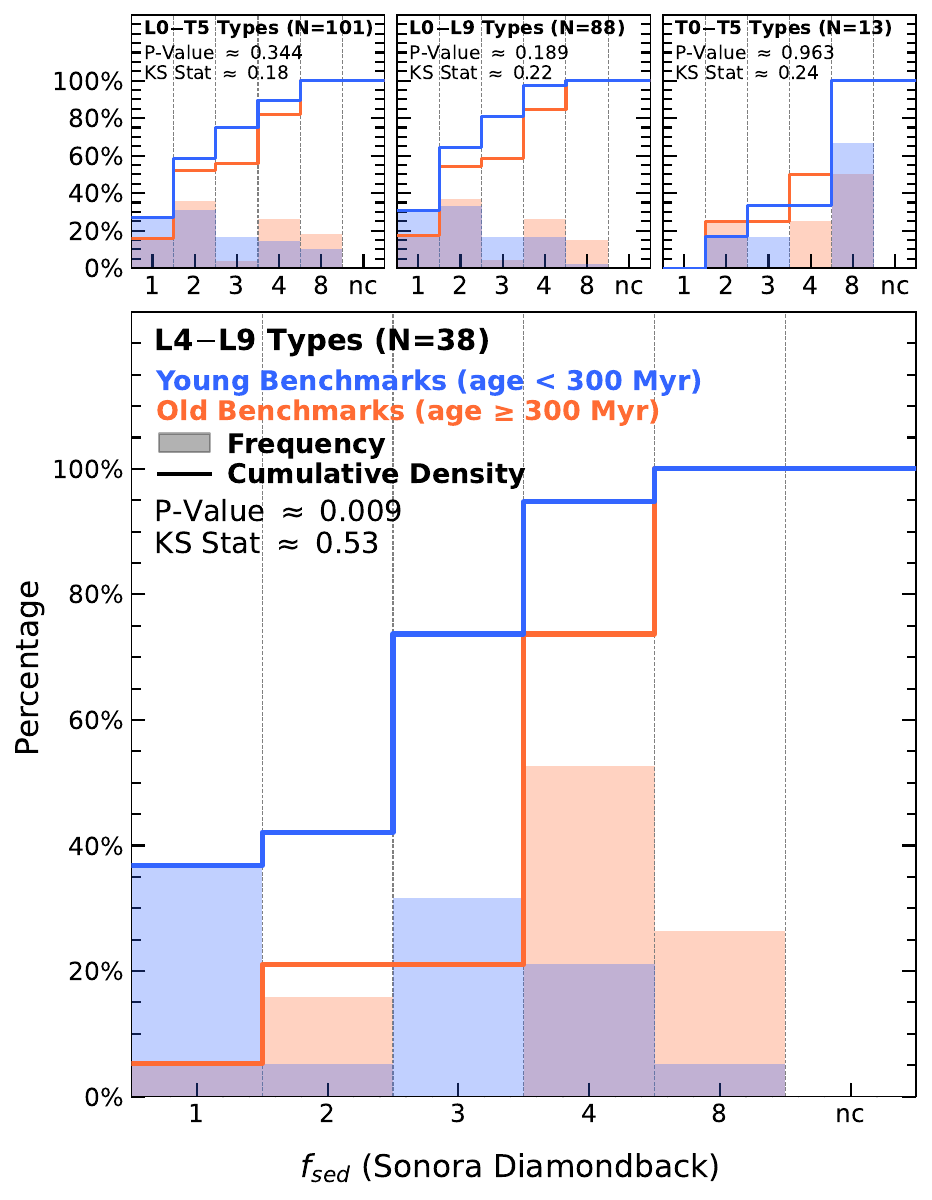}
\caption{Cumulative distributions (solid lines) of the best-fit $f_{\rm sed}$ values inferred from the \texttt{Sonora Diamondback} models for benchmark brown dwarfs in the L0--T5 (top left), L0--L9 (top middle), T0--T5 (top right), and L4--L9 (bottom) type ranges. Young ($<300$~Myr) and old ($\leqslant 300$~Myr) benchmarks are shown in blue and orange, respectively. The corresponding normalized histograms of the $f_{\rm sed}$ distributions are overlaid.}
\label{fig:fsed_ages_hist_SDB}
\end{center}
\end{figure*}

\begin{figure*}
\begin{center}
\includegraphics[width=5.in]{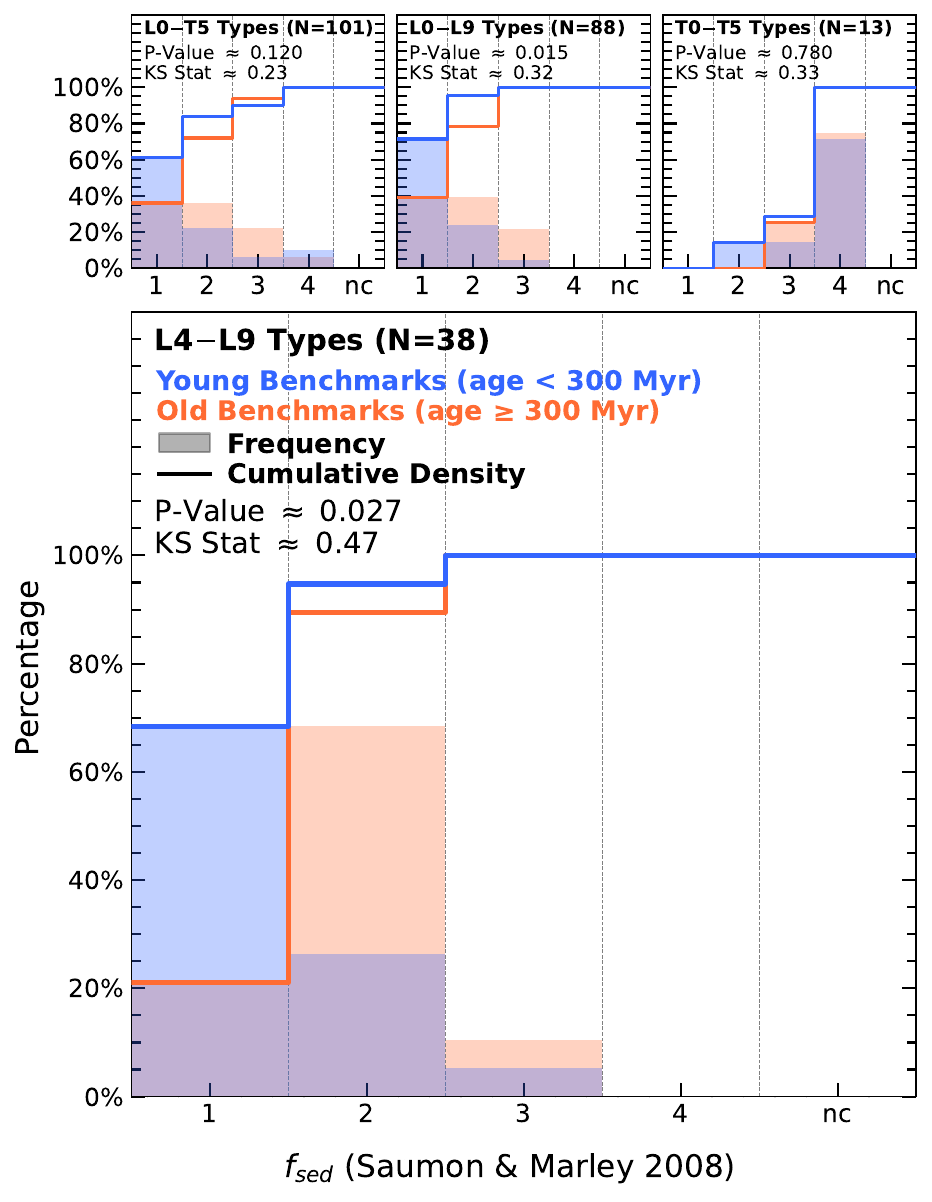}
\caption{Cumulative distributions (solid lines) of the best-fit $f_{\rm sed}$ values inferred from the \texttt{SM08} models. These panels follow the same format as Figure~\ref{fig:fsed_ages_hist_SDB}.}
\label{fig:fsed_ages_hist_SM08}
\end{center}
\end{figure*}

\subsection{Population-level trends in $f_{\rm sed}$ and age dependence near the L/T transition} 
\label{subsec:fsed_spt}

\subsubsection{$f_{\rm sed}$ vs. Spectral Type}
\label{subsubsec:fsed_spt} 

Figure~\ref{fig:fsed_spt} presents the derived $f_{\rm sed}$ values for our targets as a function of spectral type. As described by \cite{2001ApJ...556..872A}, $f_{\rm sed}$ quantifies the sedimentation efficiency of condensate clouds relative to turbulent mixing, with higher values corresponding to thinner clouds and lower opacities. The spectroscopically inferred $f_{\rm sed}$ values in our sample generally fall within 1--4 (based on \texttt{Sonora Diamondback}) or 1--3 (\texttt{SM08}) from late-M through late-L types, followed by a systematic increase toward the T dwarfs. This trend suggests that condensate clouds first emerge near the M/L transition, become increasingly thin (or patchy) across the L/T transition, and largely dissipate by the late-T types \citep[e.g.,][]{2002ApJ...571L.151B, 2002ApJ...568..335M, 2002ApJ...575..264T, 2004AJ....127.3553K, 2010ApJ...723L.117M, 2013ApJ...768..121A, 2016ApJ...833...96L, 2017ApJ...848...83L, 2022MNRAS.513.5701S}. This trend also aligns with earlier spectral-modeling studies by \cite{2008ApJ...678.1372C} and \cite{2009ApJ...702..154S}, who used customized versions of the \texttt{SM08} models and similar $\chi^{2}$-based fitting approaches to analyze the spectra of field L and T dwarfs. Our work extends these efforts by nearly an order of magnitude in sample size and leverages benchmark systems with independently measured properties, rather than relying on field-age non-benchmark objects alone.

\subsubsection{Age Dependence of $f_{\rm sed}$ Near the L/T Transition}
\label{subsubsec:age_fsed}

The known ages of our benchmark targets allow us to investigate whether the relationship between $f_{\rm sed}$ and spectral type (Section~\ref{subsubsec:fsed_spt}) differs between young ($<300$~Myr) and old ($\geqslant 300$~Myr) populations. To assess potential age dependence, we first consider samples spanning broad spectral type ranges of L0--T5 and L0--L9. As shown in Figure~\ref{fig:fsed_ages_hist_SDB}, we perform Kolmogorov-Smirnov (KS) tests (using \texttt{scipy.stats.ks2\_samp}) to compare the distributions of best-fit $f_{\rm sed}$ values derived from the \texttt{Sonora Diamondback} models for the young and old groups. These tests yield $p$-values of $\approx 0.34$ (L0--T5) and $\approx 0.19$ (L0--L9), indicating no statistically significant difference in $f_{\rm sed}$ across these broad spectral type ranges.

In contrast, when restricting the analysis to the L4--L9 subset, the KS test yields a $p$-value of 0.009 and a KS statistic of 0.53, pointing to a statistically significant difference in $f_{\rm sed}$ distributions between young and old objects at late-L spectral types (Figure~\ref{fig:fsed_ages_hist_SDB}). No such age dependence is seen in the T0--T5 range, where the $p$-value is nearly 1.0. We note that the age coverage of the young and old benchmarks is comparable between the L4--L9 and T0--T5 subsets: for L4--L9, the young and old benchmarks span ages of 24--200~Myr and 0.55--9.5~Gyr, respectively, while for T0--T5 the corresponding age ranges are 24--200~Myr and 0.3--7.3~Gyr. This similarity in age coverage suggests that the significant age dependence of $f_{\rm sed}$ observed in L4--L9, but not in T0--T5, is unlikely to arise from any differences in the underlying age distributions of the two subsamples.

We repeat the same KS tests using the best-fit $f_{\rm sed}$ values derived from the \texttt{SM08} models and obtain broadly consistent results. The resulting $p$-values are 0.12, 0.03, and 0.78 for the L0--T5, L4--L9, and T0--T5 spectral type ranges, respectively, indicating that the age dependence of $f_{\rm sed}$ is statistically significant for L4--L9 dwarfs, but not across the L0--T5 or the T0--T5 ranges. However, in contrast to the \texttt{Sonora Diamondback}-based analysis, the \texttt{SM08}-based $f_{\rm sed}$ values also suggest a statistically significant age dependence across the full L0--L9 range, with a $p$-value of 0.015 and a KS statistic of 0.32. 

In summary, our KS tests reveal an age dependence in cloud sedimentation efficiency at the late-L end of the L/T transition: younger L4--L9 benchmarks have systematically lower $f_{\rm sed}$ values than their older counterparts, whereas no such trend is observed among T0--T5 objects. Because age and surface gravity are tightly correlated for brown dwarfs, this age-dependent pattern also points to an underlying dependence of cloud properties on surface gravity. Our results directly demonstrate that the cloud properties vary with age and surface gravity, offering an explanation for the observed gravity-dependent photometric properties of brown dwarfs and self-luminous exoplanets near the L/T transition \citep[e.g.,][]{2006ApJ...651.1166M, 2007ApJ...654..570L, 2011ApJ...735L..39B, 2015ApJ...810..158F, 2016ApJ...833...96L, 2023ApJ...959...63S, 2023AJ....166..198Z}. Moreover, as noted by \citet{2020ApJ...891..171Z}, such age dependence of photometry is most pronounced among mid- to late-L dwarfs and diminishes toward mid-T types, closely matching the spectral type ranges over which we do and do not observe the age dependence in $f_{\rm sed}$. 

In addition, our findings are consistent with the results of \cite{2023MNRAS.523.4739S}, who showed that the 8--11~$\mu$m absorption features of silicate clouds exhibit systematically different shapes between low-gravity ($\log{(g)} \lesssim 4.5$~dex) and high-gravity ($\gtrsim 5$~dex) brown dwarfs across L3--L7 types. In particular, low-gravity objects tend to have larger dust grains composed of heavier molecules, whereas such clouds have largely dissipated in high-gravity dwarfs, implying that low-gravity L dwarfs have systematically lower $f_{\rm sed}$.

\begin{figure*}
\begin{center}
\includegraphics[width=7.in]{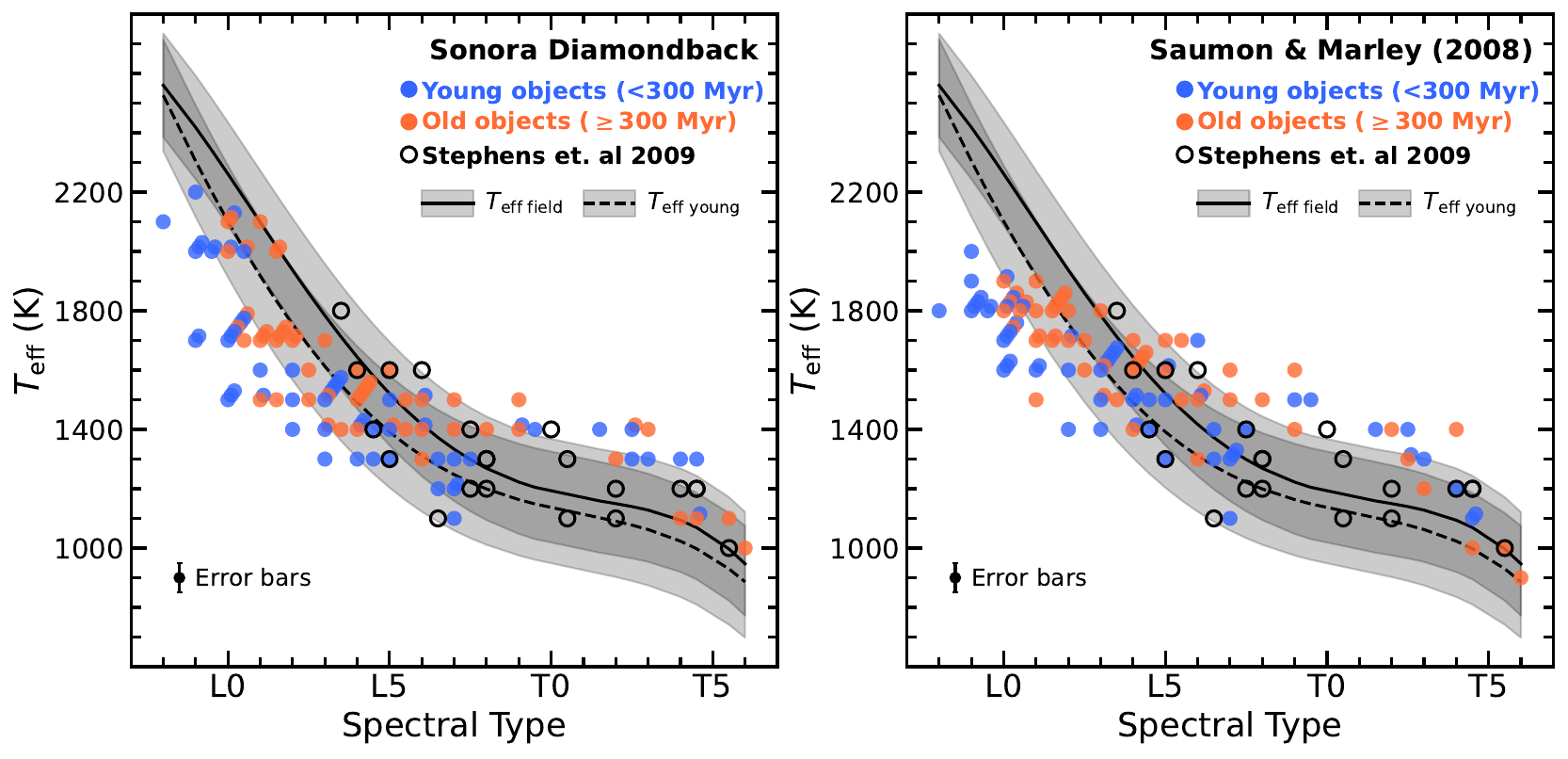}
\caption{Our best-fit effective temperatures for young ($<300$~Myr; blue) and old ($\geqslant 300$~Myr; orange) benchmarks based on the \texttt{Sonora Diamondback} models (left) and the \texttt{SM08} models (right). The empirical $T_{\rm eff}$--spectral type relation established for young and field-age objects by \cite{2023ApJ...959...63S} is shown as dashed and solid lines, respectively. Spectral fitting results from \cite{2009ApJ...702..154S} are overlaid as open black circles. For targets with identical spectral types and best-fit $T_{\rm eff}$ values, data points are slightly offset toward the upper right in these panels.}
\label{fig:teff_spt}
\end{center}
\end{figure*}

\subsection{$T_{\rm eff}$ vs. Spectral Type}
\label{subsec:teff_spt}

Figure~\ref{fig:teff_spt} presents the spectroscopically inferred $T_{\rm eff}$ values derived using both the \texttt{Sonora Diamondback} and \texttt{SM08} atmospheric models, shown as a function of spectral type. These results are compared with the empirical $T_{\rm eff}$--spectral type relations from \cite{2023ApJ...959...63S}. As summarized in Section~5.3 of \cite{2021ApJ...921...95Z}, the empirical relations are primarily established using thermal evolution models. Therefore, discrepancies between our fitted $T_{\rm eff}$ values and the empirical relations reflect differences between atmospheric and evolution model predictions, with the latter generally presumed to be relatively more reliable (Sections~\ref{sec:intro} and \ref{subsec:calib}). 

For benchmarks older than 300~Myr, the $T_{\rm eff}$ values inferred from both atmospheric model grids exhibit a level of scatter comparable to that of the empirical field-age relation but show systematic offsets as a function of spectral type. In particular, the fitted $T_{\rm eff}$ values are systematically cooler by up to about 300~K at L0--L4 types and hotter by up to 350~K at L9--T4 types. These offsets suggest the presence of non-negligible systematic errors in the fitted $T_{\rm eff}$ from atmospheric models when applied to low-resolution ($R\sim$80--250) 0.8--2.5~$\mu$m spectra. Such deviations are less pronounced when comparing the fitted $T_{\rm eff}$ from \cite{2009ApJ...702..154S} to the same $T_{\rm eff}$--spectral type empirical relation (Figure~\ref{fig:teff_spt}). In that study, the authors used customized versions of the \texttt{SM08} models to fit the spectrophotometry of 14 mid-L to mid-T dwarfs over a significantly broader wavelength range (0.8--14.5~$\mu$m). Such broad wavelength coverage likely helps mitigate modeling systematic errors that are more severe when fitting only the narrower wavelength coverage of IRTF/SpeX spectra used in this work.

For L-type benchmarks younger than 300~Myr, both \texttt{Sonora Diamondback} and \texttt{SM08} atmospheric models similarly infer systematically cooler $T_{\rm eff}$ values compared to older benchmarks at the same spectral types. This is consistent with the known age- and gravity-dependent atmospheric properties of L dwarfs \citep[e.g.,][]{2006ApJ...651.1166M, 2007ApJ...654..570L, 2013ApJ...774...55B, 2013ApJ...777L..20L, 2015ApJ...810..158F, 2016ApJ...833...96L, 2016ApJS..225...10F, 2023ApJ...959...63S, 2023AJ....166..198Z}. Additionally, relative to the empirical $T_{\rm eff}$--spectral type relation for young objects from \citet{2023ApJ...959...63S}, the fitted $T_{\rm eff}$ values of these young benchmarks exhibit increased scatter and larger negative offsets, reaching up to about 600~K. 

Taken together, the systematic discrepancies between spectroscopically inferred $T_{\rm eff}$ values and empirical relations, observed for both old and young benchmarks, demonstrate that atmospheric spectral fits are subject to significant, model-dependent systematics that persist across both \texttt{Sonora Diamondback} and \texttt{SM08} atmospheric models. In the following subsection, we quantify the modeling systematic errors of these model grids by systematically comparing the atmospheric and evolution model predictions.

\begin{figure*}
\begin{center}
\includegraphics[width=6.in]{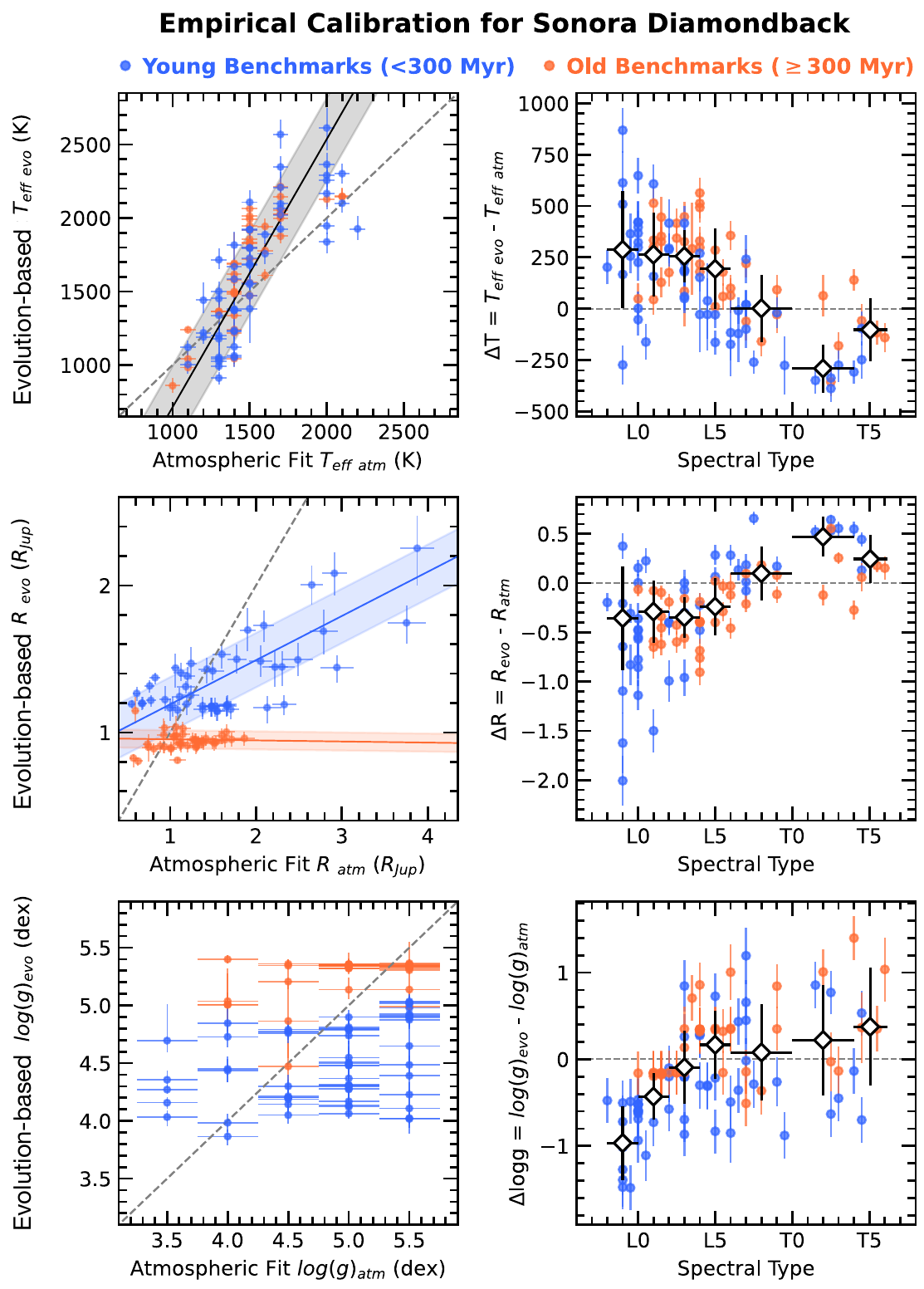}
\caption{{\it Left panels:} Comparison between spectroscopically inferred values of $T_{\rm eff}$ (top), $R$ (middle), and $\log{(g)}$ (bottom), based on \texttt{Sonora Diamondback} atmospheric models, with those derived from evolution models. Benchmarks younger and older than 300~Myr are shown in blue and orange, respectively. Linear fits to the atmospheric-evolution parameter relationships are overlaid, including a single fit of the full sample for $T_{\rm eff}$, and separate fits for $R$ of the young and old subsets. {\it Right panels:} Differences between atmospheric- and evolution-based parameters are shown as a function of spectral type. Open diamonds indicate the mean and RMS of the parameter differences within individual spectral type bins from M8 to L6, as summarized in Table~\ref{tab:evo_calb}.}
\label{fig:evo_calib_sdb}
\end{center}
\end{figure*}

\begin{figure*}
\begin{center}
\includegraphics[width=6.in]{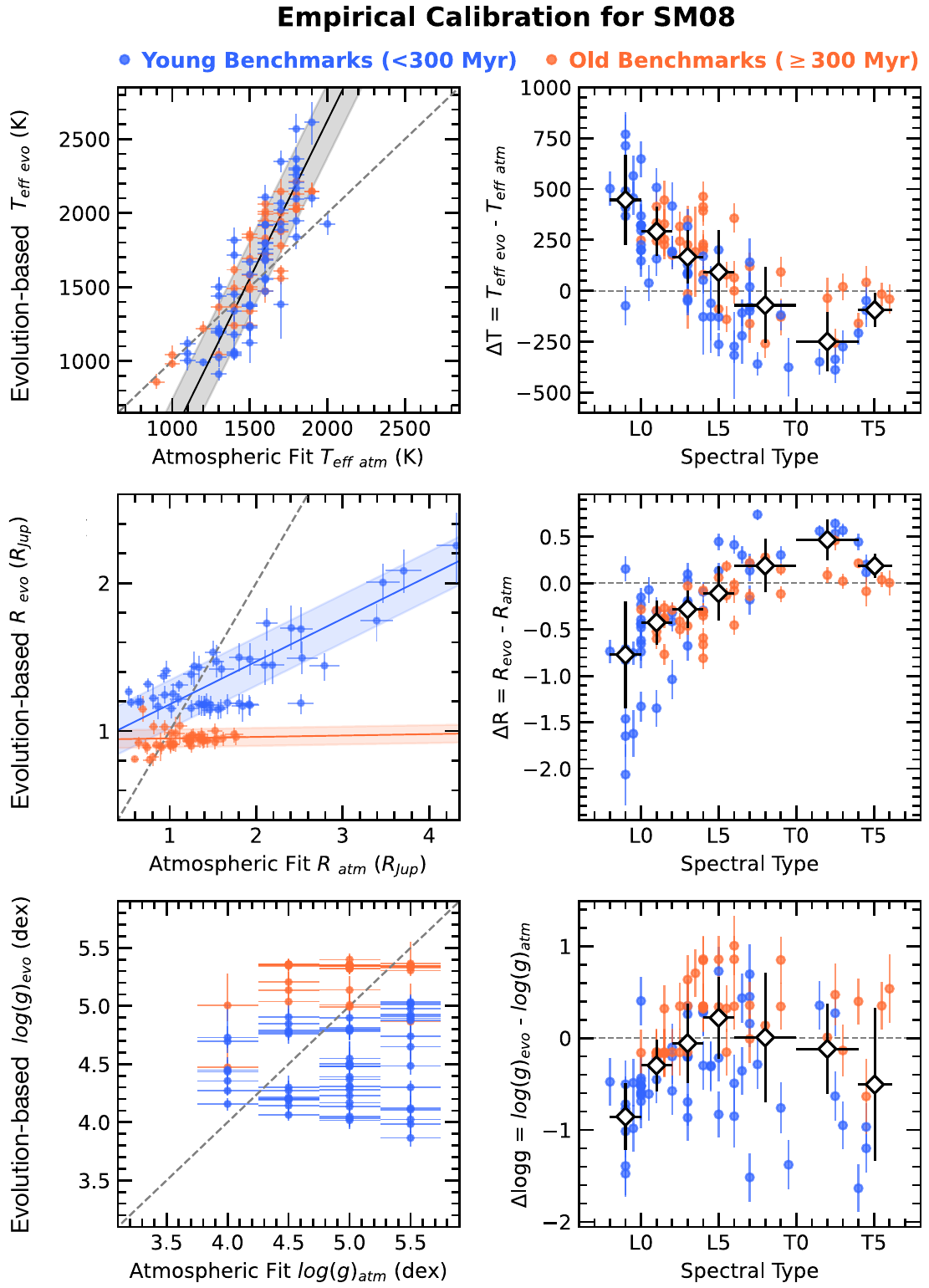}
\caption{Comparison between spectroscopically inferred values of $T_{\rm eff}$ (top), $R$ (middle), and $\log{(g)}$ (bottom), based on \texttt{SM08} atmospheric models, with those derived from the \texttt{Sonora Diamondback} evolution models. These panels follow the same format as Figure~\ref{fig:evo_calib_sdb}. }
\label{fig:evo_calib_sm08}
\end{center}
\end{figure*}

\subsection{Empirical calibration of properties inferred by \texttt{Sonora Diamondback} and \texttt{SM08} atmospheric models}
\label{subsec:calib}

Benchmark brown dwarfs and planetary-mass objects with independently determined ages offer valuable opportunities to empirically calibrate the atmospheric parameters inferred from spectral fitting against predictions from thermal evolution models. Although evolution models are not entirely free of systematic uncertainties \citep[e.g.,][]{2009ApJ...692..729D, 2014ApJ...790..133D, 2020AJ....160..196B}, their predictions may be regarded as more reliable than those derived from atmospheric spectral fitting. This is because evolution models are less sensitive to wavelength-dependent systematics in synthetic spectra of atmospheric models, such as uncertainties in molecular opacities and line lists and treatments of cloud effects, which can greatly impact the accuracy of model fits to the observed spectra. Moreover, evolution-based parameters are constrained using independently known ages of benchmarks, providing external anchors that are not available in spectral fitting. Consequently, discrepancies between atmospheric- and evolution-based parameters provide a powerful means to quantify systematic errors in atmospheric model predictions (e.g., \citealt{2006MNRAS.368.1281P, 2008ApJ...689..436L, 2020ApJ...891..171Z, 2021ApJ...916...53Z, 2023ApJ...959...63S, 2024ApJ...961..121H}). 

As shown in Figures~\ref{fig:evo_calib_sdb} and \ref{fig:evo_calib_sm08}, we compare the spectroscopically inferred values of $T_{\rm eff}$, $R$, and $\log{(g)}$ --- derived from both the \texttt{Sonora Diamondback} and \texttt{SM08} models (Section~\ref{sec:FM}) --- with corresponding parameters inferred from the \texttt{Sonora Diamondback} evolution models (Section~\ref{sec:evo}). We summarize the comparison results for each parameter, provide empirical calibration relations, and quantify the discrepancies as a function of spectral type. These results provide key context for interpreting \texttt{Sonora Diamondback}-based and \texttt{SM08}-based spectral fits in future studies.

\subsubsection{Effective temperature}
\label{subsubsec:teff}

The relationship between $T_{\rm eff}$ values derived from atmospheric ($T_{\rm eff,atm}$) and evolution ($T_{\rm eff,evo}$) models is well described by linear relations, derived using orthogonal distance regression (via \texttt{scipy.odr}) with uncertainties propagated in both axes. For $T_{\rm eff}$ values inferred from the \texttt{Sonora Diamondback} atmospheric models (Figure~\ref{fig:evo_calib_sdb}), we obtain 
\begin{equation}
\label{eq:teff_calib_sdb}
\begin{aligned} 
T_{\rm eff, evo} &= 1.82 \times T_{\rm eff, atm}\ {\rm (SDB)} - 1110.74\ {\rm K}, \\
&{\rm (RMS} = 280 {\rm K}, \ T_{\rm eff,atm}\ {\rm (SDB)} \in [1000, 2200]\ {\rm K)}
\end{aligned} 
\end{equation}
For $T_{\rm eff}$ values inferred from the \texttt{SM08} atmospheric models (Figure~\ref{fig:evo_calib_sm08}), we find
\begin{equation}
\label{eq:teff_calib_sm08}
\begin{aligned} 
T_{\rm eff, evo} &= 2.13 \times T_{\rm eff, atm}\ {\rm (SM08)} - 1639.37\ {\rm K}, \\
&{\rm (RMS} = 249 {\rm K}, \ T_{\rm eff,atm}\ {\rm (SM08)} \in [900, 2200]\ {\rm K)}
\end{aligned} 
\end{equation}

Additionally, the temperature difference $\Delta T_{\rm eff} = T_{\rm eff, evo} - T_{\rm eff, atm}$ exhibits similar trends as a function of spectral type for both \texttt{Sonora Diamondback} and \texttt{SM08} model grids. Table~\ref{tab:evo_calb} summarizes the mean values and root-mean-squares (RMSs) of $\Delta T_{\rm eff}$ in spectral type bins from M8 to T6. Near L0, the fitted $T_{\rm eff}$ values are 100--800~K cooler than evolution-based estimates, with offsets up to 500~K across the L0--L5 range. The atmospheric- and evolution-based $T_{\rm eff}$ are generally consistent over L5 to T0 (i.e., $\Delta T_{\rm eff} \approx 0$), while $T_{\rm eff, atm}$ becomes systematically hotter than $T_{\rm eff,evo}$ by up to 350~K for T0--T5 types (Table~\ref{tab:evo_calb}). 

At T0--T5 types, our results connect smoothly with those reported for the late-T regime reported by \cite{2020ApJ...891..171Z} and \cite{2021ApJ...916...53Z}, who analyzed T4--T9 benchmarks and found that spectral fits using the \texttt{Sonora Bobcat} atmospheric models \citep{2021ApJ...920...85M} yield $T_{\rm eff}$ values that are 0--150~K hotter than evolution-based estimates. These \texttt{Sonora Bobcat} models were developed by the same group as \texttt{Sonora Diamondback} but assume cloud-free atmospheres. Furthermore, unlike the results of \cite{2024ApJ...961..121H}, who reported clustering of best-fit $T_{\rm eff}$ values near 1800~K for M8--L6 dwarfs using \texttt{BT-Settl} models \citep{2012RSPTA.370.2765A, 2015A&A...577A..42B}, our results with \texttt{Sonora Diamondback} and \texttt{SM08} show no such clustering behavior.

\begin{figure*}
\begin{center}
\includegraphics[width=7.in]{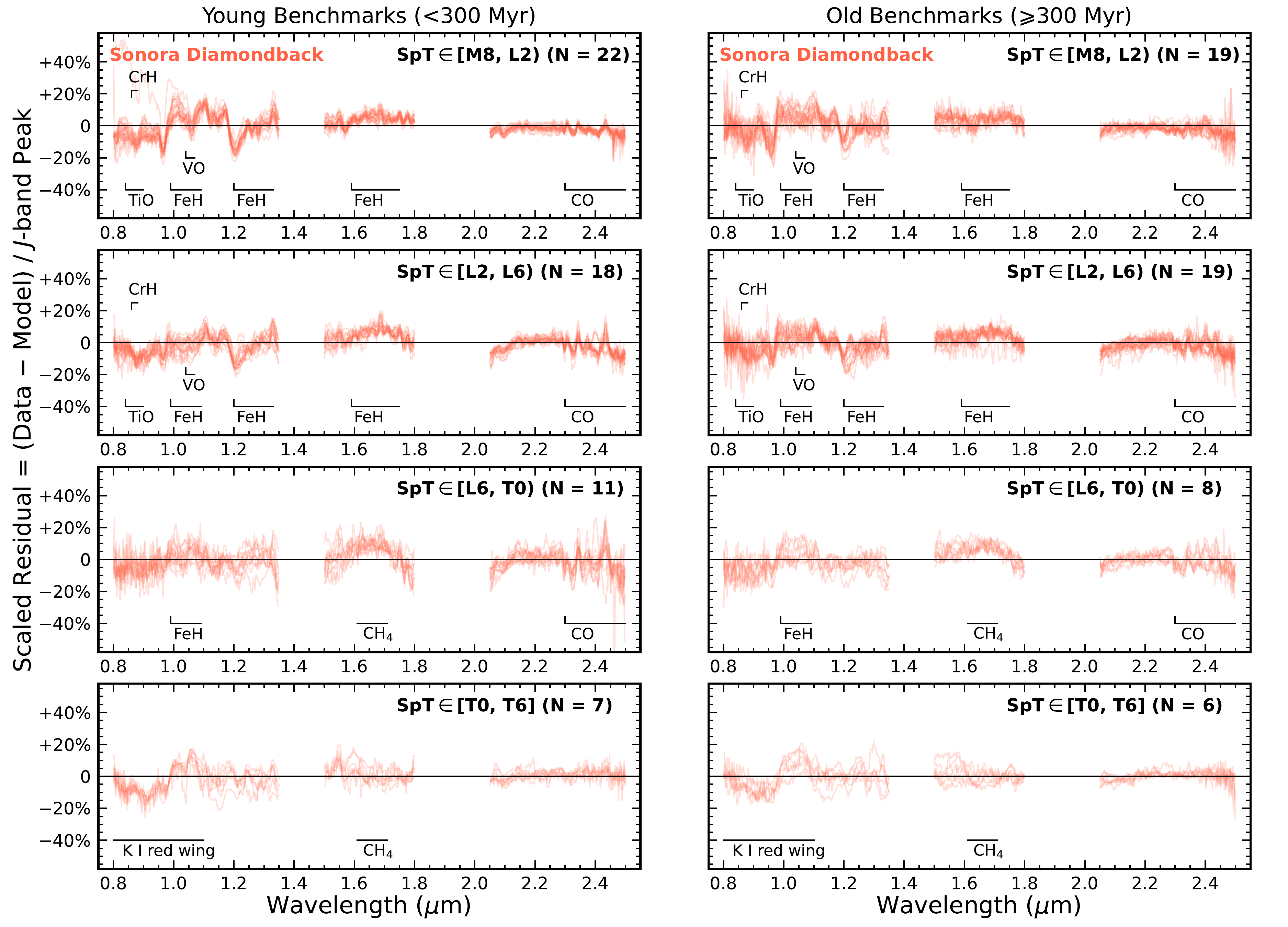}
\caption{Stacked spectral residuals of our young ($<300$~Myr; left) and old ($\geqslant 300$~Myr; right) benchmark targets with M8--L2, L2--L6, L6--T0, and T0--T6 spectral types based on \texttt{Sonora Diamondback} model fits. We highlight the absorption bands of key atmospheric species whose wavelengths line up with several prominent residual features.}
\label{fig:bin_res_sdb}
\end{center}
\end{figure*}

\subsubsection{Radius}
\label{subsubsec:radius}

The relationship between atmospheric- and evolution-based $R$ bifurcates into two distinct linear trends depending on age. For young ($<300$~Myr) and old ($\geqslant 300$~Myr) benchmarks, we derive the following empirical calibrations for radii inferred from the \texttt{Sonora Diamondback} atmospheric models (Figure~\ref{fig:evo_calib_sdb}):
\begin{equation} 
\label{eq:calib_R_sdb}
\begin{aligned} 
R_{\rm evo} &{\rm (young)} = 0.301 \times R_{\text{atm}}\ {\rm (SDB)} + 0.890\ {\rm R_{\rm Jup}}, \\
&{\rm (RMS} = 0.185 {\rm R_{\rm Jup}}, \ R_{\text{atm}}\ {\rm (SDB)} \in [0.551, 3.88]\ {\rm {\rm R_{\rm Jup}})} \\
R_{\rm evo} &{\rm (old)} = -0.008 \times R_{\text{atm}}\ {\rm (SDB)} + 0.961\ {\rm R_{\rm Jup}},  \\
&{\rm (RMS} = 0.062 {\rm R_{\rm Jup}}, \ R_{\text{atm}}\ {\rm (SDB)} \in [0.571, 1.86]\ {\rm {\rm R_{\rm Jup}})} \\
\end{aligned} 
\end{equation} 
For radii inferred from the \texttt{SM08} atmospheric models (Figure~\ref{fig:evo_calib_sm08}), we obtain
\begin{equation} 
\label{eq:calib_R_sm08}
\begin{aligned} 
R_{\rm evo} &{\rm (young)} = 0.290 \times R_{\text{atm}}\ {\rm (SM08)} + 0.891\ {\rm R_{\rm Jup}}, \\
&{\rm (RMS} = 0.164 {\rm R_{\rm Jup}}, \ R_{\text{atm}}\ {\rm (SM08)} \in [0.527, 4.32]\ {\rm {\rm R_{\rm Jup}})} \\
R_{\rm evo} &{\rm (old)} = -0.009 \times R_{\text{atm}}\ {\rm (SM08)} + 0.942\ {\rm R_{\rm Jup}},  \\
&{\rm (RMS} = 0.061 {\rm R_{\rm Jup}}, \ R_{\text{atm}}\ {\rm (SM08)} \in [0.596, 1.77]\ {\rm {\rm R_{\rm Jup}})} \\
\end{aligned} 
\end{equation} 

The relatively flat relation for older benchmarks reflects the expected similar radii around $1$~R$_{\rm Jup}$ for ultracool dwarfs at field ages. Table~\ref{tab:evo_calb} summarizes the radius difference, $\Delta R = R_{\rm evo} - R_{\rm atm}$, across spectral type bins for both \texttt{Sonora Diamondback} and \texttt{SM08} model grids.

\subsubsection{Logarithmic surface gravity}
\label{subsubsec:logg}

The atmospheric- and evolution-based estimates of $\log{(g)}$ show significant scatter across the benchmark sample for both the \texttt{Sonora Diamondback} and \texttt{SM08} model grids. Therefore, no meaningful polynomial calibration relations can be derived. Nonetheless, we report the typical difference, $\Delta \log{(g)} = \log{(g)}_{\rm evo} - \log{(g)}_{\rm atm}$, as a function of spectral type in Table~\ref{tab:evo_calb}. The large dispersion in $\log{(g)}$ reflects the well-known difficulty of robustly constraining surface gravity through spectral fitting.

\subsubsection{Notes on spectroscopically inferred mass}
\label{subsubsec:mass}

The significant and systematic differences between atmospheric-based and evolution-based predictions of radius (Section~\ref{subsubsec:radius}) and $\log{(g)}$ imply that masses of brown dwarfs and planetary-mass objects --- derived from spectroscopic $R$ and $\log{(g)}$ --- should be treated with caution. According to Figures~\ref{fig:evo_calib_sdb}--\ref{fig:evo_calib_sm08}, when compared against evolution-model predictions, spectroscopically inferred mass estimates may be overestimated by up to an order of magnitude near the M/L transition and underestimated by up to an order of magnitude for T0--T5 dwarfs.

\begin{figure*}
\begin{center}
\includegraphics[width=7.in]{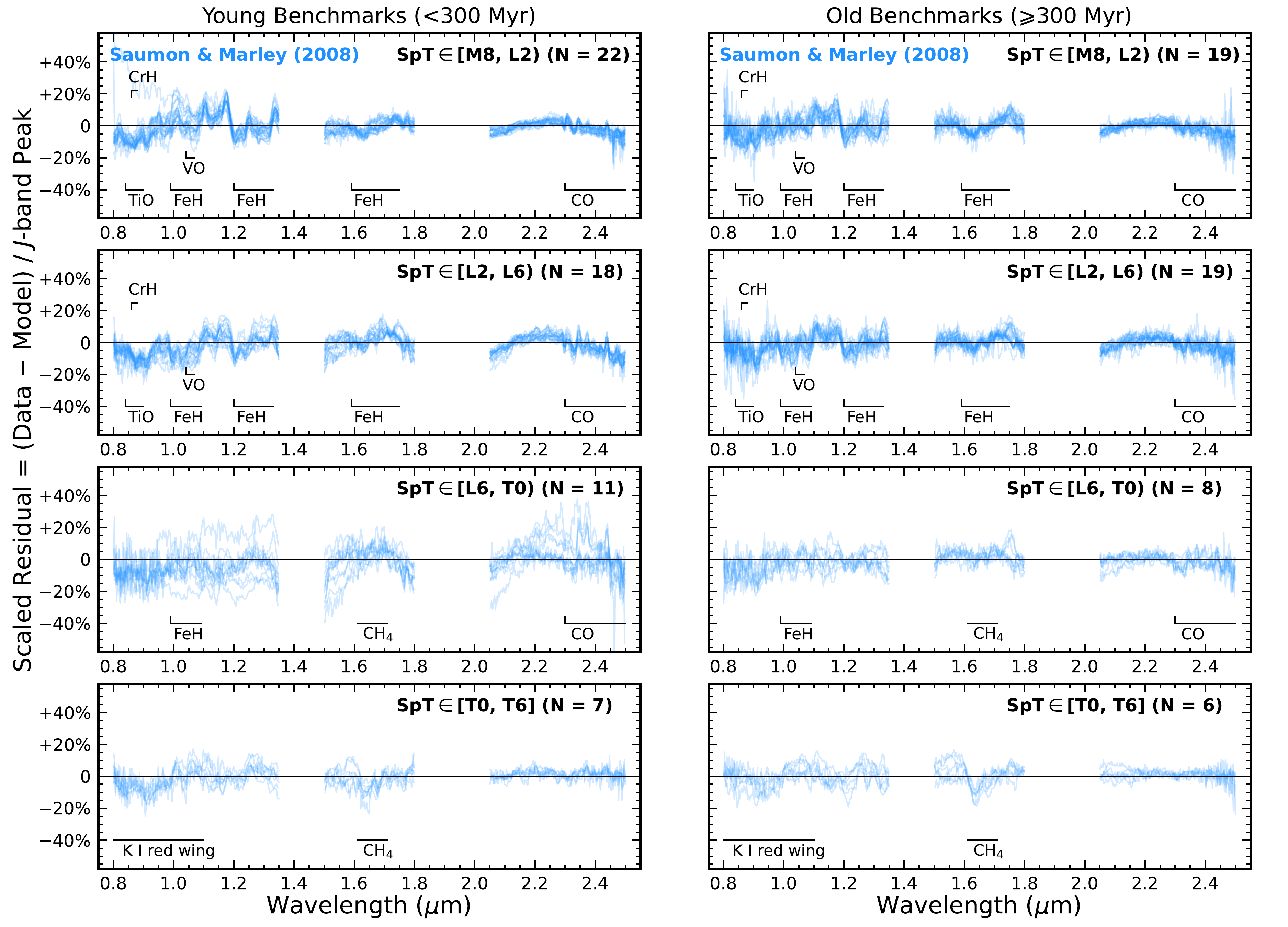}
\caption{Stacked spectral residuals of our young ($<300$~Myr; left) and old ($\geqslant 300$~Myr; right) benchmark targets based on \texttt{Sonora Diamondback} model fits. These panels follow the same format as Figure~\ref{fig:bin_res_sdb}.} 
\label{fig:bin_res_sm08}
\end{center}
\end{figure*}

\begin{figure*}
\begin{center}
\includegraphics[width=7. in]{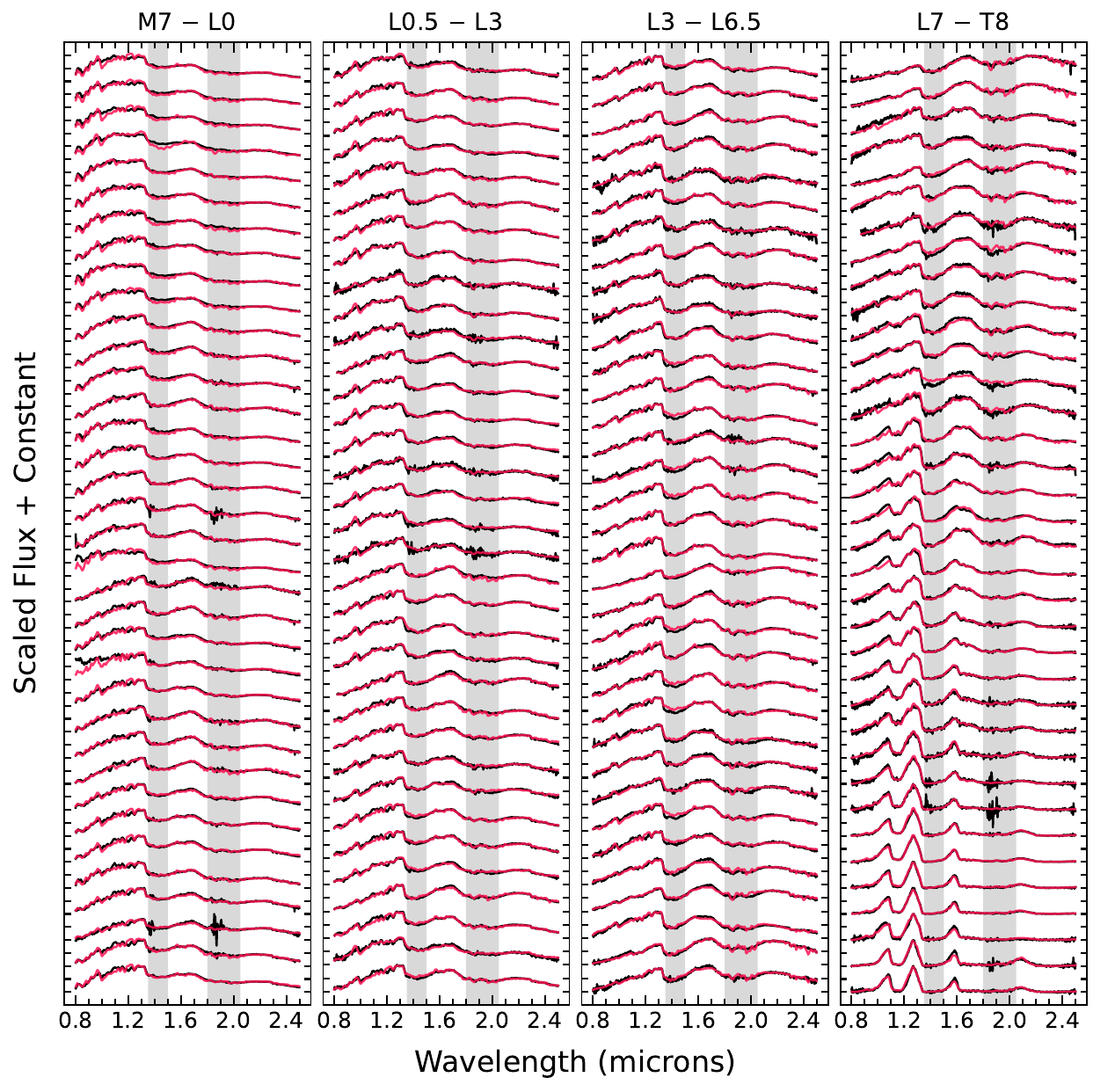}
\caption{The observed (black) and best-fit model spectra (purple), using the \texttt{Sonora Diamondback} models with an additional interstellar extinction term.}
\label{fig:ext_spec_sdb}
\end{center}
\end{figure*}

\subsection{Stacked Residuals Between Observed and Modeled Spectral Fluxes}
\label{subsec:stacked_residuals}

We compute and stack the residual spectra for the 111 M8--T6 benchmarks (after removing binaries and objects whose best-fit $T_{\rm eff}$ values lie at the boundaries of parameter space; Section~\ref{sec:FM}) to identify systematic discrepancies between observed and modeled spectral fluxes. This approach was first introduced by \cite{2021ApJ...921...95Z}, who applied it to a sample of 55 T7--T9 brown dwarfs to evaluate the performance of the \texttt{Sonora Bobcat} atmospheric models \citep{2021ApJ...920...85M}. Here, we extend their analysis to a much broader M8--T6 spectral type range using our benchmark sample and spectral fits based on both the  \texttt{Sonora Diamondback} (Section~\ref{subsubsec:sdb}) and \texttt{SM08} (Section~\ref{subsubsec:sm08}) atmospheric models.

\subsubsection{Residuals of the \texttt{Sonora Diamondback} Model Fits}
\label{subsubsec:sdb}

As shown in Figures~\ref{fig:bin_res_sdb}, the stacked residuals of the \texttt{Sonora Diamondback} model fits reveal persistent, wavelength-dependent mismatches between the observed and modeled spectral fluxes across the near-infrared. For M8--L2 benchmarks, the fitted model spectra tend to overpredict fluxes at 0.8--1.0~$\mu$m, 1.2--1.3~$\mu$m, and 2.3--2.5~$\mu$m, while underpredicting flux at 1.0--1.2~$\mu$m and 1.6--1.8~$\mu$m. These positive and negative residual structures align with major molecular absorption bands in ultracool atmospheres, including TiO, CrH, FeH, VO, and CO. Many of these features persist through the L2--L6 types, although those associated with TiO, VO, CrH, and FeH progressively weaken toward L6--T6 types, consistent with the depletion of these species from cooling near-infrared photospheres. 

For M8--L6 benchmarks, residuals in the 0.8--1.0~$\mu$m region suggest that the models underpredict the abundance of gas-phase TiO molecules, highlighting uncertainties in titanium condensation chemistry models in ultracool atmospheres \citep[e.g.,][]{1999ApJ...519..793L, 1999ApJ...512..843B, 2001ApJ...556..357A}. These objects also show positive residuals in FeH absorption bands near 1.0-1.1~$\mu$m and 1.6--1.8~$\mu$m, and negative residuals around 1.2--1.3~$\mu$m, pointing to systematic mismatches between the observed FeH opacities and those adopted by models. In particular, the adopted FeH opacities in the $H$ band \citep{2010AJ....140..919H} appear too weak to reproduce the observed fluxes. Since FeH spectral features are known to be sensitive to surface gravity \citep[e.g.,][]{2013ApJ...772...79A}, these systematic FeH-related residuals likely contribute to the difficulty of robustly constraining $\log{(g)}$ and mass from spectral fitting for late-M and L dwarfs (Figures~\ref{fig:evo_calib_sdb} and \ref{fig:evo_calib_sm08}).

For T0--T6 benchmarks, substantial residuals between 0.8 and 1.1~$\mu$m likely arise from the pressure-broadened wings of the K  \textsc{i} doublet in the optical, whose opacities and condensation models remain uncertain \citep[e.g.,][]{2020A&A...637A..38P, 2025AJ....169....9Z}. 

Additionally, in the $H$ band, the dominant source of residuals likely evolve with spectral type (Figure~\ref{fig:bin_res_sdb}). For M8--L6 objects, they are primarily linked to FeH absorption, but gradually transition to being shaped by CH$_{4}$ absorption at L6--T0 types. The fitted models underpredict the fluxes in the $H$ band, suggesting an overabundance of CH$_{4}$ in the modeled atmospheres. This feature diminishes among the T0--T6 benchmarks.

Finally, the overall residual patterns are broadly similar between young ($<300$~Myr) and old ($\geqslant 300$~Myr) benchmarks. However, young objects exhibit marginally stronger residuals at 1.2--1.3~$\mu$m and 1.6--1.8~$\mu$m across M8--L6 types. These features align with the gravity-dependent FeH absorption bands \citep[e.g.,][]{2013ApJ...772...79A}, highlighting the increased difficulty of accurately modeling low-gravity atmospheres within this spectral type range.

\begin{figure*}[t]
\begin{center}
\includegraphics[width=7in]{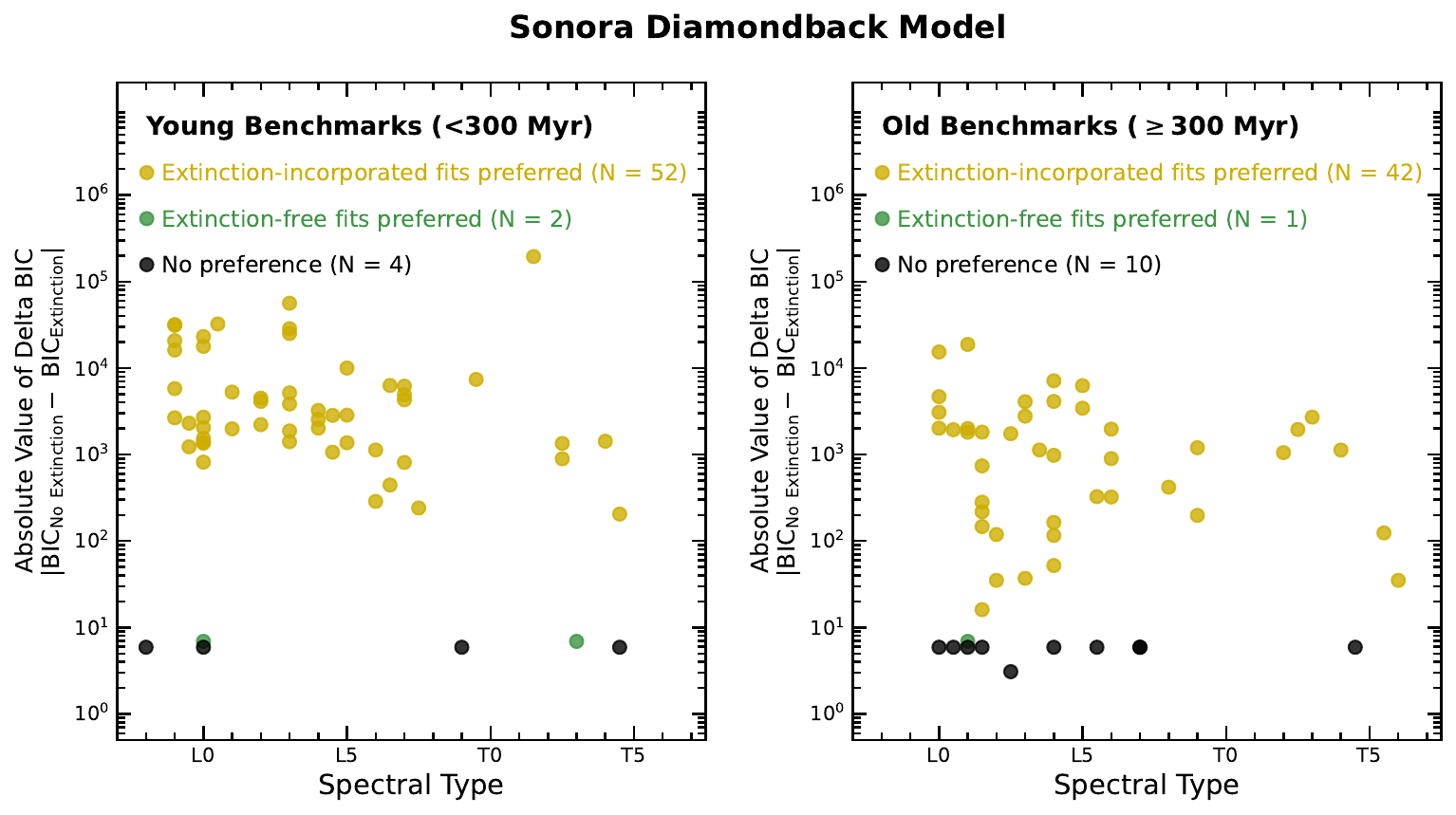}
\includegraphics[width=7in]{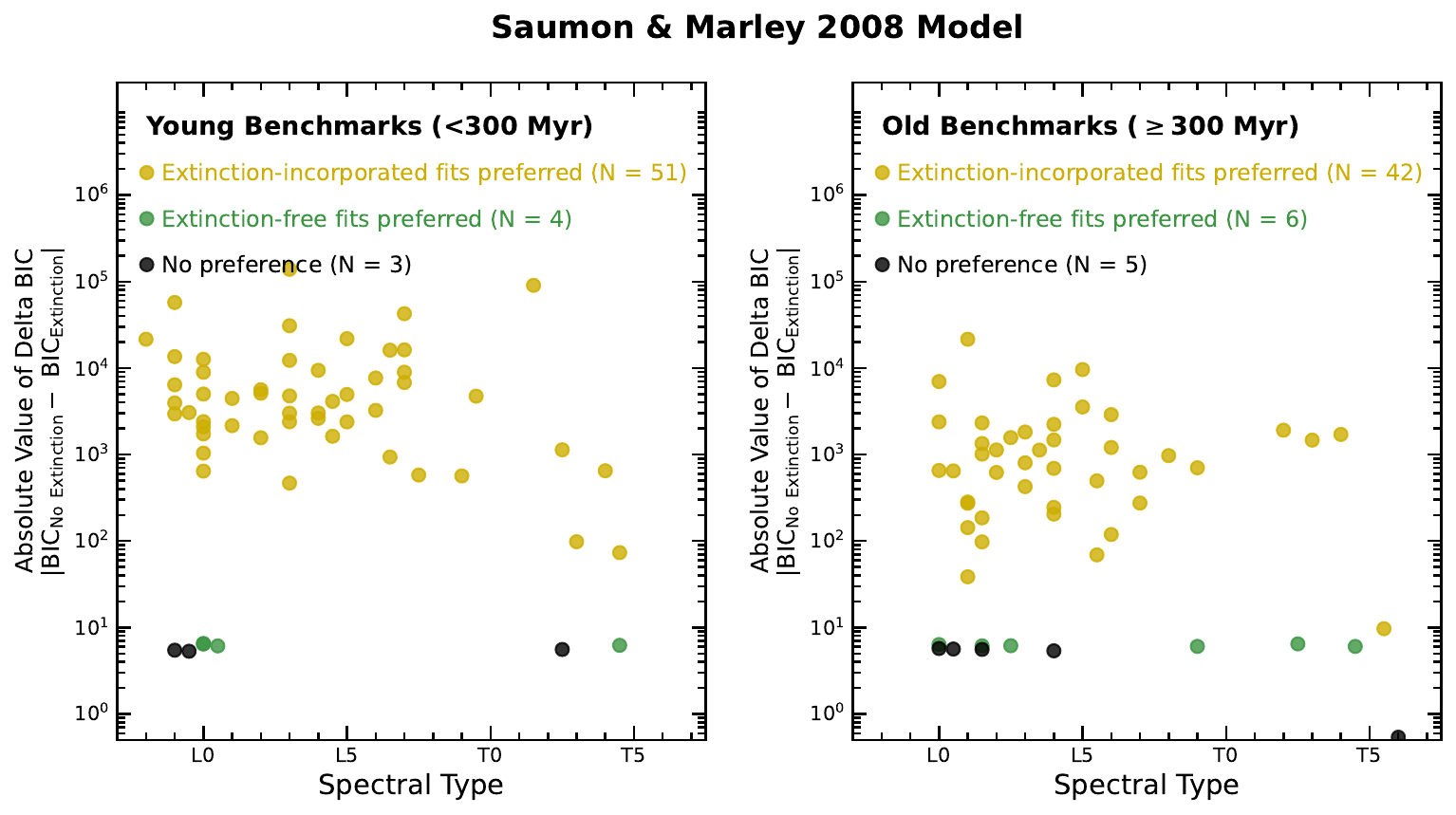}
\caption{Absolute difference in BIC between spectral fits performed with and without an additional reddening/extinction term, using the \texttt{Sonora Diamondback} (top) and \texttt{SM08} (bottom) atmospheric models, shown as a function of spectral type. Objects are shown in yellow where spectral fits including the extinction term are statistically preferred (i.e., BIC$_{\rm No\ Extinction} -$ BIC$_{\rm Extinction} > 6$), in green where fits excluding the extinction term are strongly preferred (i.e., BIC$_{\rm No\ Extinction} - $ BIC$_{\rm Extinction} < -6$), and in black where no statistically significant preference is found ($|$BIC$_{\rm No\ Extinction} -$ BIC$_{\rm Extinction} | < 6$).}
\label{fig:bic_ext_sdb}
\end{center}
\end{figure*}

\begin{figure*}[t]
\begin{center}
\includegraphics[width=7in]{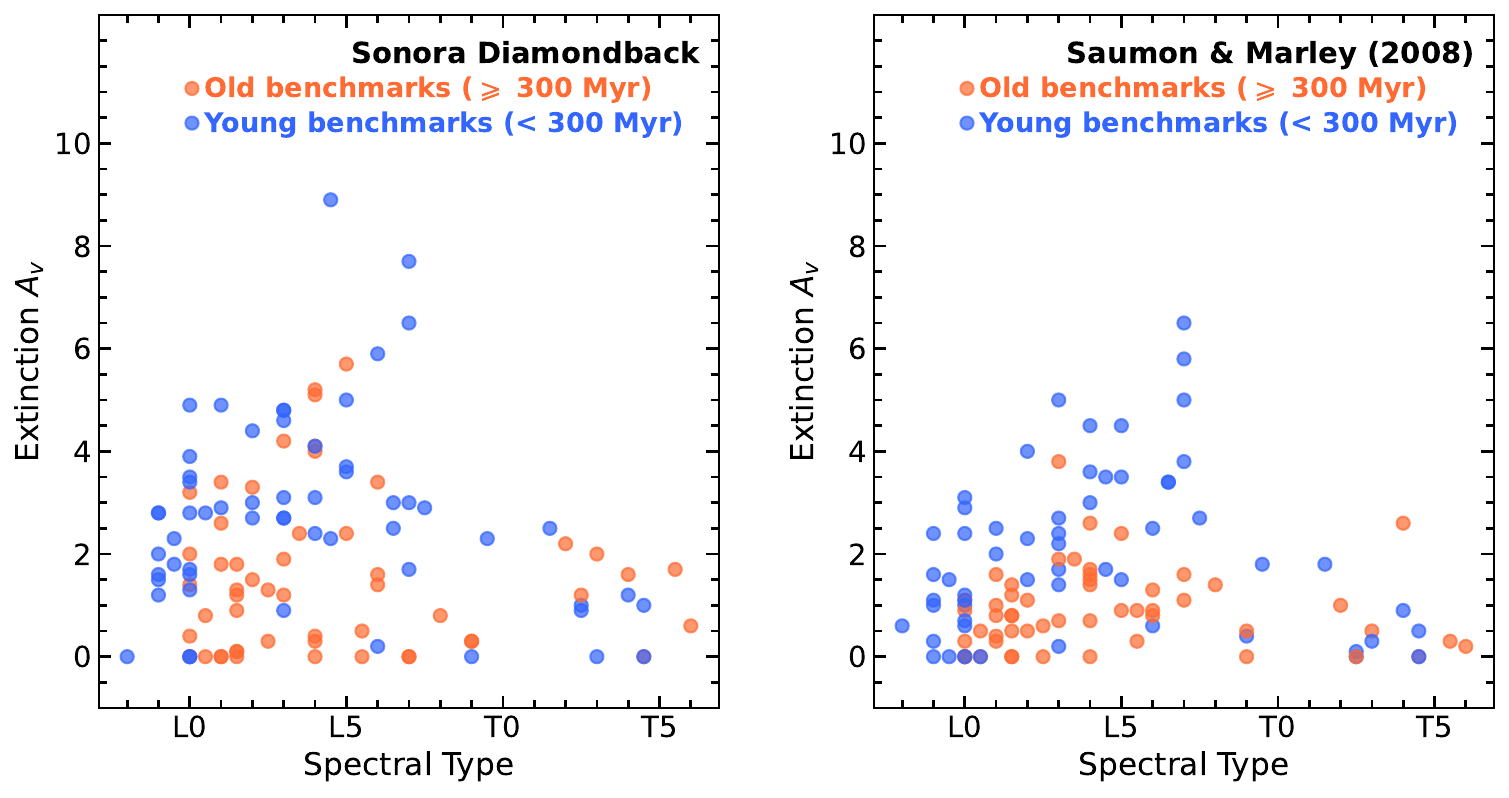}
\caption{Best-fit extinction coefficient from spectral fitting using the \texttt{Sonora Diamondback} (left) and the \texttt{SM08} (right) atmospheric models, plotted against spectral type. Young ($<300$~Myr) and old ($\geq$ 300 Myr) benchmarks are shown in blue and orange, respectively. }
\label{fig:ext_spt}
\end{center}
\end{figure*}

\subsubsection{Residuals of the \texttt{SM08} Model Fits}
\label{subsubsec:sm08}

As shown in Figure~\ref{fig:bin_res_sm08}, the stacked residuals of the \texttt{SM08} model fits exhibit similar wavelength-dependent features to those seen for the \texttt{Sonora Diamondback} models (Figure~\ref{fig:bin_res_sdb}). These features align with absorption bands of TiO, CrH, FeH, VO, CO, CH$_{4}$, and the pressure-broadened wings of the K \textsc{i} doublet. A notable difference is that the \texttt{SM08} fits exhibit a pronounced negative residual near 1.6~$\mu$m for both M8--L6 and T0--T6 benchmarks, which is likely associated with FeH absorption at earlier types and CH4 absorption at later types. This feature is absent in the \texttt{Sonroa Diamondback} fits, plausibly reflecting improvements in the adopted line lists of FeH \citep[][]{2010AJ....140..919H} and CH$_{4}$ \citep[][]{2014MNRAS.440.1649Y, 2020ApJS..247...55H, 2021ApJS..254...34G} in the more recent \texttt{Sonora Diamondback} models relative to \texttt{SM08}.

The wavelength-dependent residuals seen in both atmospheric model grids underscore the need to improve opacity databases and rainout chemistry for key atomic and molecular species in ultracool atmospheres, particularly in the context of cloud formation and dissipation, to inform future generations of atmospheric models.

\subsection{Repeated Forward-Modeling Analysis with an Additional Interstellar Extinction Term}
\label{subsec:ism}

Given the systematic discrepancies between the observed spectra and both the \texttt{Sonora Diamondback} and \texttt{SM08} model fits, we test whether introducing an interstellar extinction term as an additional free parameter in our spectral fits can help mitigate these mismatches. This approach was first explored by \citet{2020A&A...633A.124P}, who showed that adding such an extinction term improved the quality of fits for three young late-M dwarfs (5--11~Myr) with the \texttt{BT-Settl} atmospheric models \citep{2012RSPTA.370.2765A}. Subsequent work by \citet{2024ApJ...966L..11P} confirmed this finding for an L6 dwarf using multiple atmospheric models, and \citet{2024ApJ...961..121H} extended the analysis to a significantly larger sample of 90 young ($\sim 10-200$~Myr) M7--L9 dwarfs with \texttt{BT-Settl}. \cite{2024ApJ...961..121H} verified the conclusions in previous works and demonstrated that the fitted extinction term, quantified by the $V$-band extinction ($A_{V}$), increases systematically from late-M to L2 types and maintains high values near L2 ($A_{V} \approx 5$~mag). They attributed this trend to incomplete treatment of cloud effects in the \texttt{BT-Settl} models near the M/L boundary. In all these studies, as in our work, the extinction parameter, $A_{V}$, is not interpreted as genuine interstellar reddening --- particularly since the targets are not expected to have significant extinction from the interstellar medium (ISM) --- but rather as a simplified proxy for ``residual cloud effects'' that are not fully captured by the atmospheric models \citep[see also][]{2016ApJ...830...96H}. 

We repeat our forward-modeling analysis using both the \texttt{Sonora Diamondback} and \texttt{SM08} grids, incorporating the extinction law presented by \citet{2011ApJ...737..103S}. The procedure follows Section~\ref{sec:FM}, but we additionally multiply the model spectra by a factor of $10^{-0.4 \times A_{\lambda}}$ before evaluating Equation~\ref{G_k}. Here, $A_{\lambda}$ is computed from a grid of $A_{V}$ values spanning 0 to 10~mag in 0.1~mag increments, assuming $R_V = 3.1$. Figure~\ref{fig:ext_spec_sdb} presents the best-fit \texttt{Sonora Diamondback} model spectra of all 111 benchmarks, with the inferred parameters from both model grids summarized in Table~\ref{tab:ext}.

To evaluate the relative performance of the spectral fits with, and without the extinction term, we compute the Bayesian Information Criterion \citep[BIC;][]{1978AnSta...6..461S, 2007MNRAS.377L..74L} for the best-fit model of each scenario:
\begin{equation}
{\rm BIC} = \chi_{\rm min}^{2} + k \ln{N}
\label{eq:bic}
\end{equation}
where $\chi_{\rm min}^{2}$ is the $\chi^{2}$ value of the best-fit model, $k$ is the number of free parameters, and $N$ is the number of wavelength pixels in a given object's IRTF/SpeX spectrum. We compare the BIC values between two scenarios via $\Delta$BIC$=$BIC$_{\rm No\ Extinction} -$BIC$_{\rm Extinction}$. Following \cite{Kass1995}, we consider the spectral fits with (or without) the extinction term to be strongly preferred if $\Delta$BIC is above 6 (or below $-6$).

As shown in Figure~\ref{fig:bic_ext_sdb}, spectral fits that incorporate an interstellar-medium-like extinction term are strongly preferred for the majority of both young and old benchmarks, based on comparisons of BIC values. These results support the findings of \citet{2020A&A...633A.124P}, \citet{2024ApJ...966L..11P}, and \citet{2024ApJ...961..121H}, and expand them to the broader M8--T6 spectral type range. 

Moreover, as shown in Figure~\ref{fig:ext_spt}, the fitted $A_{V}$ values are systematically higher for L0--L7 dwarfs compared to those of M8--L0 and L8--T6 dwarfs. These high $A_{V}$ values suggest that additional reddening effects should be incorporated into the atmospheric models. This interpretation is consistent with recent studies that indicate the presence of high-altitude clouds in ultracool dwarf atmospheres \citep[e.g.,][]{2021ApJ...920..146L, 2025A&A...703A..79M, 2025AJ....170...64Z}, which are not currently included in the assumptions of \texttt{Sonora Diamondback} and \texttt{SM08} models. Moreover, among L0--L7 dwarfs, young benchmarks tend to require higher $A_{V}$ values ($\approx$2--9~mag based on \texttt{Sonora Diamondback} or $\approx$1--7~mag based on \texttt{SM08}) than their older counterparts ($\approx$0--6~mag based on \texttt{Sonora Diamondback} or $\approx$0--4~mag based on \texttt{SM08}). This further highlights the challenges of modeling clouds at younger ages and lower surface gravities.

Finally, we caution that the physical parameters derived from these extinction-included fits --- though yielding improved spectral fitting quality --- are based on artificially reddened and dimmed atmospheric models rather than self-consistent cloudy models. As such, we do not recommend adopting these parameters as the characteristics of the targets. The purpose of the analysis in this section is not to obtain more accurate atmospheric parameters, but rather to demonstrate the potential missing opacity sources and cloud physics in the \texttt{Sonora Diamondback} and \texttt{SM08 }model assumptions.

\section{Summary} 
\label{sec:summary}

We have systematically assessed the performance of two major grids of cloudy atmospheric models --- \texttt{Sonora Diamondback} and \texttt{SM08} --- by analyzing a uniform set of low-resolution ($R\sim$80--250) 0.8--2.5~$\mu$m spectra of 142 benchmark brown dwarfs and planetary-mass objects spanning the late-M, L, and T spectral sequence. These benchmarks include wide-orbit companions to stars and kinematic members of young moving groups, with independently determined ages ranging from 10~Myr to 10~Gyr. The main findings of our work are summarized as follows:

\begin{enumerate}
\item[1.] We derived the physical properties of all benchmark targets using the \texttt{Sonora Diamondback hybrid-grav} evolution models, based on their bolometric luminosities and independently known ages. These evolution-based properties --- $T_{\rm eff,evo}$, $\log{(g)_{\rm evo}}$, $R_{\rm evo}$, and $M_{\rm evo}$ --- serve as physically motivated anchors for evaluating the performance of atmospheric models.

\item[2.] We performed forward-model spectral fitting of the low-resolution ($R \approx$ 80--250) near-infrared (0.8--2.5~$\mu$m) spectra of all benchmarks using synthetic spectra from both the \texttt{SM08} and \texttt{Sonora Diamondback} atmospheric models, deriving $T_{\rm eff,atm}$, $\log{(g)_{\rm atm}}$, $R_{\rm atm}$, $M_{\rm atm}$, [M/H]$_{\rm atm}$, and $f_{\rm sed}$. For the majority of benchmarks, the parameters inferred from the two atmospheric model grids are typically consistent within one grid spacing. 

\item[3.] The best-fit cloud sedimentation efficiency, $f_{\rm sed}$, based on both \texttt{Sonora Diamondback} and \texttt{SM08} models increases systematically from M8 to T6 spectral types. We identify a statistically significant, population-level age dependence of $f_{\rm sed}$ among L4--L9 dwarfs, with young benchmarks ($<300$~Myr) in this spectral type range exhibiting systematically lower $f_{\rm sed}$ values than their older counterparts, with $p$-values of 0.009 and 0.027 based on KS tests applied to the \texttt{Sonora Diamondback} and \texttt{SM08} atmospheric model fits, respectively. No such age dependence is observed across the broader L0--T5 sample or within the T0--T5 subset. This trend directly demonstrates that the cloud properties vary with age and surface gravity at the late-L end of the L/T transition, offering an explanation for the observed gravity-dependent photometric properties of brown dwarfs at these types.

\item[4.] We identified systematic discrepancies between best-fit $T_{\rm eff}$ values and those predicted by empirical $T_{\rm eff}$--spectral type relations. These offsets point to non-negligible systematic errors in atmospheric spectral fits. To empirically calibrate the spectroscopically inferred parameters, we compared $T_{\rm eff}$, $\log{(g)}$, and $R$ derived from both \texttt{Sonora Diamondback} and \texttt{SM08} atmospheric models against those inferred from the \texttt{Sonora Diamondback} evolution models. Assuming the evolution-based parameters are more reliable, we derived empirical correction polynomials to convert fitted $T_{\rm eff,atm}$ and $R_{\rm atm}$ values to their calibrated ones (Equations~\ref{eq:teff_calib_sdb}--\ref{eq:calib_R_sm08}). Also, we computed typical offsets between atmospheric- and evolution-based parameters for $T_{\rm eff}$, $R$, and $\log{(g)}$ as a function of spectral type (Table~\ref{tab:evo_calb}). These calibrations provide practical means for anchoring future (and past) spectral fitting analyses using either the \texttt{Sonora Diamondback} or \texttt{SM08} atmospheric model grids.

\item[5.] By stacking residuals between the observed and modeled spectra across the full benchmark sample, we identified persistent, wavelength-dependent discrepancies that align with the absorption bands of key atomic and molecular species in ultracool atmospheres, including TiO, CrH, FeH, VO, CO, CH$_{4}$, and the pressure-broadened wings of the K~\textsc{i} doublet. For M8--L6 objects, atmospheric models tend to overpredict fluxes in the TiO absorption band at 0.8--1.0~$\mu$m, highlighting uncertainties in titanium condensation chemistry models. Also, the fitted models tend to underpredict fluxes near 1.0--1.1~$\mu$m and 1.6--1.8~$\mu$m, while overpredicting fluxes near 1.2--1.3~$\mu$m. These features are associated with FeH absorption bands, which are sensitive to surface gravity. These systematic FeH-related residuals likely contribute to the difficulty of robustly constraining $\log{(g)}$ and mass from spectral fitting for late-M and L dwarfs. For L6--T0 objects, the fitted models overpredict fluxes in the $H$-band, suggesting an overabundance of CH$_{4}$ in the modeled atmospheres. For T0--T6 types, substantial residuals between 0.8--1.1~$\mu$m arise from the pressure-broadened wings of the K~\textsc{i} doublet, whose opacities and condensation models remain uncertain. Together, these residuals highlight the need for improved opacity databases and more accurate treatments of rainout chemistry in future models, particularly in the context of cloudy atmospheres.

\item[6.] We find that the quality of spectral fits using both the \texttt{Sonora Diamondback} and \texttt{SM08} atmospheric models improves substantially for the vast majority of benchmarks when an interstellar extinction term is included as a free parameter in the forward-modeling analysis. Similar findings have been reported for late-M and L dwarfs with other model grids based on a handful of individual objects \citep[][]{2020A&A...633A.124P, 2024ApJ...966L..11P} and a larger sample \citep[90 objects;][]{2024ApJ...961..121H}. Our work extends these findings to a broader sample of 142 benchmarks spanning late-M, L, and T types. The fitted $A_{V}$ values are particularly high among L0--L7 dwarfs compared to earlier- and later-type benchmarks, suggesting that current models may be missing certain sources of opacity. Furthermore, within this spectral type range, young benchmarks ($<300$~Myr) tend to exhibit higher fitted $A_{V}$ values than older ones, highlighting the increasing complexity of modeling clouds at younger ages and lower surface gravities. 

\end{enumerate}

In this work, we focus on assessing the performance of cloudy atmospheric models when applied to low-resolution 0.8--2.5~$\mu$m spectra, a dataset that remains widely used in studies of brown dwarfs and directly imaged exoplanets. An important next step will be to extend this benchmarking analysis to spectra with different wavelength coverage and/or higher spectral resolution, using atmospheric model grids that adopt different assumptions (e.g., disequilibrium chemistry, cloud microphysics). Such analyses will quantify systematic errors in inferred atmospheric parameters in a manner tailored to different wavelength ranges, spectral resolutions, and model assumptions, thereby providing key context for future work that applies atmospheric models to datasets with similar properties. Notably, the growing population of ultracool dwarfs expected from ongoing and upcoming surveys \citep[e.g.,][]{2025ApJ...991...84D, 2025A&A...697A...7M, 2025arXiv250322497Z, 2025AJ....170..360Z} such as Vera C. Rubin Observatory’s Legacy Survey of Space and Time \citep[][]{2019ApJ...873..111I}, Euclid \citep[][]{2025A&A...697A...1E}, and SPHEREx \citep{2025arXiv251102985B} will substantially expand the available sample for such expanded benchmarking analysis in the near future.

\vspace{10pt}

We thank the anonymous referee for providing suggestions that improved the manuscript. Z.Z. acknowledges support from the NASA Hubble Fellowship grant HST-HF2-51522.001-A during the early stages of this work. This work has benefitted from The UltracoolSheet,\footnote{\url{http://bit.ly/UltracoolSheet}} maintained by Will Best, Trent Dupuy, Michael Liu, Aniket Sanghi, Rob Siverd, and Zhoujian Zhang, and developed from compilations by \cite{2012ApJS..201...19D}, \cite{2013Sci...341.1492D}, \cite{2016ApJ...833...96L}, \cite{2018ApJS..234....1B}, \cite{2021AJ....161...42B}, \cite{2023ApJ...959...63S}, and \cite{2023AJ....166..103S}.

\facilities{IRTF (SpeX)}

\software{\texttt{astropy} \citep{2013A&A...558A..33A, 2018AJ....156..123A, 2022ApJ...935..167A}, \texttt{ipython} \citep{PER-GRA:2007}, \texttt{numpy} \citep{2020NumPy-Array}, \texttt{scipy} \citep{2020SciPy-NMeth}, \texttt{matplotlib} \citep{Hunter:2007}.}

\appendix

\section{Parameter Comparison Between \texttt{hybrid-grav} and \texttt{hybrid} Versions of \texttt{Sonora Diamondback} Evolution Models}
\label{app:hybrid_grav_vs_hybrid}

\begin{figure*}[b!]
\begin{center}
\includegraphics[width=5.in]{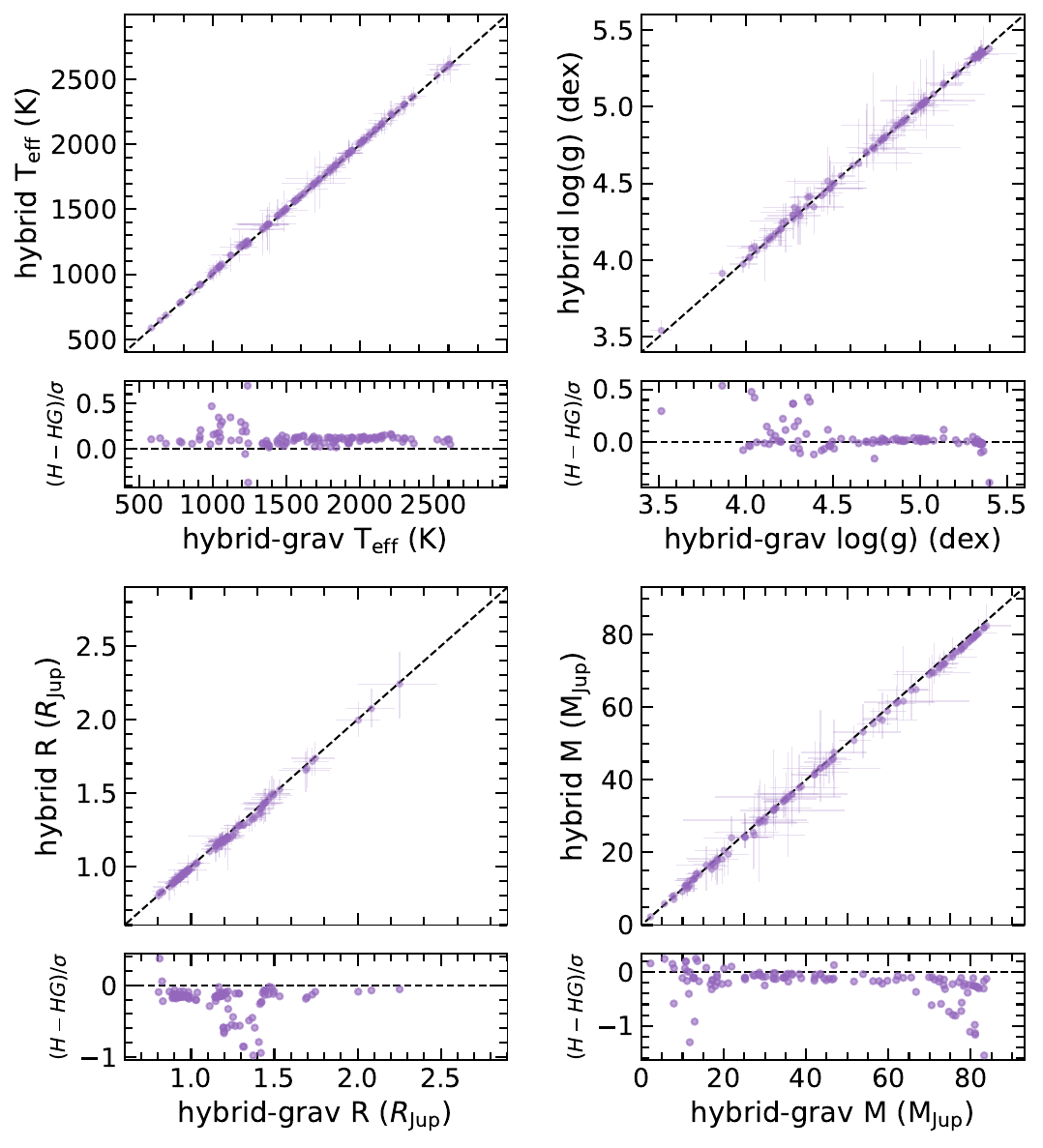}
\caption{Comparison of parameters derived from the \texttt{Sonora Diamondback} evolution model framework using the \texttt{hybrid} and \texttt{hybrid-grav} models. Each panel shows a 1:1 plot above its residual (\texttt{hybrid}-\texttt{hybrid-grav}) normalized by the standard deviation of the difference. Parameters shown are effective temperature (orange), surface gravity (blue), radius (green), and mass (pink).}
\label{fig:hybrid_grav_vs_hybrid}
\end{center}
\end{figure*}

Figure~\ref{fig:hybrid_grav_vs_hybrid} compares the physical parameters of our benchmark targets inferred by the \texttt{hybrid-grav} and \texttt{hybrid} versions of \texttt{Sonora Diamondback} evolution models. The averaged parameter differences (i.e., \texttt{hybrid-grav} minus \texttt{hybrid}) are $10 \pm 105$K in $T_{\rm eff}$, $0.004 \pm 0.148$ in $\log{(g)}$, $-0.02 \pm 0.10$ in $R$, and $-0.7 \pm 7.0$ in $M$.. These discrepancies are not statistically significant.

\section{Comparison of the Quality of Spectral Fits Between the \texttt{Sonora Diamondback} and \texttt{SM08} Models}
\label{app:compare}

\begin{figure*}[t]
\begin{center}
\includegraphics[width=7in]{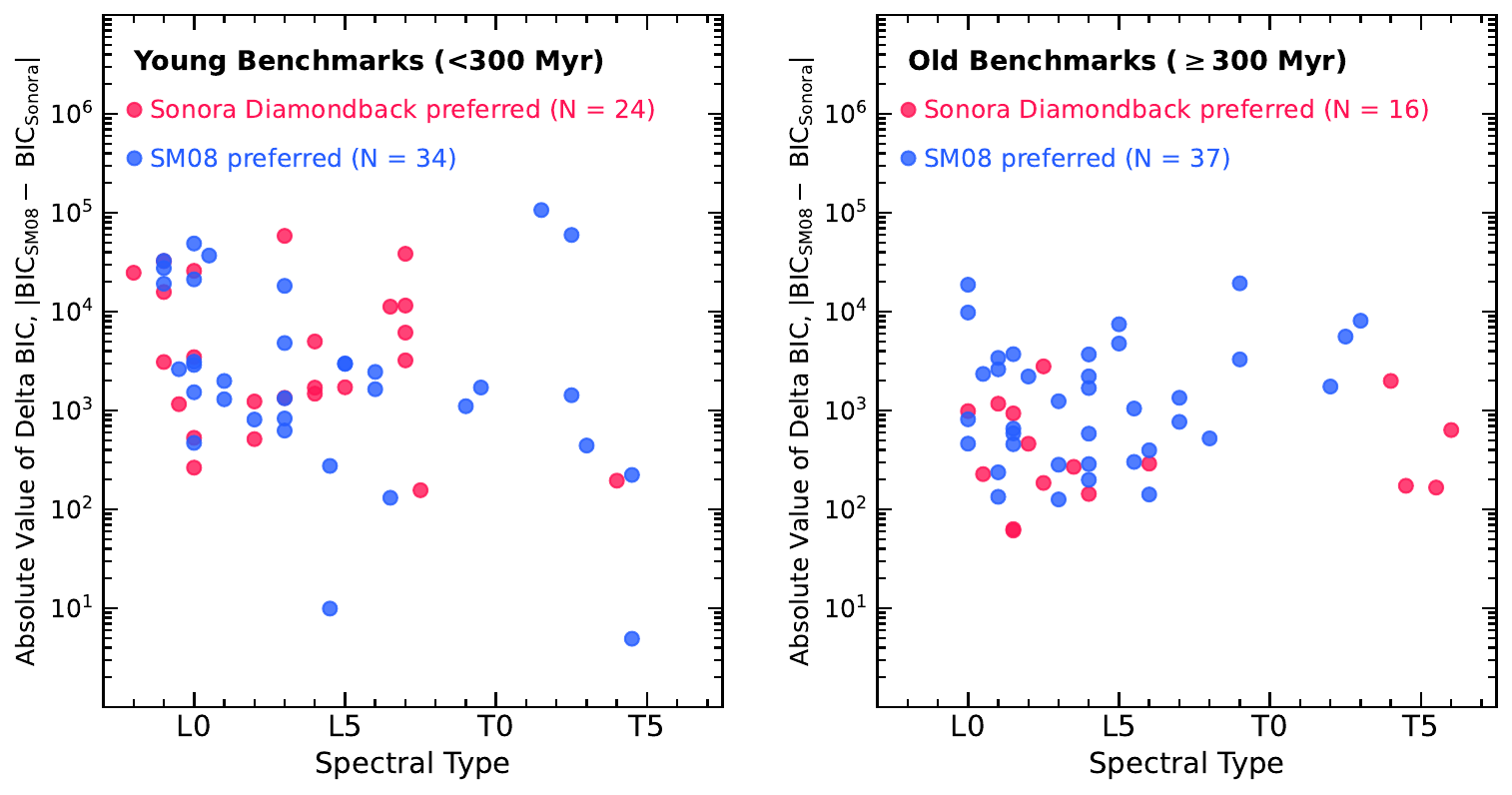}
\caption{Absolute values of $\Delta$BIC$=$BIC$_{\rm SM08} -$BIC$_{\rm Sonora}$ for benchmarks younger (left) and older (right) than 300~Myr. Red and blue points indicate objects for which the \texttt{Sonora Diamondback} and \texttt{SM08} models are preferred, respectively. }
\label{fig:compare_sonora_sm08}
\end{center}
\end{figure*}

We compare the relative performance of the \texttt{Sonora Diamondback} and \texttt{SM08} model grids across the full benchmark sample by computing the BIC value for the best-fit model from each grid (Equation~\ref{eq:bic}). For each target, we define $\Delta$BIC$=$BIC$_{\rm SM08} -$BIC$_{\rm Sonora}$. Following \cite{Kass1995}, we consider the \texttt{Sonora Diamondback} (or \texttt{SM08}) model grid to be strongly preferred if $\Delta$BIC is above 6 (or below $-6$). As shown in Figure~\ref{fig:compare_sonora_sm08}, the \texttt{Sonora Diamondback} models provide statistically better fits than \texttt{SM08} for only 40 out of 111 benchmark brown dwarfs (i.e., $36\%$ of the full sample). This result is noteworthy given that the \texttt{Sonora Diamondback} models incorporate substantial theoretical advances since the development of \texttt{SM08}, including updated atomic and molecular opacities. Our closer examination of the spectral fits for individual benchmarks reveals that neither \texttt{SM08} nor the \texttt{Sonora Diamondback} model grid provides a clearly superior fit to the observed spectra across all wavelength regions for most objects. 

Our results highlight the ongoing challenges in modeling cloudy atmospheres. Physical processes such as atmospheric inhomogeneity and the presence of high-altitude clouds \citep[e.g.,][]{2010ApJ...723L.117M, 2014ApJ...785..158R, 2020ApJ...903...15L, 2021ApJ...920..146L, 2025A&A...703A..79M, 2025AJ....170...64Z} --- effects that are not incorporated in \texttt{SM08} and \texttt{Sonora Diamondback} model grids --- may play important roles in shaping the observed spectra. These findings underscore the need for continued development of next-generation cloudy atmospheric models for exoplanets and brown dwarfs.

{
 
\begin{longrotatetable}

\begin{deluxetable*}{lcccccrccccccc}
\setlength{\tabcolsep}{3pt}

\tablecaption{Spectral type, astrometry, and photometry of benchmarks.} \label{tab:sample_dat}

\tablehead{ \multicolumn{1}{l}{Name\tablenotemark{\scriptsize a}} & \multicolumn{1}{c}{Disc. Ref} & \multicolumn{2}{c}{Spectral Type} & \multicolumn{1}{c}{} & \multicolumn{4}{c}{Astrometry} & \multicolumn{1}{c}{} & \multicolumn{4}{c}{Photometry}  \\
\cline{3-4} \cline{6-9} \cline{11-14} 
\multicolumn{1}{l}{} & \multicolumn{1}{l}{} & \multicolumn{1}{c}{SpT}  & \multicolumn{1}{c}{Ref.} & \multicolumn{1}{c}{} & \multicolumn{1}{c}{RA} & \multicolumn{1}{c}{DEC} & \multicolumn{1}{c}{Parallax} & \multicolumn{1}{c}{Ref. Plx.} & \multicolumn{1}{c}{} & \multicolumn{1}{c}{$J_{\rm MKO}$} & \multicolumn{1}{c}{$H_{\rm MKO}$} &  \multicolumn{1}{c}{$K_{\rm MKO}$} & \multicolumn{1}{c}{Ref.} \\
\multicolumn{1}{l}{} & \multicolumn{1}{l}{} & \multicolumn{1}{c}{}  & \multicolumn{1}{l}{} & \multicolumn{1}{c}{} & \multicolumn{1}{c}{(hh:mm:ss.s)} & \multicolumn{1}{c}{(dd:mm:ss.s)} & \multicolumn{1}{c}{(mas)} & \multicolumn{1}{l}{} & \multicolumn{1}{c}{} & \multicolumn{1}{c}{(mag)} & \multicolumn{1}{c}{(mag)} &  \multicolumn{1}{c}{(mag)} & \multicolumn{1}{l}{} }
\startdata
\multicolumn{14}{c}{Young Benchmarks (Age $<$ 300 Myr)} \\
\hline
1RXS J235133.3+312720 B$\star$$^{b}$ & 16 & L0 & 16 & & 23:51:33.9 & 31:27:23.0 & $23.13\pm0.02$ & 5 & & $15.333\pm0.100$ & $14.637\pm0.050$ & $13.930\pm0.050$ & 14;17 \\
2MASS J01415823-4633574$^{c}$ & 70 & L0 & 1;29 & & 01:41:58.2 & $-$46:33:57.2 & $27.30\pm0.40$ & 52 & & $14.782\pm0.043$ & $13.958\pm0.032$ & $13.065\pm0.032$ & 14 \\
2MASS J03231002-4631237 & 105 & L0 & 29 & & 03:23:10.0 & $-$46:31:23.5 & $23.60\pm0.80$ & 51 & & $15.350\pm0.070$ & $14.407\pm0.050$ & $13.670\pm0.050$ & 14 \\
2MASS J03264225-2102057$^{c}$ & 59 & L5 & 46 & & 03:26:42.3 & $-$21:02:06.0 &  $\dots$ & $\dots$  & & $15.970\pm0.010$ & $14.851\pm0.051$ & $13.854\pm0.051$ & 97;115 \\
2MASS J03552337+1133437 & 104 & L5 & 29 & & 03:55:23.4 & 11:33:43.9 & $109.60\pm0.70$ & 51 & & $13.940\pm0.010$ & $12.601\pm0.021$ & $11.502\pm0.001$ & 14;40;81 \\
2MASS J04433761+0002051 & 65 & M9 & 46 & & 04:43:37.6 & 00:02:05.3 & $47.41\pm0.19$ & 51 & & $12.447\pm0.026$ & $11.858\pm0.021$ & $11.176\pm0.021$ & 14 \\
2MASS J05184616-2756457 & 28 & L1 & 1;45 & & 05:18:46.2 & $-$27:56:45.6 & $17.30\pm0.80$ & 51 & & $15.188\pm0.043$ & $14.359\pm0.040$ & $13.573\pm0.040$ & 14 \\
\enddata
\end{deluxetable*} 
\end{longrotatetable}

}

\clearpage

{
\startlongtable
\begin{deluxetable*}{lccccccccc}

\setlength{\tabcolsep}{5pt}
\tablecaption{Physical properties of our sample based on the Sonora Diamondback \textsc{hybrid-grav} Thermal Evolution Models }\label{tab:evo}
\tablehead{\multicolumn{6}{c}{} & \multicolumn{4}{c}{Sonora Diamondback \textsc{hybrid-grav}} \\
\cline{7-10}
\multicolumn{1}{l}{Object} & \multicolumn{1}{c}{Age\tablenotemark{\scriptsize a}} & \multicolumn{1}{c}{Ref. Age} & \multicolumn{1}{c}{log($L_{bol}/L_{\odot}$)} & \multicolumn{1}{c}{Ref. $L_{bol}$} & \multicolumn{1}{c}{} & \multicolumn{1}{c}{$T_{eff}$} & \multicolumn{1}{c}{log(g)}  & \multicolumn{1}{c}{R} & \multicolumn{1}{c}{M}  \\
\multicolumn{1}{l}{} & \multicolumn{1}{c}{(Gyr)} & \multicolumn{1}{l}{} & \multicolumn{1}{c}{(dex)} & \multicolumn{1}{l}{} & \multicolumn{1}{l}{} & \multicolumn{1}{c}{(K)} & \multicolumn{1}{c}{(dex)} & \multicolumn{1}{c}{($R_{Jup}$)} & \multicolumn{1}{c}{($M_{Jup}$)}
}
\startdata
\multicolumn{10}{c}{Young Benchmarks (Age $<$ 300 Myr)} \\
1RXS J235133.3+312720 B & $0.149^{+0.051}_{-0.019}$ & 2 & $-3.72\pm0.02$ & 25 & & $1947^{+57}_{-61}$ & $4.88^{+0.08}_{-0.07}$ & $1.18^{+0.07}_{-0.06}$ & $42.0^{+5.7}_{-2.8}$ \\
2MASS J01415823-4633574 & $0.045^{+0.004}_{-0.004}$ & 2 & $-3.57\pm0.03$ & 25 & & $1921^{+83}_{-92}$ & $4.48^{+0.09}_{-0.21}$ & $1.45^{+0.13}_{-0.11}$ & $25.1^{+2.5}_{-6.6}$ \\
2MASS J03231002-4631237 & $0.045^{+0.004}_{-0.004}$ & 2 & $-3.66\pm0.03$ & 25 & & $1799^{+83}_{-74}$ & $4.31^{+0.19}_{-0.12}$ & $1.49^{+0.10}_{-0.11}$ & $18.4^{+7.1}_{-3.1}$ \\
2MASS J03264225-2102057 & $0.149^{+0.051}_{-0.019}$ & 2 & $-4.17\pm0.05$ & 25 & & $1501^{+62}_{-69}$ & $4.73^{+0.09}_{-0.10}$ & $1.19^{+0.08}_{-0.06}$ & $30.2^{+4.2}_{-2.8}$ \\
2MASS J03552337+1133437$^{b}$ & $0.149^{+0.051}_{-0.019}$ & 2 & $\dots$ & $\dots$ & & $\dots$ & $\dots$ & $\dots$ & $\dots$ \\
2MASS J04433761+0002051 & $0.024^{+0.003}_{-0.003}$ & 2 & $-3.22\pm0.02$ & 25 & & $2167^{+110}_{-90}$ & $4.23^{+0.20}_{-0.11}$ & $1.70^{+0.15}_{-0.15}$ & $20.1^{+7.1}_{-2.6}$ \\
2MASS J05184616-2756457 & $0.042^{+0.006}_{-0.004}$ & 2 & $-3.41\pm0.04$ & 25 & & $2108^{+77}_{-80}$ & $4.55^{+0.06}_{-0.08}$ & $1.44^{+0.10}_{-0.09}$ & $29.5^{+2.3}_{-2.3}$ \\
\enddata
\end{deluxetable*}
}

\clearpage

{
 
\begin{longrotatetable}

\begin{deluxetable*}{lccccccccccccc}
\setlength{\tabcolsep}{3pt}

\setlength{\tabcolsep}{3pt}
\tablecaption{Spectroscopically inferred physical properties} \label{tab:chi2}
\tablehead{ \multicolumn{3}{l}{}  & \multicolumn{1}{c}{} & \multicolumn{4}{c}{\texttt{SM08}} & \multicolumn{1}{c}{} & \multicolumn{5}{c}{\texttt{Sonora Diamondback}} \\
\cline{5-8} \cline{10-14} 
\multicolumn{1}{l}{Object} & \multicolumn{1}{c}{Slit Width} & \multicolumn{1}{c}{Spec. Ref} & \multicolumn{1}{c}{} & \multicolumn{1}{c}{$T_{\rm eff}$}  & \multicolumn{1}{c}{log($g$)} & \multicolumn{1}{c}{$f_{\rm sed}$} & \multicolumn{1}{c}{$R$} & \multicolumn{1}{c}{} & \multicolumn{1}{c}{$T_{\rm eff}$}  & \multicolumn{1}{c}{log($g$)} & \multicolumn{1}{c}{$f_{\rm sed}$} & \multicolumn{1}{c}{[M/H]} & \multicolumn{1}{c}{$R$} \\
\multicolumn{1}{l}{}  & \multicolumn{1}{c}{(arcsec)} & \multicolumn{2}{c}{} & \multicolumn{1}{c}{(K)}  & \multicolumn{1}{c}{(dex)} & \multicolumn{1}{c}{} & \multicolumn{1}{c}{($R_{\rm Jup}$)} & \multicolumn{1}{c}{} & \multicolumn{1}{c}{(K)}  & \multicolumn{1}{c}{(dex)} & \multicolumn{1}{c}{} & \multicolumn{1}{c}{(dex)} & \multicolumn{1}{c}{($R_{\rm Jup}$)}
}
\startdata
\multicolumn{14}{c}{Young Benchmarks (Age $<$ 300 Myr)} \\
\hline
1RXS J235133.3+312720 B & 0.3 & 6 & & $1800.0^{+50}_{-50}$ & $5.5^{+0.25}_{-0.25}$ & $1^{+0.5}_{-0.5}$ & $1.33^{+0.07}_{-0.07}$ & & $2000^{+50}_{-50}$ & $5.5^{+0.25}_{-0.25}$ & $1^{+0.5}_{-0.5}$ & $0.5^{+0.25}_{-0.25}$ & $1.03^{+0.05}_{-0.05}$ \\
2MASS J01415823-4633574 & 0.5 & 30 & & $1600.0^{+50}_{-50}$ & $5.0^{+0.25}_{-0.25}$ & $1^{+0.5}_{-0.5}$ & $2.10^{+0.13}_{-0.13}$ & & $1500^{+50}_{-50}$ & $5.0^{+0.25}_{-0.25}$ & $2^{+0.5}_{-0.5}$ & $0.5^{+0.25}_{-0.25}$ & $2.22^{+0.15}_{-0.15}$ \\
2MASS J03231002-4631237 & 0.5 & 25 & & $1600.0^{+50}_{-50}$ & $5.0^{+0.25}_{-0.25}$ & $1^{+0.5}_{-0.5}$ & $1.93^{+0.12}_{-0.12}$ & & $1500^{+50}_{-50}$ & $5.0^{+0.25}_{-0.25}$ & $1^{+0.5}_{-0.5}$ & $0.5^{+0.25}_{-0.25}$ & $2.05^{+0.14}_{-0.14}$ \\
2MASS J03264225-2102057 & 0.5 & 27 & & $1300.0^{+50}_{-50}$ & $4.0^{+0.25}_{-0.25}$ & $1^{+0.5}_{-0.5}$ & $\dots$ & & $1300^{+50}_{-50}$ & $4.0^{+0.25}_{-0.25}$ & $1^{+0.5}_{-0.5}$ & $0.0^{+0.25}_{-0.25}$ & $\dots$ \\
2MASS J03552337+1133437\tablenotemark{\scriptsize b} & 0.5 & 3 & & $1300.0^{+50}_{-50}$ & $4.0^{+0.25}_{-0.25}$ & $1^{+0.5}_{-0.5}$ & $1.57^{+0.12}_{-0.12}$ & & $1200^{+50}_{-50}$ & $4.0^{+0.25}_{-0.25}$ & $1^{+0.5}_{-0.5}$ & $0.5^{+0.25}_{-0.25}$ & $1.89^{+0.16}_{-0.16}$ \\
2MASS J04433761+0002051 & 0.5 & 16 & & $1800.0^{+50}_{-50}$ & $5.0^{+0.25}_{-0.25}$ & $1^{+0.5}_{-0.5}$ & $2.41^{+0.13}_{-0.13}$ & & $2000^{+50}_{-50}$ & $5.5^{+0.25}_{-0.25}$ & $1^{+0.5}_{-0.5}$ & $0.5^{+0.25}_{-0.25}$ & $1.90^{+0.10}_{-0.10}$ \\
2MASS J05184616-2756457 & 0.5 & 2 & & $1600.0^{+50}_{-50}$ & $5.0^{+0.25}_{-0.25}$ & $1^{+0.5}_{-0.5}$ & $2.79^{+0.17}_{-0.17}$ & & $1500^{+50}_{-50}$ & $5.0^{+0.25}_{-0.25}$ & $2^{+0.5}_{-0.5}$ & $0.0^{+0.25}_{-0.25}$ & $2.94^{+0.20}_{-0.20}$ \\
\enddata
\end{deluxetable*}
\end{longrotatetable}
}

\clearpage

{
\setlength{\tabcolsep}{6pt}
\begin{deluxetable*}{cccrcccrcccrc}
\tablecaption{Systematic Difference in Physical Parameters Inferred from Atmospheric and Evolution Models  \label{tab:evo_calb}}
\tablehead{ \multicolumn{1}{c}{Range} & \multicolumn{1}{c}{} & \multicolumn{3}{c}{$\Delta T$ = $T_{\rm evo}-T_{\rm atm}$} & \multicolumn{1}{c}{} & \multicolumn{3}{c}{$\Delta R = R_{\rm evo}-R_{\rm atm}$} & \multicolumn{1}{c}{} & \multicolumn{3}{c}{$\Delta\log {(g)} = \log{(g)}_{\rm evo} - \log{(g)}_{\rm atm}$} \\
\cline{3-5} \cline{7-9} \cline{11-13}
\multicolumn{2}{c}{} & \multicolumn{1}{c}{N} & \multicolumn{1}{c}{Mean}  & \multicolumn{1}{c}{RMS} & \multicolumn{1}{c}{} & \multicolumn{1}{c}{N} & \multicolumn{1}{c}{Mean}  & \multicolumn{1}{c}{RMS} & \multicolumn{1}{c}{}  & \multicolumn{1}{c}{N} & \multicolumn{1}{c}{Mean}  & \multicolumn{1}{c}{RMS} \\
\multicolumn{2}{c}{} & \multicolumn{1}{c}{}  & \multicolumn{1}{c}{(K)}  & \multicolumn{1}{c}{(K)}  & \multicolumn{1}{c}{} & \multicolumn{1}{c}{} & \multicolumn{1}{c}{($R_{\rm Jup}$)} & \multicolumn{1}{c}{($R_{\rm Jup}$)} & \multicolumn{1}{c}{}  & \multicolumn{1}{c}{}  & \multicolumn{1}{c}{(dex)}  & \multicolumn{1}{c}{(dex)} 
} 
\startdata
\multicolumn{13}{c}{\texttt{Sonora Diamondback}} \\
\hline
M8 $\leq$ SpT < L0 & & 9 & $287$ & $286$ & & 9 & $-0.36$ & $0.53$ & & 9 & $-0.97$ & $0.43$ \\
L0 $\leq$ SpT < L2 & & 22 & $264$ & $203$ & & 22 & $-0.29$ & $0.32$ & & 22 & $-0.43$ & $0.27$ \\
L2 $\leq$ SpT < L4 & & 17 & $256$ & $128$ & & 15 & $-0.35$ & $0.20$ & & 17 & $-0.09$ & $0.42$ \\
L4 $\leq$ SpT < L6 & & 19 & $194$ & $196$ & & 15 & $-0.24$ & $0.29$ & & 19 & $0.17$ & $0.40$ \\
L6 $\leq$ SpT < T0 & & 19 & $2$ & $164$ & & 15 & $0.10$ & $0.27$ & & 19 & $0.08$ & $0.56$ \\
T0 $\leq$ SpT < T4 & & 7 & $-291$ & $117$ & & 7 & $0.47$ & $0.20$ & & 7 & $0.22$ & $0.64$ \\
T4 $\leq$ SpT $\leq$ T6 & & 7 & $-102$ & $154$ & & 7 & $0.24$ & $0.25$ & & 7 & $0.38$ & $0.68$ \\
\enddata
\end{deluxetable*}

}

\clearpage

{
\startlongtable
\begin{deluxetable*}{lcccccccccccc}

\setlength{\tabcolsep}{2pt}
\tablecaption{Spectroscopically inferred physical properties with additional ISM-like extinction incorporated.}\label{tab:ext}

\tablehead{ \multicolumn{1}{c}{Object} & \multicolumn{6}{c}{\texttt{Sonora Diamondback}} & \multicolumn{1}{c}{} & \multicolumn{5}{c}{\texttt{SM08}} \\
\cline{2-7} \cline{9-13} 
\multicolumn{1}{c}{} & \multicolumn{1}{c}{$T_{eff}$}  & \multicolumn{1}{c}{log(g)} & \multicolumn{1}{c}{f$_{sed}$} & \multicolumn{1}{c}{[M/H]} & \multicolumn{1}{c}{R} & \multicolumn{1}{c}{A$_{V}$}  & \multicolumn{1}{c}{} & \multicolumn{1}{c}{T$_{eff}$}  & \multicolumn{1}{c}{log(g)} & \multicolumn{1}{c}{f$_{sed}$} & \multicolumn{1}{c}{R} & \multicolumn{1}{c}{A$_{V}$} \\
\multicolumn{1}{c}{} & \multicolumn{1}{c}{(K)}  & \multicolumn{1}{c}{(dex)} & \multicolumn{1}{c}{} & \multicolumn{1}{c}{(dex)} & \multicolumn{1}{c}{(R$_{Jup}$)} & \multicolumn{1}{c}{(mag)} & \multicolumn{1}{c}{} & \multicolumn{1}{c}{(K)}  & \multicolumn{1}{c}{(dex)} & \multicolumn{1}{c}{} & \multicolumn{1}{c}{(R$_{Jup}$)} & \multicolumn{1}{c}{(mag)} 
}
\startdata
\multicolumn{13}{c}{Young Benchmarks (Age $<$ 300 Myr)} \\
\hline
1RXS J235133.3+312720 B & $2000^{+50}_{-50}$ & $5.5^{+0.25}_{-0.25}$ & $1^{+0.5}_{-0.5}$ & $0.5^{+0.25}_{-0.25}$ & $1.03^{+0.05}_{-0.05}$ & 0.0 & & $1800^{+50}_{-50}$ & $5.5^{+0.25}_{-0.25}$ & $1^{+0.5}_{-0.5}$ & $1.33^{+0.07}_{-0.07}$ & 0.0 \\
2MASS J01415823-4633574 & $2400^{+50}_{-50}$ & $3.5^{+0.25}_{-0.25}$ & $1^{+0.5}_{-0.5}$ & $0.5^{+0.25}_{-0.25}$ & $1.25^{+0.05}_{-0.05}$ & 4.9 & & $1600^{+50}_{-50}$ & $5.0^{+0.25}_{-0.25}$ & $1^{+0.5}_{-0.5}$ & $2.1^{+0.13}_{-0.13}$ & 0.0 \\
2MASS J03231002-4631237 & $2000^{+50}_{-50}$ & $5.5^{+0.25}_{-0.25}$ & $1^{+0.5}_{-0.5}$ & $0.5^{+0.25}_{-0.25}$ & $1.43^{+0.07}_{-0.07}$ & 3.4 & & $1800^{+50}_{-50}$ & $5.5^{+0.25}_{-0.25}$ & $1^{+0.5}_{-0.5}$ & $1.82^{+0.1}_{-0.1}$ & 3.1 \\
2MASS J03264225-2102057 & $1500^{+50}_{-50}$ & $5.0^{+0.25}_{-0.25}$ & $4^{+2}_{-0.5}$ & $0.5^{+0.25}_{-0.25}$ & $\dots$ & 3.6 & & $1600^{+50}_{-50}$ & $4.5^{+0.25}_{-0.25}$ & $2^{+0.5}_{-0.5}$ & $\dots$ & 4.5 \\
2MASS J03552337+1133437\tablenotemark{\scriptsize b} & $1500^{+50}_{-50}$ & $5.0^{+0.25}_{-0.25}$ & $1^{+0.5}_{-0.5}$ & $0.5^{+0.25}_{-0.25}$ & $1.51^{+0.1}_{-0.1}$ & 5.2 & & $1500^{+50}_{-50}$ & $4.5^{+0.25}_{-0.25}$ & $1^{+0.5}_{-0.5}$ & $1.53^{+0.1}_{-0.1}$ & 4.9 \\
2MASS J04433761+0002051 & $2400^{+50}_{-50}$ & $3.5^{+0.25}_{-0.25}$ & $1^{+0.5}_{-0.5}$ & $0.5^{+0.25}_{-0.25}$ & $1.58^{+0.07}_{-0.07}$ & 2.0 & & $2400^{+50}_{-50}$ & $4.0^{+0.25}_{-0.25}$ & $2^{+0.5}_{-0.5}$ & $1.52^{+0.06}_{-0.06}$ & 1.6 \\
2MASS J05184616-2756457 & $2000^{+50}_{-50}$ & $5.5^{+0.25}_{-0.25}$ & $1^{+0.5}_{-0.5}$ & $0.5^{+0.25}_{-0.25}$ & $1.99^{+0.1}_{-0.1}$ & 2.9 & & $1800^{+50}_{-50}$ & $5.0^{+0.25}_{-0.25}$ & $1^{+0.5}_{-0.5}$ & $2.46^{+0.14}_{-0.14}$ & 2.5 \\
\enddata
\tablecomments{Adopted uncertainties.} 
\tablenotetext{a}{The best-fit T$_{\mathrm{eff}}$ for these objects touch the model grid boundary, possibly indicating the spectra are not well-suited for the models and derived parameters should be considered with caution.}
\tablenotetext{b}{The binarity of these objects has been confirmed or suggested by adaptive optics observations (see note b of Table \ref{tab:sample_dat})}
\end{deluxetable*}
}

\bibliographystyle{aasjournal}
\bibliography{ms}

@dataset{ultracoolsheet,
  author       = {Best, William M. J. and Dupuy, Trent J. and Liu, Michael C. and Sanghi, Aniket and Siverd, Robert J. and Zhang, Zhoujian},
  title        = {{The UltracoolSheet: Photometry, Astrometry, Spectroscopy, and Multiplicity for 4000+ Ultracool Dwarfs and Imaged Exoplanets}},
  month        = jul,
  year         = 2024,
  publisher    = {Zenodo},
  version      = {2.0.0},
  doi          = {10.5281/zenodo.10573247},
  url          = {https://doi.org/10.5281/zenodo.10573247}
}

@ARTICLE{2003PASP..115..362R,
       author = {{Rayner}, J.~T. and {Toomey}, D.~W. and {Onaka}, P.~M. and {Denault}, A.~J. and {Stahlberger}, W.~E. and {Vacca}, W.~D. and {Cushing}, M.~C. and {Wang}, S.},
        title = "{SpeX: A Medium-Resolution 0.8-5.5 Micron Spectrograph and Imager for the NASA Infrared Telescope Facility}",
      journal = {\pasp},
         year = 2003,
        month = mar,
       volume = {115},
       number = {805},
        pages = {362-382},
          doi = {10.1086/367745},
       adsurl = {https://ui.adsabs.harvard.edu/abs/2003PASP..115..362R}
}

@ARTICLE{2008ApJ...689.1327S,
       author = {{Saumon}, D. and {Marley}, Mark S.},
        title = "{The Evolution of L and T Dwarfs in Color-Magnitude Diagrams}",
      journal = {\apj},
         year = 2008,
        month = dec,
       volume = {689},
       number = {2},
        pages = {1327-1344},
          doi = {10.1086/592734},
archivePrefix = {arXiv},
       eprint = {0808.2611},
 primaryClass = {astro-ph},
       adsurl = {https://ui.adsabs.harvard.edu/abs/2008ApJ...689.1327S}
}

@ARTICLE{2006MNRAS.367..454H,
       author = {{Hewett}, P.~C. and {Warren}, S.~J. and {Leggett}, S.~K. and {Hodgkin}, S.~T.},
        title = "{The UKIRT Infrared Deep Sky Survey ZY JHK photometric system: passbands and synthetic colours}",
      journal = {\mnras},
         year = 2006,
        month = apr,
       volume = {367},
       number = {2},
        pages = {454-468},
          doi = {10.1111/j.1365-2966.2005.09969.x},
archivePrefix = {arXiv},
       eprint = {astro-ph/0601592},
 primaryClass = {astro-ph},
       adsurl = {https://ui.adsabs.harvard.edu/abs/2006MNRAS.367..454H}
}

@ARTICLE{2007MNRAS.379.1599L,
       author = {{Lawrence}, A. and {Warren}, S.~J. and {Almaini}, O. and {Edge}, A.~C. and {Hambly}, N.~C. and {Jameson}, R.~F. and {Lucas}, P. and {Casali}, M. and {Adamson}, A. and {Dye}, S. and {Emerson}, J.~P. and {Foucaud}, S. and {Hewett}, P. and {Hirst}, P. and {Hodgkin}, S.~T. and {Irwin}, M.~J. and {Lodieu}, N. and {McMahon}, R.~G. and {Simpson}, C. and {Smail}, I. and {Mortlock}, D. and {Folger}, M.},
        title = "{The UKIRT Infrared Deep Sky Survey (UKIDSS)}",
      journal = {\mnras},
         year = 2007,
        month = aug,
       volume = {379},
       number = {4},
        pages = {1599-1617},
          doi = {10.1111/j.1365-2966.2007.12040.x},
archivePrefix = {arXiv},
       eprint = {astro-ph/0604426},
 primaryClass = {astro-ph},
       adsurl = {https://ui.adsabs.harvard.edu/abs/2007MNRAS.379.1599L}
}

@ARTICLE{2023AJ....166..198Z,
       author = {{Zhang}, Zhoujian and {Molli{\`e}re}, Paul and {Hawkins}, Keith and {Manea}, Catherine and {Fortney}, Jonathan J. and {Morley}, Caroline V. and {Skemer}, Andrew and {Marley}, Mark S. and {Bowler}, Brendan P. and {Carter}, Aarynn L. and {Franson}, Kyle and {Maas}, Zachary G. and {Sneden}, Christopher},
        title = "{ELemental abundances of Planets and brown dwarfs Imaged around Stars (ELPIS). I. Potential Metal Enrichment of the Exoplanet AF Lep b and a Novel Retrieval Approach for Cloudy Self-luminous Atmospheres}",
      journal = {\aj},
         year = 2023,
        month = nov,
       volume = {166},
       number = {5},
          eid = {198},
        pages = {198},
          doi = {10.3847/1538-3881/acf768},
archivePrefix = {arXiv},
       eprint = {2309.02488},
 primaryClass = {astro-ph.EP},
       adsurl = {https://ui.adsabs.harvard.edu/abs/2023AJ....166..198Z}
}

@ARTICLE{2021ApJ...916...53Z,
       author = {{Zhang}, Zhoujian and {Liu}, Michael C. and {Marley}, Mark S. and {Line}, Michael R. and {Best}, William M.~J.},
        title = "{Uniform Forward-modeling Analysis of Ultracool Dwarfs. I. Methodology and Benchmarking}",
      journal = {\apj},
         year = 2021,
        month = jul,
       volume = {916},
       number = {1},
          eid = {53},
        pages = {53},
          doi = {10.3847/1538-4357/abf8b2},
archivePrefix = {arXiv},
       eprint = {2011.12294},
 primaryClass = {astro-ph.SR},
       adsurl = {https://ui.adsabs.harvard.edu/abs/2021ApJ...916...53Z}
}

@ARTICLE{2021ApJ...921...95Z,
       author = {{Zhang}, Zhoujian and {Liu}, Michael C. and {Marley}, Mark S. and {Line}, Michael R. and {Best}, William M.~J.},
        title = "{Uniform Forward-modeling Analysis of Ultracool Dwarfs. II. Atmospheric Properties of 55 Late-T Dwarfs}",
      journal = {\apj},
         year = 2021,
        month = nov,
       volume = {921},
       number = {1},
          eid = {95},
        pages = {95},
          doi = {10.3847/1538-4357/ac0af7},
archivePrefix = {arXiv},
       eprint = {2105.05256},
 primaryClass = {astro-ph.SR},
       adsurl = {https://ui.adsabs.harvard.edu/abs/2021ApJ...921...95Z}
}

@ARTICLE{2024ApJ...961..121H,
       author = {{Hurt}, Spencer A. and {Liu}, Michael C. and {Zhang}, Zhoujian and {Phillips}, Mark and {Allers}, Katelyn N. and {Deacon}, Niall R. and {Aller}, Kimberly M. and {Best}, William M.~J.},
        title = "{Uniform Forward-modeling Analysis of Ultracool Dwarfs. III. Late-M and L Dwarfs in Young Moving Groups, the Pleiades, and the Hyades}",
      journal = {\apj},
         year = 2024,
        month = jan,
       volume = {961},
       number = {1},
          eid = {121},
        pages = {121},
          doi = {10.3847/1538-4357/ad0b12},
archivePrefix = {arXiv},
       eprint = {2311.04268},
 primaryClass = {astro-ph.SR},
       adsurl = {https://ui.adsabs.harvard.edu/abs/2024ApJ...961..121H}
}

@ARTICLE{2008Sci...322.1348M,
       author = {{Marois}, Christian and {Macintosh}, Bruce and {Barman}, Travis and {Zuckerman}, B. and {Song}, Inseok and {Patience}, Jennifer and {Lafreni{\`e}re}, David and {Doyon}, Ren{\'e}},
        title = "{Direct Imaging of Multiple Planets Orbiting the Star HR 8799}",
      journal = {Science},
         year = 2008,
        month = nov,
       volume = {322},
       number = {5906},
        pages = {1348},
          doi = {10.1126/science.1166585},
archivePrefix = {arXiv},
       eprint = {0811.2606},
 primaryClass = {astro-ph},
       adsurl = {https://ui.adsabs.harvard.edu/abs/2008Sci...322.1348M}
}

@ARTICLE{2011ApJ...735L..39B,
       author = {{Barman}, Travis S. and {Macintosh}, Bruce and {Konopacky}, Quinn M. and {Marois}, Christian},
        title = "{The Young Planet-mass Object 2M1207b: A Cool, Cloudy, and Methane-poor Atmosphere}",
      journal = {\apjl},
         year = 2011,
        month = jul,
       volume = {735},
       number = {2},
          eid = {L39},
        pages = {L39},
          doi = {10.1088/2041-8205/735/2/L39},
archivePrefix = {arXiv},
       eprint = {1106.1201},
 primaryClass = {astro-ph.SR},
       adsurl = {https://ui.adsabs.harvard.edu/abs/2011ApJ...735L..39B}
}

@ARTICLE{2006ApJ...651.1166M,
       author = {{Metchev}, Stanimir A. and {Hillenbrand}, Lynne A.},
        title = "{HD 203030B: An Unusually Cool Young Substellar Companion near the L/T Transition}",
      journal = {\apj},
         year = 2006,
        month = nov,
       volume = {651},
       number = {2},
        pages = {1166-1176},
          doi = {10.1086/507836},
archivePrefix = {arXiv},
       eprint = {astro-ph/0607514},
 primaryClass = {astro-ph},
       adsurl = {https://ui.adsabs.harvard.edu/abs/2006ApJ...651.1166M}
}

@ARTICLE{2015ApJ...810..158F,
       author = {{Filippazzo}, Joseph C. and {Rice}, Emily L. and {Faherty}, Jacqueline and {Cruz}, Kelle L. and {Van Gordon}, Mollie M. and {Looper}, Dagny L.},
        title = "{Fundamental Parameters and Spectral Energy Distributions of Young and Field Age Objects with Masses Spanning the Stellar to Planetary Regime}",
      journal = {\apj},
         year = 2015,
        month = sep,
       volume = {810},
       number = {2},
          eid = {158},
        pages = {158},
          doi = {10.1088/0004-637X/810/2/158},
archivePrefix = {arXiv},
       eprint = {1508.01767},
 primaryClass = {astro-ph.SR},
       adsurl = {https://ui.adsabs.harvard.edu/abs/2015ApJ...810..158F}
}

@ARTICLE{2019MNRAS.483..480V,
       author = {{Vos}, Johanna M. and {Biller}, Beth A. and {Bonavita}, Mariangela and {Eriksson}, Simon and {Liu}, Michael C. and {Best}, William M.~J. and {Metchev}, Stanimir and {Radigan}, Jacqueline and {Allers}, Katelyn N. and {Janson}, Markus and {Buenzli}, Esther and {Dupuy}, Trent J. and {Bonnefoy}, Micka{\"e}l and {Manjavacas}, Elena and {Brandner}, Wolfgang and {Crossfield}, Ian and {Deacon}, Niall and {Henning}, Thomas and {Homeier}, Derek and {Kopytova}, Taisiya and {Schlieder}, Joshua},
        title = "{A search for variability in exoplanet analogues and low-gravity brown dwarfs}",
      journal = {\mnras},
         year = 2019,
        month = feb,
       volume = {483},
       number = {1},
        pages = {480-502},
          doi = {10.1093/mnras/sty3123},
archivePrefix = {arXiv},
       eprint = {1811.08370},
 primaryClass = {astro-ph.SR},
       adsurl = {https://ui.adsabs.harvard.edu/abs/2019MNRAS.483..480V}
}

@ARTICLE{2011ApJ...737..103S,
       author = {{Schlafly}, Edward F. and {Finkbeiner}, Douglas P.},
        title = "{Measuring Reddening with Sloan Digital Sky Survey Stellar Spectra and Recalibrating SFD}",
      journal = {\apj},
         year = 2011,
        month = aug,
       volume = {737},
       number = {2},
          eid = {103},
        pages = {103},
          doi = {10.1088/0004-637X/737/2/103},
archivePrefix = {arXiv},
       eprint = {1012.4804},
 primaryClass = {astro-ph.GA},
       adsurl = {https://ui.adsabs.harvard.edu/abs/2011ApJ...737..103S}
}

@ARTICLE{2017ApJ...848...83L,
       author = {{Line}, Michael R. and {Marley}, Mark S. and {Liu}, Michael C. and {Burningham}, Ben and {Morley}, Caroline V. and {Hinkel}, Natalie R. and {Teske}, Johanna and {Fortney}, Jonathan J. and {Freedman}, Richard and {Lupu}, Roxana},
        title = "{Uniform Atmospheric Retrieval Analysis of Ultracool Dwarfs. II. Properties of 11 T dwarfs}",
      journal = {\apj},
         year = 2017,
        month = oct,
       volume = {848},
       number = {2},
          eid = {83},
        pages = {83},
          doi = {10.3847/1538-4357/aa7ff0},
archivePrefix = {arXiv},
       eprint = {1612.02809},
 primaryClass = {astro-ph.SR},
       adsurl = {https://ui.adsabs.harvard.edu/abs/2017ApJ...848...83L}
}

@ARTICLE{2015ARA&A..53..279M,
       author = {{Marley}, M.~S. and {Robinson}, T.~D.},
        title = "{On the Cool Side: Modeling the Atmospheres of Brown Dwarfs and Giant Planets}",
      journal = {\araa},
         year = 2015,
        month = aug,
       volume = {53},
        pages = {279-323},
          doi = {10.1146/annurev-astro-082214-122522},
archivePrefix = {arXiv},
       eprint = {1410.6512},
 primaryClass = {astro-ph.EP},
       adsurl = {https://ui.adsabs.harvard.edu/abs/2015ARA&A..53..279M}
}

@ARTICLE{2006ApJ...639.1095B,
       author = {{Burgasser}, Adam J. and {Burrows}, Adam and {Kirkpatrick}, J. Davy},
        title = "{A Method for Determining the Physical Properties of the Coldest Known Brown Dwarfs}",
      journal = {\apj},
         year = 2006,
        month = mar,
       volume = {639},
       number = {2},
        pages = {1095-1113},
          doi = {10.1086/499344},
archivePrefix = {arXiv},
       eprint = {astro-ph/0510707},
 primaryClass = {astro-ph},
       adsurl = {https://ui.adsabs.harvard.edu/abs/2006ApJ...639.1095B}
}

@ARTICLE{2023ApJ...951L..48B,
       author = {{Beiler}, Samuel A. and {Cushing}, Michael C. and {Kirkpatrick}, J. Davy and {Schneider}, Adam C. and {Mukherjee}, Sagnick and {Marley}, Mark S.},
        title = "{The First JWST Spectral Energy Distribution of a Y Dwarf}",
      journal = {\apjl},
         year = 2023,
        month = jul,
       volume = {951},
       number = {2},
          eid = {L48},
        pages = {L48},
          doi = {10.3847/2041-8213/ace32c},
archivePrefix = {arXiv},
       eprint = {2306.11807},
 primaryClass = {astro-ph.SR},
       adsurl = {https://ui.adsabs.harvard.edu/abs/2023ApJ...951L..48B}
}

@ARTICLE{2014A&A...562A.127B,
       author = {{Bonnefoy}, M. and {Chauvin}, G. and {Lagrange}, A. -M. and {Rojo}, P. and {Allard}, F. and {Pinte}, C. and {Dumas}, C. and {Homeier}, D.},
        title = "{A library of near-infrared integral field spectra of young M-L dwarfs}",
      journal = {\aap},
         year = 2014,
        month = feb,
       volume = {562},
          eid = {A127},
        pages = {A127},
          doi = {10.1051/0004-6361/201118270},
archivePrefix = {arXiv},
       eprint = {1306.3709},
 primaryClass = {astro-ph.SR},
       adsurl = {https://ui.adsabs.harvard.edu/abs/2014A&A...562A.127B}
}

@ARTICLE{2021ApJ...920...85M,
       author = {{Marley}, Mark S. and {Saumon}, Didier and {Visscher}, Channon and {Lupu}, Roxana and {Freedman}, Richard and {Morley}, Caroline and {Fortney}, Jonathan J. and {Seay}, Christopher and {Smith}, Adam J.~R.~W. and {Teal}, D.~J. and {Wang}, Ruoyan},
        title = "{The Sonora Brown Dwarf Atmosphere and Evolution Models. I. Model Description and Application to Cloudless Atmospheres in Rainout Chemical Equilibrium}",
      journal = {\apj},
         year = 2021,
        month = oct,
       volume = {920},
       number = {2},
          eid = {85},
        pages = {85},
          doi = {10.3847/1538-4357/ac141d},
archivePrefix = {arXiv},
       eprint = {2107.07434},
 primaryClass = {astro-ph.SR},
       adsurl = {https://ui.adsabs.harvard.edu/abs/2021ApJ...920...85M}
}

@ARTICLE{2015ApJ...812..128C,
       author = {{Czekala}, Ian and {Andrews}, Sean M. and {Mandel}, Kaisey S. and {Hogg}, David W. and {Green}, Gregory M.},
        title = "{Constructing a Flexible Likelihood Function for Spectroscopic Inference}",
      journal = {\apj},
         year = 2015,
        month = oct,
       volume = {812},
       number = {2},
          eid = {128},
        pages = {128},
          doi = {10.1088/0004-637X/812/2/128},
archivePrefix = {arXiv},
       eprint = {1412.5177},
 primaryClass = {astro-ph.SR},
       adsurl = {https://ui.adsabs.harvard.edu/abs/2015ApJ...812..128C}
}

@ARTICLE{2013ApJ...772...79A,
       author = {{Allers}, K.~N. and {Liu}, Michael C.},
        title = "{A Near-infrared Spectroscopic Study of Young Field Ultracool Dwarfs}",
      journal = {\apj},
         year = 2013,
        month = aug,
       volume = {772},
       number = {2},
          eid = {79},
        pages = {79},
          doi = {10.1088/0004-637X/772/2/79},
archivePrefix = {arXiv},
       eprint = {1305.4418},
 primaryClass = {astro-ph.SR},
       adsurl = {https://ui.adsabs.harvard.edu/abs/2013ApJ...772...79A}
}

@ARTICLE{2014ApJ...794..143B,
       author = {{Bardalez Gagliuffi}, Daniella C. and {Burgasser}, Adam J. and {Gelino}, Christopher R. and {Looper}, Dagny L. and {Nicholls}, Christine P. and {Schmidt}, Sarah J. and {Cruz}, Kelle and {West}, Andrew A. and {Gizis}, John E. and {Metchev}, Stanimir},
        title = "{SpeX Spectroscopy of Unresolved Very Low Mass Binaries. II. Identification of 14 Candidate Binaries with Late-M/Early-L and T Dwarf Components}",
      journal = {\apj},
         year = 2014,
        month = oct,
       volume = {794},
       number = {2},
          eid = {143},
        pages = {143},
          doi = {10.1088/0004-637X/794/2/143},
archivePrefix = {arXiv},
       eprint = {1408.3089},
 primaryClass = {astro-ph.SR},
       adsurl = {https://ui.adsabs.harvard.edu/abs/2014ApJ...794..143B}
}

@ARTICLE{2013ApJ...777...84B,
       author = {{Best}, William M.~J. and {Liu}, Michael C. and {Magnier}, Eugene A. and {Aller}, Kimberly M. and {Deacon}, Niall R. and {Dupuy}, Trent J. and {Redstone}, Joshua and {Burgett}, W.~S. and {Chambers}, K.~C. and {Hodapp}, K.~W. and {Kaiser}, N. and {Kudritzki}, R. -P. and {Morgan}, J.~S. and {Price}, P.~A. and {Tonry}, J.~L. and {Wainscoat}, R.~J.},
        title = "{A Search for L/T Transition Dwarfs with Pan-STARRS1 and WISE: Discovery of Seven Nearby Objects Including Two Candidate Spectroscopic Variables}",
      journal = {\apj},
         year = 2013,
        month = nov,
       volume = {777},
       number = {2},
          eid = {84},
        pages = {84},
          doi = {10.1088/0004-637X/777/2/84},
archivePrefix = {arXiv},
       eprint = {1309.0503},
 primaryClass = {astro-ph.SR},
       adsurl = {https://ui.adsabs.harvard.edu/abs/2013ApJ...777...84B}
}

@ARTICLE{2015ApJ...814..118B,
       author = {{Best}, William M.~J. and {Liu}, Michael C. and {Magnier}, Eugene A. and {Deacon}, Niall R. and {Aller}, Kimberly M. and {Redstone}, Joshua and {Burgett}, W.~S. and {Chambers}, K.~C. and {Draper}, P. and {Flewelling}, H. and {Hodapp}, K.~W. and {Kaiser}, N. and {Metcalfe}, N. and {Tonry}, J.~L. and {Wainscoat}, R.~J. and {Waters}, C.},
        title = "{A Search for L/T Transition Dwarfs with Pan-STARRS1 and WISE. II. L/T Transition Atmospheres and Young Discoveries}",
      journal = {\apj},
         year = 2015,
        month = dec,
       volume = {814},
       number = {2},
          eid = {118},
        pages = {118},
          doi = {10.1088/0004-637X/814/2/118},
archivePrefix = {arXiv},
       eprint = {1612.02824},
 primaryClass = {astro-ph.SR},
       adsurl = {https://ui.adsabs.harvard.edu/abs/2015ApJ...814..118B}
}

@ARTICLE{2012ApJ...753..142B,
       author = {{Bowler}, Brendan P. and {Liu}, Michael C. and {Shkolnik}, Evgenya L. and {Dupuy}, Trent J. and {Cieza}, Lucas A. and {Kraus}, Adam L. and {Tamura}, Motohide},
        title = "{Planets around Low-mass Stars (PALMS). I. A Substellar Companion to the Young M Dwarf 1RXS J235133.3+312720}",
      journal = {\apj},
         year = 2012,
        month = jul,
       volume = {753},
       number = {2},
          eid = {142},
        pages = {142},
          doi = {10.1088/0004-637X/753/2/142},
archivePrefix = {arXiv},
       eprint = {1205.2084},
 primaryClass = {astro-ph.SR},
       adsurl = {https://ui.adsabs.harvard.edu/abs/2012ApJ...753..142B}
}

@ARTICLE{2006ApJ...637.1067B,
       author = {{Burgasser}, Adam J. and {Geballe}, T.~R. and {Leggett}, S.~K. and {Kirkpatrick}, J. Davy and {Golimowski}, David A.},
        title = "{A Unified Near-Infrared Spectral Classification Scheme for T Dwarfs}",
      journal = {\apj},
         year = 2006,
        month = feb,
       volume = {637},
       number = {2},
        pages = {1067-1093},
          doi = {10.1086/498563},
archivePrefix = {arXiv},
       eprint = {astro-ph/0510090},
 primaryClass = {astro-ph},
       adsurl = {https://ui.adsabs.harvard.edu/abs/2006ApJ...637.1067B}
}

@ARTICLE{2010ApJ...710.1142B,
       author = {{Burgasser}, Adam J. and {Cruz}, Kelle L. and {Cushing}, Michael and {Gelino}, Christopher R. and {Looper}, Dagny L. and {Faherty}, Jacqueline K. and {Kirkpatrick}, J. Davy and {Reid}, I. Neill},
        title = "{SpeX Spectroscopy of Unresolved Very Low Mass Binaries. I. Identification of 17 Candidate Binaries Straddling the L Dwarf/T Dwarf Transition}",
      journal = {\apj},
         year = 2010,
        month = feb,
       volume = {710},
       number = {2},
        pages = {1142-1169},
          doi = {10.1088/0004-637X/710/2/1142},
archivePrefix = {arXiv},
       eprint = {0912.3808},
 primaryClass = {astro-ph.SR},
       adsurl = {https://ui.adsabs.harvard.edu/abs/2010ApJ...710.1142B}
}

@ARTICLE{2006AJ....131.2722C,
       author = {{Chiu}, K. and {Fan}, X. and {Leggett}, S.~K. and {Golimowski}, D.~A. and {Zheng}, W. and {Geballe}, T.~R. and {Schneider}, D.~P. and {Brinkmann}, J.},
        title = "{Seventy-One New L and T Dwarfs from the Sloan Digital Sky Survey}",
      journal = {\aj},
         year = 2006,
        month = jun,
       volume = {131},
       number = {5},
        pages = {2722-2736},
          doi = {10.1086/501431},
archivePrefix = {arXiv},
       eprint = {astro-ph/0601089},
 primaryClass = {astro-ph},
       adsurl = {https://ui.adsabs.harvard.edu/abs/2006AJ....131.2722C}
}

@ARTICLE{2014ApJ...792..119D,
       author = {{Deacon}, Niall R. and {Liu}, Michael C. and {Magnier}, Eugene A. and {Aller}, Kimberly M. and {Best}, William M.~J. and {Dupuy}, Trent and {Bowler}, Brendan P. and {Mann}, Andrew W. and {Redstone}, Joshua A. and {Burgett}, William S. and {Chambers}, Kenneth C. and {Draper}, Peter W. and {Flewelling}, H. and {Hodapp}, Klaus W. and {Kaiser}, Nick and {Kudritzki}, Rolf-Peter and {Morgan}, Jeff S. and {Metcalfe}, Nigel and {Price}, Paul A. and {Tonry}, John L. and {Wainscoat}, Richard J.},
        title = "{Wide Cool and Ultracool Companions to Nearby Stars from Pan-STARRS 1}",
      journal = {\apj},
         year = 2014,
        month = sep,
       volume = {792},
       number = {2},
          eid = {119},
        pages = {119},
          doi = {10.1088/0004-637X/792/2/119},
archivePrefix = {arXiv},
       eprint = {1407.2938},
 primaryClass = {astro-ph.SR},
       adsurl = {https://ui.adsabs.harvard.edu/abs/2014ApJ...792..119D}
}

@ARTICLE{2017MNRAS.467.1126D,
       author = {{Deacon}, N.~R. and {Magnier}, E.~A. and {Liu}, Michael C. and {Schlieder}, Joshua E. and {Aller}, Kimberly M. and {Best}, William M.~J. and {Bowler}, Brendan P. and {Burgett}, W.~S. and {Chambers}, K.~C. and {Draper}, P.~W. and {Flewelling}, H. and {Hodapp}, K.~W. and {Kaiser}, N. and {Metcalfe}, N. and {Sweeney}, W.~E. and {Wainscoat}, R.~J. and {Waters}, C.},
        title = "{2MASS 0213+3648 C: A wide T3 benchmark companion to an active, old M dwarf binary}",
      journal = {\mnras},
         year = 2017,
        month = may,
       volume = {467},
       number = {1},
        pages = {1126-1139},
          doi = {10.1093/mnras/stx065},
archivePrefix = {arXiv},
       eprint = {1701.03104},
 primaryClass = {astro-ph.SR},
       adsurl = {https://ui.adsabs.harvard.edu/abs/2017MNRAS.467.1126D}
}

@ARTICLE{2012ApJS..201...19D,
       author = {{Dupuy}, Trent J. and {Liu}, Michael C.},
        title = "{The Hawaii Infrared Parallax Program. I. Ultracool Binaries and the L/T Transition}",
      journal = {\apjs},
         year = 2012,
        month = aug,
       volume = {201},
       number = {2},
          eid = {19},
        pages = {19},
          doi = {10.1088/0067-0049/201/2/19},
archivePrefix = {arXiv},
       eprint = {1201.2465},
 primaryClass = {astro-ph.SR},
       adsurl = {https://ui.adsabs.harvard.edu/abs/2012ApJS..201...19D}
}

@ARTICLE{2010AJ....139..176F,
       author = {{Faherty}, Jacqueline K. and {Burgasser}, Adam J. and {West}, Andrew A. and {Bochanski}, John J. and {Cruz}, Kelle L. and {Shara}, Michael M. and {Walter}, Frederick M.},
        title = "{The Brown Dwarf Kinematics Project. II. Details on Nine Wide Common Proper Motion Very Low Mass Companions to Nearby Stars}",
      journal = {\aj},
         year = 2010,
        month = jan,
       volume = {139},
       number = {1},
        pages = {176-194},
          doi = {10.1088/0004-6256/139/1/176},
archivePrefix = {arXiv},
       eprint = {0911.1363},
 primaryClass = {astro-ph.SR},
       adsurl = {https://ui.adsabs.harvard.edu/abs/2010AJ....139..176F}
}

@ARTICLE{2016ApJS..225...10F,
       author = {{Faherty}, Jacqueline K. and {Riedel}, Adric R. and {Cruz}, Kelle L. and {Gagne}, Jonathan and {Filippazzo}, Joseph C. and {Lambrides}, Erini and {Fica}, Haley and {Weinberger}, Alycia and {Thorstensen}, John R. and {Tinney}, C.~G. and {Baldassare}, Vivienne and {Lemonier}, Emily and {Rice}, Emily L.},
        title = "{Population Properties of Brown Dwarf Analogs to Exoplanets}",
      journal = {\apjs},
         year = 2016,
        month = jul,
       volume = {225},
       number = {1},
          eid = {10},
        pages = {10},
          doi = {10.3847/0067-0049/225/1/10},
archivePrefix = {arXiv},
       eprint = {1605.07927},
 primaryClass = {astro-ph.SR},
       adsurl = {https://ui.adsabs.harvard.edu/abs/2016ApJS..225...10F}
}

@ARTICLE{2014ApJ...785L..14G,
       author = {{Gagn{\'e}}, Jonathan and {Faherty}, Jacqueline K. and {Cruz}, Kelle and {Lafreni{\`e}re}, David and {Doyon}, Ren{\'e} and {Malo}, Lison and {Artigau}, {\'E}tienne},
        title = "{The Coolest Isolated Brown Dwarf Candidate Member of TWA}",
      journal = {\apjl},
         year = 2014,
        month = apr,
       volume = {785},
       number = {1},
          eid = {L14},
        pages = {L14},
          doi = {10.1088/2041-8205/785/1/L14},
archivePrefix = {arXiv},
       eprint = {1403.3120},
 primaryClass = {astro-ph.SR},
       adsurl = {https://ui.adsabs.harvard.edu/abs/2014ApJ...785L..14G}
}

@ARTICLE{2015ApJS..219...33G,
       author = {{Gagn{\'e}}, Jonathan and {Faherty}, Jacqueline K. and {Cruz}, Kelle L. and {Lafreni{\'e}re}, David and {Doyon}, Ren{\'e} and {Malo}, Lison and {Burgasser}, Adam J. and {Naud}, Marie-Eve and {Artigau}, {\'E}tienne and {Bouchard}, Sandie and {Gizis}, John E. and {Albert}, Lo{\"\i}c},
        title = "{BANYAN. VII. A New Population of Young Substellar Candidate Members of Nearby Moving Groups from the BASS Survey}",
      journal = {\apjs},
         year = 2015,
        month = aug,
       volume = {219},
       number = {2},
          eid = {33},
        pages = {33},
          doi = {10.1088/0067-0049/219/2/33},
archivePrefix = {arXiv},
       eprint = {1506.07712},
 primaryClass = {astro-ph.SR},
       adsurl = {https://ui.adsabs.harvard.edu/abs/2015ApJS..219...33G}
}

@ARTICLE{2012AJ....144...94G,
       author = {{Gizis}, John E. and {Faherty}, Jacqueline K. and {Liu}, Michael C. and {Castro}, Philip J. and {Shaw}, John D. and {Vrba}, Frederick J. and {Harris}, Hugh C. and {Aller}, Kimberly M. and {Deacon}, Niall R.},
        title = "{Discovery of an Unusually Red L-type Brown Dwarf}",
      journal = {\aj},
         year = 2012,
        month = oct,
       volume = {144},
       number = {4},
          eid = {94},
        pages = {94},
          doi = {10.1088/0004-6256/144/4/94},
archivePrefix = {arXiv},
       eprint = {1207.4012},
 primaryClass = {astro-ph.SR},
       adsurl = {https://ui.adsabs.harvard.edu/abs/2012AJ....144...94G}
}

@ARTICLE{2010ApJS..190..100K,
       author = {{Kirkpatrick}, J. Davy and {Looper}, Dagny L. and {Burgasser}, Adam J. and {Schurr}, Steven D. and {Cutri}, Roc M. and {Cushing}, Michael C. and {Cruz}, Kelle L. and {Sweet}, Anne C. and {Knapp}, Gillian R. and {Barman}, Travis S. and {Bochanski}, John J. and {Roellig}, Thomas L. and {McLean}, Ian S. and {McGovern}, Mark R. and {Rice}, Emily L.},
        title = "{Discoveries from a Near-infrared Proper Motion Survey Using Multi-epoch Two Micron All-Sky Survey Data}",
      journal = {\apjs},
         year = 2010,
        month = sep,
       volume = {190},
       number = {1},
        pages = {100-146},
          doi = {10.1088/0067-0049/190/1/100},
archivePrefix = {arXiv},
       eprint = {1008.3591},
 primaryClass = {astro-ph.SR},
       adsurl = {https://ui.adsabs.harvard.edu/abs/2010ApJS..190..100K}
}

@ARTICLE{2011ApJS..197...19K,
       author = {{Kirkpatrick}, J. Davy and {Cushing}, Michael C. and {Gelino}, Christopher R. and {Griffith}, Roger L. and {Skrutskie}, Michael F. and {Marsh}, Kenneth A. and {Wright}, Edward L. and {Mainzer}, A. and {Eisenhardt}, Peter R. and {McLean}, Ian S. and {Thompson}, Maggie A. and {Bauer}, James M. and {Benford}, Dominic J. and {Bridge}, Carrie R. and {Lake}, Sean E. and {Petty}, Sara M. and {Stanford}, S.~A. and {Tsai}, Chao-Wei and {Bailey}, Vanessa and {Beichman}, Charles A. and {Bloom}, Joshua S. and {Bochanski}, John J. and {Burgasser}, Adam J. and {Capak}, Peter L. and {Cruz}, Kelle L. and {Hinz}, Philip M. and {Kartaltepe}, Jeyhan S. and {Knox}, Russell P. and {Manohar}, Swarnima and {Masters}, Daniel and {Morales-Calder{\'o}n}, Maria and {Prato}, Lisa A. and {Rodigas}, Timothy J. and {Salvato}, Mara and {Schurr}, Steven D. and {Scoville}, Nicholas Z. and {Simcoe}, Robert A. and {Stapelfeldt}, Karl R. and {Stern}, Daniel and {Stock}, Nathan D. and {Vacca}, William D.},
        title = "{The First Hundred Brown Dwarfs Discovered by the Wide-field Infrared Survey Explorer (WISE)}",
      journal = {\apjs},
         year = 2011,
        month = dec,
       volume = {197},
       number = {2},
          eid = {19},
        pages = {19},
          doi = {10.1088/0067-0049/197/2/19},
archivePrefix = {arXiv},
       eprint = {1108.4677},
 primaryClass = {astro-ph.SR},
       adsurl = {https://ui.adsabs.harvard.edu/abs/2011ApJS..197...19K}
}

@ARTICLE{2004AJ....127.3553K,
       author = {{Knapp}, G.~R. and {Leggett}, S.~K. and {Fan}, X. and {Marley}, M.~S. and {Geballe}, T.~R. and {Golimowski}, D.~A. and {Finkbeiner}, D. and {Gunn}, J.~E. and {Hennawi}, J. and {Ivezi{\'c}}, Z. and {Lupton}, R.~H. and {Schlegel}, D.~J. and {Strauss}, M.~A. and {Tsvetanov}, Z.~I. and {Chiu}, K. and {Hoversten}, E.~A. and {Glazebrook}, K. and {Zheng}, W. and {Hendrickson}, M. and {Williams}, C.~C. and {Uomoto}, A. and {Vrba}, F.~J. and {Henden}, A.~A. and {Luginbuhl}, C.~B. and {Guetter}, H.~H. and {Munn}, J.~A. and {Canzian}, B. and {Schneider}, Donald P. and {Brinkmann}, J.},
        title = "{Near-Infrared Photometry and Spectroscopy of L and T Dwarfs: The Effects of Temperature, Clouds, and Gravity}",
      journal = {\aj},
         year = 2004,
        month = jun,
       volume = {127},
       number = {6},
        pages = {3553-3578},
          doi = {10.1086/420707},
archivePrefix = {arXiv},
       eprint = {astro-ph/0402451},
 primaryClass = {astro-ph},
       adsurl = {https://ui.adsabs.harvard.edu/abs/2004AJ....127.3553K}
}

@ARTICLE{2016ApJ...833...96L,
       author = {{Liu}, Michael C. and {Dupuy}, Trent J. and {Allers}, Katelyn N.},
        title = "{The Hawaii Infrared Parallax Program. II. Young Ultracool Field Dwarfs}",
      journal = {\apj},
         year = 2016,
        month = dec,
       volume = {833},
       number = {1},
          eid = {96},
        pages = {96},
          doi = {10.3847/1538-4357/833/1/96},
archivePrefix = {arXiv},
       eprint = {1612.02426},
 primaryClass = {astro-ph.SR},
       adsurl = {https://ui.adsabs.harvard.edu/abs/2016ApJ...833...96L}
}

@ARTICLE{2007ApJ...654..570L,
       author = {{Luhman}, K.~L. and {Patten}, B.~M. and {Marengo}, M. and {Schuster}, M.~T. and {Hora}, J.~L. and {Ellis}, R.~G. and {Stauffer}, J.~R. and {Sonnett}, S.~M. and {Winston}, E. and {Gutermuth}, R.~A. and {Megeath}, S.~T. and {Backman}, D.~E. and {Henry}, T.~J. and {Werner}, M.~W. and {Fazio}, G.~G.},
        title = "{Discovery of Two T Dwarf Companions with the Spitzer Space Telescope}",
      journal = {\apj},
         year = 2007,
        month = jan,
       volume = {654},
       number = {1},
        pages = {570-579},
          doi = {10.1086/509073},
archivePrefix = {arXiv},
       eprint = {astro-ph/0609464},
 primaryClass = {astro-ph},
       adsurl = {https://ui.adsabs.harvard.edu/abs/2007ApJ...654..570L}
}

@ARTICLE{2017AJ....154..262M,
       author = {{Miles-P{\'a}ez}, Paulo A. and {Metchev}, Stanimir and {Luhman}, Kevin L. and {Marengo}, Massimo and {Hulsebus}, Alan},
        title = "{The Prototypical Young L/T-Transition Dwarf HD 203030B Likely Has Planetary Mass}",
      journal = {\aj},
         year = 2017,
        month = dec,
       volume = {154},
       number = {6},
          eid = {262},
        pages = {262},
          doi = {10.3847/1538-3881/aa9711},
archivePrefix = {arXiv},
       eprint = {1710.11274},
 primaryClass = {astro-ph.SR},
       adsurl = {https://ui.adsabs.harvard.edu/abs/2017AJ....154..262M}
}

@ARTICLE{2006AJ....132..891R,
       author = {{Reid}, I. Neill and {Lewitus}, E. and {Allen}, P.~R. and {Cruz}, Kelle L. and {Burgasser}, Adam J.},
        title = "{A Search for Binary Systems among the Nearest L Dwarfs}",
      journal = {\aj},
         year = 2006,
        month = aug,
       volume = {132},
       number = {2},
        pages = {891-901},
          doi = {10.1086/505626},
archivePrefix = {arXiv},
       eprint = {astro-ph/0606331},
 primaryClass = {astro-ph},
       adsurl = {https://ui.adsabs.harvard.edu/abs/2006AJ....132..891R}
}

@ARTICLE{2023ApJ...959...63S,
       author = {{Sanghi}, Aniket and {Liu}, Michael C. and {Best}, William M.~J. and {Dupuy}, Trent J. and {Siverd}, Robert J. and {Zhang}, Zhoujian and {Hurt}, Spencer A. and {Magnier}, Eugene A. and {Aller}, Kimberly M. and {Deacon}, Niall R.},
        title = "{The Hawaii Infrared Parallax Program. VI. The Fundamental Properties of 1000+ Ultracool Dwarfs and Planetary-mass Objects Using Optical to Mid-infrared Spectral Energy Distributions and Comparison to BT-Settl and ATMO 2020 Model Atmospheres}",
      journal = {\apj},
         year = 2023,
        month = dec,
       volume = {959},
       number = {1},
          eid = {63},
        pages = {63},
          doi = {10.3847/1538-4357/acff66},
archivePrefix = {arXiv},
       eprint = {2309.03082},
 primaryClass = {astro-ph.SR},
       adsurl = {https://ui.adsabs.harvard.edu/abs/2023ApJ...959...63S}
}

@ARTICLE{2014AJ....147...34S,
       author = {{Schneider}, Adam C. and {Cushing}, Michael C. and {Kirkpatrick}, J. Davy and {Mace}, Gregory N. and {Gelino}, Christopher R. and {Faherty}, Jacqueline K. and {Fajardo-Acosta}, Sergio and {Sheppard}, Scott S.},
        title = "{Discovery of the Young L Dwarf WISE J174102.78-464225.5}",
      journal = {\aj},
         year = 2014,
        month = feb,
       volume = {147},
       number = {2},
          eid = {34},
        pages = {34},
          doi = {10.1088/0004-6256/147/2/34},
archivePrefix = {arXiv},
       eprint = {1311.5941},
 primaryClass = {astro-ph.SR},
       adsurl = {https://ui.adsabs.harvard.edu/abs/2014AJ....147...34S}
}

@ARTICLE{2023AJ....166..103S,
       author = {{Schneider}, Adam C. and {Munn}, Jeffrey A. and {Vrba}, Frederick J. and {Bruursema}, Justice and {Dahm}, Scott E. and {Williams}, Stephen J. and {Liu}, Michael C. and {Dorland}, Bryan N.},
        title = "{Astrometry and Photometry for {\ensuremath{\approx}}1000 L, T, and Y Dwarfs from the UKIRT Hemisphere Survey}",
      journal = {\aj},
         year = 2023,
        month = sep,
       volume = {166},
       number = {3},
          eid = {103},
        pages = {103},
          doi = {10.3847/1538-3881/ace9bf},
archivePrefix = {arXiv},
       eprint = {2307.11882},
 primaryClass = {astro-ph.SR},
       adsurl = {https://ui.adsabs.harvard.edu/abs/2023AJ....166..103S}
}

@ARTICLE{2021AJ....161...42B,
       author = {{Best}, William M.~J. and {Liu}, Michael C. and {Magnier}, Eugene A. and {Dupuy}, Trent J.},
        title = "{A Volume-limited Sample of Ultracool Dwarfs. I. Construction, Space Density, and a Gap in the L/T Transition}",
      journal = {\aj},
         year = 2021,
        month = jan,
       volume = {161},
       number = {1},
          eid = {42},
        pages = {42},
          doi = {10.3847/1538-3881/abc893},
archivePrefix = {arXiv},
       eprint = {2010.15853},
 primaryClass = {astro-ph.SR},
       adsurl = {https://ui.adsabs.harvard.edu/abs/2021AJ....161...42B}
}

@ARTICLE{2014ApJ...790..133D,
       author = {{Dupuy}, Trent J. and {Liu}, Michael C. and {Ireland}, Michael J.},
        title = "{New Evidence for a Substellar Luminosity Problem: Dynamical Mass for the Brown Dwarf Binary Gl 417BC}",
      journal = {\apj},
         year = 2014,
        month = aug,
       volume = {790},
       number = {2},
          eid = {133},
        pages = {133},
          doi = {10.1088/0004-637X/790/2/133},
archivePrefix = {arXiv},
       eprint = {1406.1184},
 primaryClass = {astro-ph.SR},
       adsurl = {https://ui.adsabs.harvard.edu/abs/2014ApJ...790..133D}
}

@ARTICLE{2010ApJS..186...63R,
       author = {{Rice}, Emily L. and {Barman}, T. and {Mclean}, Ian S. and {Prato}, L. and {Kirkpatrick}, J. Davy},
        title = "{Physical Properties of Young Brown Dwarfs and Very Low Mass Stars Inferred from High-resolution Model Spectra}",
      journal = {\apjs},
         year = 2010,
        month = jan,
       volume = {186},
       number = {1},
        pages = {63-84},
          doi = {10.1088/0067-0049/186/1/63},
archivePrefix = {arXiv},
       eprint = {0911.3844},
 primaryClass = {astro-ph.SR},
       adsurl = {https://ui.adsabs.harvard.edu/abs/2010ApJS..186...63R}
}

@ARTICLE{2008ApJ...678.1372C,
       author = {{Cushing}, Michael C. and {Marley}, Mark S. and {Saumon}, D. and {Kelly}, Brandon C. and {Vacca}, William D. and {Rayner}, John T. and {Freedman}, Richard S. and {Lodders}, Katharina and {Roellig}, Thomas L.},
        title = "{Atmospheric Parameters of Field L and T Dwarfs}",
      journal = {\apj},
         year = 2008,
        month = may,
       volume = {678},
       number = {2},
        pages = {1372-1395},
          doi = {10.1086/526489},
archivePrefix = {arXiv},
       eprint = {0711.0801},
 primaryClass = {astro-ph},
       adsurl = {https://ui.adsabs.harvard.edu/abs/2008ApJ...678.1372C}
}

@ARTICLE{2009ApJ...702..154S,
       author = {{Stephens}, D.~C. and {Leggett}, S.~K. and {Cushing}, Michael C. and {Marley}, Mark S. and {Saumon}, D. and {Geballe}, T.~R. and {Golimowski}, David A. and {Fan}, Xiaohui and {Noll}, K.~S.},
        title = "{The 0.8-14.5 {\ensuremath{\mu}}m Spectra of Mid-L to Mid-T Dwarfs: Diagnostics of Effective Temperature, Grain Sedimentation, Gas Transport, and Surface Gravity}",
      journal = {\apj},
         year = 2009,
        month = sep,
       volume = {702},
       number = {1},
        pages = {154-170},
          doi = {10.1088/0004-637X/702/1/154},
archivePrefix = {arXiv},
       eprint = {0906.2991},
 primaryClass = {astro-ph.SR},
       adsurl = {https://ui.adsabs.harvard.edu/abs/2009ApJ...702..154S}
}

@ARTICLE{2007ApJ...667..537L,
       author = {{Leggett}, S.~K. and {Marley}, M.~S. and {Freedman}, R. and {Saumon}, D. and {Liu}, Michael C. and {Geballe}, T.~R. and {Golimowski}, D.~A. and {Stephens}, D.~C.},
        title = "{Physical and Spectral Characteristics of the T8 and Later Type Dwarfs}",
      journal = {\apj},
         year = 2007,
        month = sep,
       volume = {667},
       number = {1},
        pages = {537-548},
          doi = {10.1086/519948},
archivePrefix = {arXiv},
       eprint = {0705.2602},
 primaryClass = {astro-ph},
       adsurl = {https://ui.adsabs.harvard.edu/abs/2007ApJ...667..537L}
}

@ARTICLE{1996ApJ...465L.123A,
       author = {{Allard}, France and {Hauschildt}, Peter H. and {Baraffe}, Isabelle and {Chabrier}, Gilles},
        title = "{Synthetic Spectra and Mass Determination of the Brown Dwarf GI 229B}",
      journal = {\apjl},
         year = 1996,
        month = jul,
       volume = {465},
        pages = {L123},
          doi = {10.1086/310143},
       adsurl = {https://ui.adsabs.harvard.edu/abs/1996ApJ...465L.123A}
}

@ARTICLE{1996Sci...272.1919M,
       author = {{Marley}, M.~S. and {Saumon}, D. and {Guillot}, T. and {Freedman}, R.~S. and {Hubbard}, W.~B. and {Burrows}, A. and {Lunine}, J.~I.},
        title = "{Atmospheric, Evolutionary, and Spectral Models of the Brown Dwarf Gliese 229 B}",
      journal = {Science},
         year = 1996,
        month = jun,
       volume = {272},
       number = {5270},
        pages = {1919-1921},
          doi = {10.1126/science.272.5270.1919},
archivePrefix = {arXiv},
       eprint = {astro-ph/9606036},
 primaryClass = {astro-ph},
       adsurl = {https://ui.adsabs.harvard.edu/abs/1996Sci...272.1919M}
}

@ARTICLE{2000ApJ...541..374S,
       author = {{Saumon}, D. and {Geballe}, T.~R. and {Leggett}, S.~K. and {Marley}, M.~S. and {Freedman}, R.~S. and {Lodders}, K. and {Fegley}, Jr., B. and {Sengupta}, S.~K.},
        title = "{Molecular Abundances in the Atmosphere of the T Dwarf GL 229B}",
      journal = {\apj},
         year = 2000,
        month = sep,
       volume = {541},
       number = {1},
        pages = {374-389},
          doi = {10.1086/309410},
archivePrefix = {arXiv},
       eprint = {astro-ph/0003353},
 primaryClass = {astro-ph},
       adsurl = {https://ui.adsabs.harvard.edu/abs/2000ApJ...541..374S}
}

@ARTICLE{2001ApJ...556..373G,
       author = {{Geballe}, T.~R. and {Saumon}, D. and {Leggett}, S.~K. and {Knapp}, G.~R. and {Marley}, M.~S. and {Lodders}, K.},
        title = "{Infrared Observations and Modeling of One of the Coolest T Dwarfs: Gliese 570D}",
      journal = {\apj},
         year = 2001,
        month = jul,
       volume = {556},
       number = {1},
        pages = {373-379},
          doi = {10.1086/321575},
archivePrefix = {arXiv},
       eprint = {astro-ph/0103187},
 primaryClass = {astro-ph},
       adsurl = {https://ui.adsabs.harvard.edu/abs/2001ApJ...556..373G}
}

@ARTICLE{2006ApJ...647..552S,
       author = {{Saumon}, D. and {Marley}, M.~S. and {Cushing}, M.~C. and {Leggett}, S.~K. and {Roellig}, T.~L. and {Lodders}, K. and {Freedman}, R.~S.},
        title = "{Ammonia as a Tracer of Chemical Equilibrium in the T7.5 Dwarf Gliese 570D}",
      journal = {\apj},
         year = 2006,
        month = aug,
       volume = {647},
       number = {1},
        pages = {552-557},
          doi = {10.1086/505419},
archivePrefix = {arXiv},
       eprint = {astro-ph/0605563},
 primaryClass = {astro-ph},
       adsurl = {https://ui.adsabs.harvard.edu/abs/2006ApJ...647..552S}
}

@ARTICLE{2024ApJ...966L..11P,
       author = {{Petrus}, Simon and {Whiteford}, Niall and {Patapis}, Polychronis and {Biller}, Beth A. and {Skemer}, Andrew and {Hinkley}, Sasha and {Su{\'a}rez}, Genaro and {Palma-Bifani}, Paulina and {Morley}, Caroline V. and {Tremblin}, Pascal and {Charnay}, Benjamin and {Vos}, Johanna M. and {Wang}, Jason J. and {Stone}, Jordan M. and {Bonnefoy}, Micka{\"e}l and {Chauvin}, Ga{\"e}l and {Miles}, Brittany E. and {Carter}, Aarynn L. and {Lueber}, Anna and {Helling}, Christiane and {Sutlieff}, Ben J. and {Janson}, Markus and {Gonzales}, Eileen C. and {Hoch}, Kielan K.~W. and {Absil}, Olivier and {Balmer}, William O. and {Boccaletti}, Anthony and {Bonavita}, Mariangela and {Booth}, Mark and {Bowler}, Brendan P. and {Briesemeister}, Zackery W. and {Bryan}, Marta L. and {Calissendorff}, Per and {Cantalloube}, Faustine and {Chen}, Christine H. and {Choquet}, Elodie and {Christiaens}, Valentin and {Cugno}, Gabriele and {Currie}, Thayne and {Danielski}, Camilla and {De Furio}, Matthew and {Dupuy}, Trent J. and {Factor}, Samuel M. and {Faherty}, Jacqueline K. and {Fitzgerald}, Michael P. and {Fortney}, Jonathan J. and {Franson}, Kyle and {Girard}, Julien H. and {Grady}, Carol A. and {Henning}, Thomas and {Hines}, Dean C. and {Hood}, Callie E. and {Howe}, Alex R. and {Kalas}, Paul and {Kammerer}, Jens and {Kennedy}, Grant M. and {Kenworthy}, Matthew A. and {Kervella}, Pierre and {Kim}, Minjae and {Kitzmann}, Daniel and {Kraus}, Adam L. and {Kuzuhara}, Masayuki and {Lagage}, Pierre-Olivier and {Lagrange}, Anne-Marie and {Lawson}, Kellen and {Lazzoni}, Cecilia and {Leisenring}, Jarron M. and {Lew}, Ben W.~P. and {Liu}, Michael C. and {Liu}, Pengyu and {Llop-Sayson}, Jorge and {Lloyd}, James P. and {Macintosh}, Bruce and {M{\^a}lin}, Mathilde and {Manjavacas}, Elena and {Marino}, Sebasti{\'a}n and {Marley}, Mark S. and {Marois}, Christian and {Martinez}, Raquel A. and {Matthews}, Elisabeth C. and {Matthews}, Brenda C. and {Mawet}, Dimitri and {Mazoyer}, Johan and {McElwain}, Michael W. and {Metchev}, Stanimir and {Meyer}, Michael R. and {Millar-Blanchaer}, Maxwell A. and {Molli{\`e}re}, Paul and {Moran}, Sarah E. and {Mukherjee}, Sagnick and {Pantin}, Eric and {Perrin}, Marshall D. and {Pueyo}, Laurent and {Quanz}, Sascha P. and {Quirrenbach}, Andreas and {Ray}, Shrishmoy and {Rebollido}, Isabel and {Adams Redai}, Jea and {Ren}, Bin B. and {Rickman}, Emily and {Sallum}, Steph and {Samland}, Matthias and {Sargent}, Benjamin and {Schlieder}, Joshua E. and {Stapelfeldt}, Karl R. and {Tamura}, Motohide and {Tan}, Xianyu and {Theissen}, Christopher A. and {Uyama}, Taichi and {Vasist}, Malavika and {Vigan}, Arthur and {Wagner}, Kevin and {Ward-Duong}, Kimberly and {Wolff}, Schuyler G. and {Worthen}, Kadin and {Wyatt}, Mark C. and {Ygouf}, Marie and {Zurlo}, Alice and {Zhang}, Xi and {Zhang}, Keming and {Zhang}, Zhoujian and {Zhou}, Yifan},
        title = "{The JWST Early Release Science Program for Direct Observations of Exoplanetary Systems. V. Do Self-consistent Atmospheric Models Represent JWST Spectra? A Showcase with VHS 1256{\textendash}1257 b}",
      journal = {\apjl},
         year = 2024,
        month = may,
       volume = {966},
       number = {1},
          eid = {L11},
        pages = {L11},
          doi = {10.3847/2041-8213/ad3e7c},
archivePrefix = {arXiv},
       eprint = {2312.03852},
 primaryClass = {astro-ph.EP},
       adsurl = {https://ui.adsabs.harvard.edu/abs/2024ApJ...966L..11P}
}

@ARTICLE{2024ApJ...975...59M,
       author = {{Morley}, Caroline V. and {Mukherjee}, Sagnick and {Marley}, Mark S. and {Fortney}, Jonathan J. and {Visscher}, Channon and {Lupu}, Roxana and {Gharib-Nezhad}, Ehsan and {Thorngren}, Daniel and {Freedman}, Richard and {Batalha}, Natasha},
        title = "{The Sonora Substellar Atmosphere Models. III. Diamondback: Atmospheric Properties, Spectra, and Evolution for Warm Cloudy Substellar Objects}",
      journal = {\apj},
         year = 2024,
        month = nov,
       volume = {975},
       number = {1},
          eid = {59},
        pages = {59},
          doi = {10.3847/1538-4357/ad71d5},
archivePrefix = {arXiv},
       eprint = {2402.00758},
 primaryClass = {astro-ph.SR},
       adsurl = {https://ui.adsabs.harvard.edu/abs/2024ApJ...975...59M}
}

@ARTICLE{2013A&A...558A..33A,
   author = {{Astropy Collaboration} and {Robitaille}, T.~P. and {Tollerud}, E.~J. and 
	{Greenfield}, P. and {Droettboom}, M. and {Bray}, E. and {Aldcroft}, T. and 
	{Davis}, M. and {Ginsburg}, A. and {Price-Whelan}, A.~M. and 
	{Kerzendorf}, W.~E. and {Conley}, A. and {Crighton}, N. and 
	{Barbary}, K. and {Muna}, D. and {Ferguson}, H. and {Grollier}, F. and 
	{Parikh}, M.~M. and {Nair}, P.~H. and {Unther}, H.~M. and {Deil}, C. and 
	{Woillez}, J. and {Conseil}, S. and {Kramer}, R. and {Turner}, J.~E.~H. and 
	{Singer}, L. and {Fox}, R. and {Weaver}, B.~A. and {Zabalza}, V. and 
	{Edwards}, Z.~I. and {Azalee Bostroem}, K. and {Burke}, D.~J. and 
	{Casey}, A.~R. and {Crawford}, S.~M. and {Dencheva}, N. and 
	{Ely}, J. and {Jenness}, T. and {Labrie}, K. and {Lim}, P.~L. and 
	{Pierfederici}, F. and {Pontzen}, A. and {Ptak}, A. and {Refsdal}, B. and 
	{Servillat}, M. and {Streicher}, O.},
    title = "{Astropy: A community Python package for astronomy}",
  journal = {\aap},
archivePrefix = "arXiv",
   eprint = {1307.6212},
 primaryClass = "astro-ph.IM",
     year = 2013,
    month = oct,
   volume = 558,
      eid = {A33},
    pages = {A33},
      doi = {10.1051/0004-6361/201322068},
   adsurl = {https://ui.adsabs.harvard.edu/abs/2013A%26A...558A..33A}
}

@ARTICLE{2018AJ....156..123A,
   author = {{Astropy Collaboration} and {Price-Whelan}, A.~M. and {Sip{\H o}cz}, B.~M. and 
	{G{\"u}nther}, H.~M. and {Lim}, P.~L. and {Crawford}, S.~M. and 
	{Conseil}, S. and {Shupe}, D.~L. and {Craig}, M.~W. and {Dencheva}, N. and 
	{Ginsburg}, A. and {VanderPlas}, J.~T. and {Bradley}, L.~D. and 
	{P{\'e}rez-Su{\'a}rez}, D. and {de Val-Borro}, M. and {Aldcroft}, T.~L. and 
	{Cruz}, K.~L. and {Robitaille}, T.~P. and {Tollerud}, E.~J. and 
	{Ardelean}, C. and {Babej}, T. and {Bach}, Y.~P. and {Bachetti}, M. and 
	{Bakanov}, A.~V. and {Bamford}, S.~P. and {Barentsen}, G. and 
	{Barmby}, P. and {Baumbach}, A. and {Berry}, K.~L. and {Biscani}, F. and 
	{Boquien}, M. and {Bostroem}, K.~A. and {Bouma}, L.~G. and {Brammer}, G.~B. and 
	{Bray}, E.~M. and {Breytenbach}, H. and {Buddelmeijer}, H. and 
	{Burke}, D.~J. and {Calderone}, G. and {Cano Rodr{\'{\i}}guez}, J.~L. and 
	{Cara}, M. and {Cardoso}, J.~V.~M. and {Cheedella}, S. and {Copin}, Y. and 
	{Corrales}, L. and {Crichton}, D. and {D'Avella}, D. and {Deil}, C. and 
	{Depagne}, {\'E}. and {Dietrich}, J.~P. and {Donath}, A. and 
	{Droettboom}, M. and {Earl}, N. and {Erben}, T. and {Fabbro}, S. and 
	{Ferreira}, L.~A. and {Finethy}, T. and {Fox}, R.~T. and {Garrison}, L.~H. and 
	{Gibbons}, S.~L.~J. and {Goldstein}, D.~A. and {Gommers}, R. and 
	{Greco}, J.~P. and {Greenfield}, P. and {Groener}, A.~M. and 
	{Grollier}, F. and {Hagen}, A. and {Hirst}, P. and {Homeier}, D. and 
	{Horton}, A.~J. and {Hosseinzadeh}, G. and {Hu}, L. and {Hunkeler}, J.~S. and 
	{Ivezi{\'c}}, {\v Z}. and {Jain}, A. and {Jenness}, T. and {Kanarek}, G. and 
	{Kendrew}, S. and {Kern}, N.~S. and {Kerzendorf}, W.~E. and 
	{Khvalko}, A. and {King}, J. and {Kirkby}, D. and {Kulkarni}, A.~M. and 
	{Kumar}, A. and {Lee}, A. and {Lenz}, D. and {Littlefair}, S.~P. and 
	{Ma}, Z. and {Macleod}, D.~M. and {Mastropietro}, M. and {McCully}, C. and 
	{Montagnac}, S. and {Morris}, B.~M. and {Mueller}, M. and {Mumford}, S.~J. and 
	{Muna}, D. and {Murphy}, N.~A. and {Nelson}, S. and {Nguyen}, G.~H. and 
	{Ninan}, J.~P. and {N{\"o}the}, M. and {Ogaz}, S. and {Oh}, S. and 
	{Parejko}, J.~K. and {Parley}, N. and {Pascual}, S. and {Patil}, R. and 
	{Patil}, A.~A. and {Plunkett}, A.~L. and {Prochaska}, J.~X. and 
	{Rastogi}, T. and {Reddy Janga}, V. and {Sabater}, J. and {Sakurikar}, P. and 
	{Seifert}, M. and {Sherbert}, L.~E. and {Sherwood-Taylor}, H. and 
	{Shih}, A.~Y. and {Sick}, J. and {Silbiger}, M.~T. and {Singanamalla}, S. and 
	{Singer}, L.~P. and {Sladen}, P.~H. and {Sooley}, K.~A. and 
	{Sornarajah}, S. and {Streicher}, O. and {Teuben}, P. and {Thomas}, S.~W. and 
	{Tremblay}, G.~R. and {Turner}, J.~E.~H. and {Terr{\'o}n}, V. and 
	{van Kerkwijk}, M.~H. and {de la Vega}, A. and {Watkins}, L.~L. and 
	{Weaver}, B.~A. and {Whitmore}, J.~B. and {Woillez}, J. and 
	{Zabalza}, V. and {Astropy Contributors}},
    title = "{The Astropy Project: Building an Open-science Project and Status of the v2.0 Core Package}",
  journal = {\aj},
archivePrefix = "arXiv",
   eprint = {1801.02634},
 primaryClass = "astro-ph.IM",
     year = 2018,
    month = sep,
   volume = 156,
      eid = {123},
    pages = {123},
      doi = {10.3847/1538-3881/aabc4f},
   adsurl = {https://ui.adsabs.harvard.edu/abs/2018AJ....156..123A}
}

@Article{Hunter:2007,
  Author    = {Hunter, J. D.},
  Title     = {Matplotlib: A 2D graphics environment},
  Journal   = {Computing in Science \& Engineering},
  Volume    = {9},
  Number    = {3},
  Pages     = {90--95},
  publisher = {IEEE COMPUTER SOC},
  doi       = {10.1109/MCSE.2007.55},
  year      = 2007
}

@Article{PER-GRA:2007,
  Author    = {P\'erez, Fernando and Granger, Brian E.},
  Title     = {{IP}ython: a System for Interactive Scientific Computing},
  Journal   = {Computing in Science and Engineering},
  Volume    = {9},
  Number    = {3},
  Pages     = {21--29},
  month     = may,
  year      = 2007,
  url       = "https://ipython.org",
  ISSN      = "1521-9615",
  doi       = {10.1109/MCSE.2007.53},
  publisher = {IEEE Computer Society},
}

@Misc{scipy,
  author =    {Eric Jones and Travis Oliphant and Pearu Peterson and others},
  title =     {{SciPy}: Open source scientific tools for {Python}},
  year =      {2001},
  url = "http://www.scipy.org/",
  note = {[Online; accessed <today>]}
}

@Misc{numpy,
  author =    {Travis Oliphant},
  title =     {{NumPy}: A guide to {NumPy}},
  year =      {2006},
  howpublished = {USA: Trelgol Publishing},
  url = "http://www.numpy.org/",
  note = {[Online; accessed <today>]}
 }

@ARTICLE{2023A&A...670A..90P,
       author = {{Palma-Bifani}, P. and {Chauvin}, G. and {Bonnefoy}, M. and {Rojo}, P.~M. and {Petrus}, S. and {Rodet}, L. and {Langlois}, M. and {Allard}, F. and {Charnay}, B. and {Desgrange}, C. and {Homeier}, D. and {Lagrange}, A. -M. and {Beuzit}, J. -L. and {Baudoz}, P. and {Boccaletti}, A. and {Chomez}, A. and {Delorme}, P. and {Desidera}, S. and {Feldt}, M. and {Ginski}, C. and {Gratton}, R. and {Maire}, A. -L. and {Meyer}, M. and {Samland}, M. and {Snellen}, I. and {Vigan}, A. and {Zhang}, Y.},
        title = "{Peering into the young planetary system AB Pic. Atmosphere, orbit, obliquity, and second planetary candidate}",
      journal = {\aap},
         year = 2023,
        month = feb,
       volume = {670},
          eid = {A90},
        pages = {A90},
          doi = {10.1051/0004-6361/202244294},
archivePrefix = {arXiv},
       eprint = {2211.01474},
 primaryClass = {astro-ph.EP},
       adsurl = {https://ui.adsabs.harvard.edu/abs/2023A&A...670A..90P}
}

@ARTICLE{2021ApJ...920..146L,
       author = {{Luna}, Jessica L. and {Morley}, Caroline V.},
        title = "{Empirically Determining Substellar Cloud Compositions in the Era of the James Webb Space Telescope}",
      journal = {\apj},
         year = 2021,
        month = oct,
       volume = {920},
       number = {2},
          eid = {146},
        pages = {146},
          doi = {10.3847/1538-4357/ac1865},
archivePrefix = {arXiv},
       eprint = {2108.03161},
 primaryClass = {astro-ph.SR},
       adsurl = {https://ui.adsabs.harvard.edu/abs/2021ApJ...920..146L}
}

@ARTICLE{2024ApJ...963...73M,
       author = {{Mukherjee}, Sagnick and {Fortney}, Jonathan J. and {Morley}, Caroline V. and {Batalha}, Natasha E. and {Marley}, Mark S. and {Karalidi}, Theodora and {Visscher}, Channon and {Lupu}, Roxana and {Freedman}, Richard and {Gharib-Nezhad}, Ehsan},
        title = "{The Sonora Substellar Atmosphere Models. IV. Elf Owl: Atmospheric Mixing and Chemical Disequilibrium with Varying Metallicity and C/O Ratios}",
      journal = {\apj},
         year = 2024,
        month = mar,
       volume = {963},
       number = {1},
          eid = {73},
        pages = {73},
          doi = {10.3847/1538-4357/ad18c2},
archivePrefix = {arXiv},
       eprint = {2402.00756},
 primaryClass = {astro-ph.EP},
       adsurl = {https://ui.adsabs.harvard.edu/abs/2024ApJ...963...73M}
}

@ARTICLE{2024ApJ...961..210P,
       author = {{Phillips}, Mark W. and {Liu}, Michael C. and {Zhang}, Zhoujian},
        title = "{The Carbon-to-oxygen Ratio in Cool Brown Dwarfs and Giant Exoplanets. I. The Benchmark Late-T Dwarfs GJ 570D, HD 3651B, and Ross 458C}",
      journal = {\apj},
         year = 2024,
        month = feb,
       volume = {961},
       number = {2},
          eid = {210},
        pages = {210},
          doi = {10.3847/1538-4357/ad06ba},
archivePrefix = {arXiv},
       eprint = {2312.02001},
 primaryClass = {astro-ph.SR},
       adsurl = {https://ui.adsabs.harvard.edu/abs/2024ApJ...961..210P}
}

@ARTICLE{2025AJ....169....9Z,
       author = {{Zhang}, Zhoujian and {Mukherjee}, Sagnick and {Liu}, Michael C. and {Fortney}, Jonathan J. and {Mader}, Emily and {Best}, William M.~J. and {Dupuy}, Trent J. and {Leggett}, Sandy K. and {Karalidi}, Theodora and {Line}, Michael R. and {Marley}, Mark S. and {Morley}, Caroline V. and {Phillips}, Mark W. and {Siverd}, Robert J. and {Zalesky}, Joseph A.},
        title = "{Disequilibrium Chemistry, Diabatic Thermal Structure, and Clouds in the Atmosphere of COCONUTS-2b}",
      journal = {\aj},
         year = 2025,
        month = jan,
       volume = {169},
       number = {1},
          eid = {9},
        pages = {9},
          doi = {10.3847/1538-3881/ad8b2d},
archivePrefix = {arXiv},
       eprint = {2410.10939},
 primaryClass = {astro-ph.EP},
       adsurl = {https://ui.adsabs.harvard.edu/abs/2025AJ....169....9Z}
}

@ARTICLE{2020A&A...637A..38P,
       author = {{Phillips}, M.~W. and {Tremblin}, P. and {Baraffe}, I. and {Chabrier}, G. and {Allard}, N.~F. and {Spiegelman}, F. and {Goyal}, J.~M. and {Drummond}, B. and {H{\'e}brard}, E.},
        title = "{A new set of atmosphere and evolution models for cool T-Y brown dwarfs and giant exoplanets}",
      journal = {\aap},
         year = 2020,
        month = may,
       volume = {637},
          eid = {A38},
        pages = {A38},
          doi = {10.1051/0004-6361/201937381},
archivePrefix = {arXiv},
       eprint = {2003.13717},
 primaryClass = {astro-ph.SR},
       adsurl = {https://ui.adsabs.harvard.edu/abs/2020A&A...637A..38P}
}

@ARTICLE{2017ApJ...850..150B,
       author = {{Baudino}, Jean-Loup and {Molli{\`e}re}, Paul and {Venot}, Olivia and {Tremblin}, Pascal and {B{\'e}zard}, Bruno and {Lagage}, Pierre-Olivier},
        title = "{Toward the Analysis of JWST  Exoplanet Spectra: Identifying Troublesome Model Parameters}",
      journal = {\apj},
         year = 2017,
        month = dec,
       volume = {850},
       number = {2},
          eid = {150},
        pages = {150},
          doi = {10.3847/1538-4357/aa95be},
archivePrefix = {arXiv},
       eprint = {1710.08235},
 primaryClass = {astro-ph.EP},
       adsurl = {https://ui.adsabs.harvard.edu/abs/2017ApJ...850..150B}
}

@ARTICLE{2021ApJ...918...11L,
       author = {{Leggett}, S.~K. and {Tremblin}, Pascal and {Phillips}, Mark W. and {Dupuy}, Trent J. and {Marley}, Mark and {Morley}, Caroline and {Schneider}, Adam and {Caselden}, Dan and {Guillaume}, Colin and {Logsdon}, Sarah E.},
        title = "{Measuring and Replicating the 1-20 {\ensuremath{\mu}}m Energy Distributions of the Coldest Brown Dwarfs: Rotating, Turbulent, and Nonadiabatic Atmospheres}",
      journal = {\apj},
         year = 2021,
        month = sep,
       volume = {918},
       number = {1},
          eid = {11},
        pages = {11},
          doi = {10.3847/1538-4357/ac0cfe},
archivePrefix = {arXiv},
       eprint = {2107.00696},
 primaryClass = {astro-ph.SR},
       adsurl = {https://ui.adsabs.harvard.edu/abs/2021ApJ...918...11L}
}

@ARTICLE{2015ApJ...804L..17T,
       author = {{Tremblin}, P. and {Amundsen}, D.~S. and {Mourier}, P. and {Baraffe}, I. and {Chabrier}, G. and {Drummond}, B. and {Homeier}, D. and {Venot}, O.},
        title = "{Fingering Convection and Cloudless Models for Cool Brown Dwarf Atmospheres}",
      journal = {\apjl},
         year = 2015,
        month = may,
       volume = {804},
       number = {1},
          eid = {L17},
        pages = {L17},
          doi = {10.1088/2041-8205/804/1/L17},
archivePrefix = {arXiv},
       eprint = {1504.03334},
 primaryClass = {astro-ph.SR},
       adsurl = {https://ui.adsabs.harvard.edu/abs/2015ApJ...804L..17T}
}

@ARTICLE{2016ApJ...817L..19T,
       author = {{Tremblin}, P. and {Amundsen}, D.~S. and {Chabrier}, G. and {Baraffe}, I. and {Drummond}, B. and {Hinkley}, S. and {Mourier}, P. and {Venot}, O.},
        title = "{Cloudless Atmospheres for L/T Dwarfs and Extrasolar Giant Planets}",
      journal = {\apjl},
         year = 2016,
        month = feb,
       volume = {817},
       number = {2},
          eid = {L19},
        pages = {L19},
          doi = {10.3847/2041-8205/817/2/L19},
archivePrefix = {arXiv},
       eprint = {1601.03652},
 primaryClass = {astro-ph.EP},
       adsurl = {https://ui.adsabs.harvard.edu/abs/2016ApJ...817L..19T}
}

@ARTICLE{2020A&A...633A.124P,
       author = {{Petrus}, S. and {Bonnefoy}, M. and {Chauvin}, G. and {Babusiaux}, C. and {Delorme}, P. and {Lagrange}, A. -M. and {Florent}, N. and {Bayo}, A. and {Janson}, M. and {Biller}, B. and {Manjavacas}, E. and {Marleau}, G. -D. and {Kopytova}, T.},
        title = "{A new take on the low-mass brown dwarf companions on wide orbits in Upper-Scorpius}",
      journal = {\aap},
         year = 2020,
        month = jan,
       volume = {633},
          eid = {A124},
        pages = {A124},
          doi = {10.1051/0004-6361/201935732},
archivePrefix = {arXiv},
       eprint = {1910.00347},
 primaryClass = {astro-ph.EP},
       adsurl = {https://ui.adsabs.harvard.edu/abs/2020A&A...633A.124P}
}

@ARTICLE{2006MNRAS.368.1281P,
       author = {{Pinfield}, D.~J. and {Jones}, H.~R.~A. and {Lucas}, P.~W. and {Kendall}, T.~R. and {Folkes}, S.~L. and {Day-Jones}, A.~C. and {Chappelle}, R.~J. and {Steele}, I.~A.},
        title = "{Finding benchmark brown dwarfs to probe the substellar initial mass function as a function of time}",
      journal = {\mnras},
         year = 2006,
        month = may,
       volume = {368},
       number = {3},
        pages = {1281-1295},
          doi = {10.1111/j.1365-2966.2006.10213.x},
archivePrefix = {arXiv},
       eprint = {astro-ph/0603320},
 primaryClass = {astro-ph},
       adsurl = {https://ui.adsabs.harvard.edu/abs/2006MNRAS.368.1281P}
}

@ARTICLE{2008ApJ...689..436L,
       author = {{Liu}, Michael C. and {Dupuy}, Trent J. and {Ireland}, Michael J.},
        title = "{Keck Laser Guide Star Adaptive Optics Monitoring of 2MASS J15344984-2952274AB: First Dynamical Mass Determination of a Binary T Dwarf}",
      journal = {\apj},
         year = 2008,
        month = dec,
       volume = {689},
       number = {1},
        pages = {436-460},
          doi = {10.1086/591837},
archivePrefix = {arXiv},
       eprint = {0807.0238},
 primaryClass = {astro-ph},
       adsurl = {https://ui.adsabs.harvard.edu/abs/2008ApJ...689..436L}
}

@ARTICLE{2018ApJ...856...23G,
       author = {{Gagn{\'e}}, Jonathan and {Mamajek}, Eric E. and {Malo}, Lison and {Riedel}, Adric and {Rodriguez}, David and {Lafreni{\`e}re}, David and {Faherty}, Jacqueline K. and {Roy-Loubier}, Olivier and {Pueyo}, Laurent and {Robin}, Annie C. and {Doyon}, Ren{\'e}},
        title = "{BANYAN. XI. The BANYAN {\ensuremath{\Sigma}} Multivariate Bayesian Algorithm to Identify Members of Young Associations with 150 pc}",
      journal = {\apj},
         year = 2018,
        month = mar,
       volume = {856},
       number = {1},
          eid = {23},
        pages = {23},
          doi = {10.3847/1538-4357/aaae09},
archivePrefix = {arXiv},
       eprint = {1801.09051},
 primaryClass = {astro-ph.SR},
       adsurl = {https://ui.adsabs.harvard.edu/abs/2018ApJ...856...23G}
}

@ARTICLE{2020ApJ...891..171Z,
       author = {{Zhang}, Zhoujian and {Liu}, Michael C. and {Hermes}, J.~J. and {Magnier}, Eugene A. and {Marley}, Mark S. and {Tremblay}, Pier-Emmanuel and {Tucker}, Michael A. and {Do}, Aaron and {Payne}, Anna V. and {Shappee}, Benjamin J.},
        title = "{COol Companions ON Ultrawide orbiTS (COCONUTS). I. A High-gravity T4 Benchmark around an Old White Dwarf and a Re-examination of the Surface-gravity Dependence of the L/T Transition}",
      journal = {\apj},
         year = 2020,
        month = mar,
       volume = {891},
       number = {2},
          eid = {171},
        pages = {171},
          doi = {10.3847/1538-4357/ab765c},
archivePrefix = {arXiv},
       eprint = {2002.05723},
 primaryClass = {astro-ph.SR},
       adsurl = {https://ui.adsabs.harvard.edu/abs/2020ApJ...891..171Z}
}

@ARTICLE{2021ApJ...911....7Z,
       author = {{Zhang}, Zhoujian and {Liu}, Michael C. and {Best}, William M.~J. and {Dupuy}, Trent J. and {Siverd}, Robert J.},
        title = "{The Hawaii Infrared Parallax Program. V. New T-dwarf Members and Candidate Members of Nearby Young Moving Groups}",
      journal = {\apj},
         year = 2021,
        month = apr,
       volume = {911},
       number = {1},
          eid = {7},
        pages = {7},
          doi = {10.3847/1538-4357/abe3fa},
archivePrefix = {arXiv},
       eprint = {2102.05045},
 primaryClass = {astro-ph.EP},
       adsurl = {https://ui.adsabs.harvard.edu/abs/2021ApJ...911....7Z}
}

@ARTICLE{2022ApJ...935...15Z,
       author = {{Zhang}, Zhoujian and {Liu}, Michael C. and {Morley}, Caroline V. and {Magnier}, Eugene A. and {Tucker}, Michael A. and {Vanderbosch}, Zachary P. and {Do}, Aaron and {Shappee}, Benjamin J.},
        title = "{COol Companions ON Ultrawide orbiTS (COCONUTS). III. A Very Red L6 Benchmark Brown Dwarf around a Young M5 Dwarf}",
      journal = {\apj},
         year = 2022,
        month = aug,
       volume = {935},
       number = {1},
          eid = {15},
        pages = {15},
          doi = {10.3847/1538-4357/ac7ce9},
archivePrefix = {arXiv},
       eprint = {2207.02865},
 primaryClass = {astro-ph.SR},
       adsurl = {https://ui.adsabs.harvard.edu/abs/2022ApJ...935...15Z}
}

@ARTICLE{2024AJ....167..253R,
       author = {{Rothermich}, Austin and {Faherty}, Jacqueline K. and {Bardalez-Gagliuffi}, Daniella and {Schneider}, Adam C. and {Kirkpatrick}, J. Davy and {Meisner}, Aaron M. and {Burgasser}, Adam J. and {Kuchner}, Marc and {Allers}, Katelyn and {Gagn{\'e}}, Jonathan and {Caselden}, Dan and {Calamari}, Emily and {Popinchalk}, Mark and {Su{\'a}rez}, Genaro and {Gerasimov}, Roman and {Aganze}, Christian and {Softich}, Emma and {Hsu}, Chin-Chun and {Karpoor}, Preethi and {Theissen}, Christopher A. and {Rees}, Jon and {Cecilio-Flores-Elie}, Rosario and {Cushing}, Michael C. and {Marocco}, Federico and {Casewell}, Sarah and {Bickle}, Thomas P. and {Hamlet}, Les and {Allen}, Michaela B. and {Beaulieu}, Paul and {Colin}, Guillaume and {Gantier}, Jean Marc and {Gramaize}, Leopold and {Jalowiczor}, Peter and {Kabatnik}, Martin and {Kiwy}, Frank and {Martin}, David W. and {Pendrill}, Billy and {Pumphrey}, Ben and {Sainio}, Arttu and {Schumann}, Jorg and {Stevnbak}, Nikolaj and {Sun}, Guoyou and {Tanner}, Christopher and {Thakur}, Vinod and {Thevenot}, Melina and {Wedracki}, Zbigniew},
        title = "{89 New Ultracool Dwarf Comoving Companions Identified with the Backyard Worlds: Planet 9 Citizen Science Project}",
      journal = {\aj},
         year = 2024,
        month = jun,
       volume = {167},
       number = {6},
          eid = {253},
        pages = {253},
          doi = {10.3847/1538-3881/ad324e},
archivePrefix = {arXiv},
       eprint = {2403.04592},
 primaryClass = {astro-ph.SR},
       adsurl = {https://ui.adsabs.harvard.edu/abs/2024AJ....167..253R}
}

@ARTICLE{2024arXiv241204597B,
       author = {{Bravo}, Alexia and {Schneider}, Adam C. and {Casewell}, Sarah and {Rothermich}, Austin and {Faherty}, Jacqueline K. and {French}, Jenni R. and {Bickle}, Thomas P. and {Meisner}, Aaron M. and {Kirkpatrick}, J. Davy and {Kuchner}, Marc J. and {Burgasser}, Adam J. and {Marocco}, Federico and {Debes}, John H. and {Sainio}, Arttu and {Gramaize}, L{\'e}opold and {Kiwy}, Frank and {Jalowiczor}, Peter A. and {Abdullahi}, Awab},
        title = "{New Ultracool Companions to Nearby White Dwarfs}",
      journal = {arXiv e-prints},
         year = 2024,
        month = dec,
          eid = {arXiv:2412.04597},
        pages = {arXiv:2412.04597},
          doi = {10.48550/arXiv.2412.04597},
archivePrefix = {arXiv},
       eprint = {2412.04597},
 primaryClass = {astro-ph.SR},
       adsurl = {https://ui.adsabs.harvard.edu/abs/2024arXiv241204597B}
}

@ARTICLE{2024ApJ...963...67C,
       author = {{Calamari}, Emily and {Faherty}, Jacqueline K. and {Visscher}, Channon and {Gemma}, Marina E. and {Burningham}, Ben and {Rothermich}, Austin},
        title = "{Predicting Cloud Conditions in Substellar Mass Objects Using Ultracool Dwarf Companions}",
      journal = {\apj},
         year = 2024,
        month = mar,
       volume = {963},
       number = {1},
          eid = {67},
        pages = {67},
          doi = {10.3847/1538-4357/ad1f6d},
archivePrefix = {arXiv},
       eprint = {2401.11038},
 primaryClass = {astro-ph.SR},
       adsurl = {https://ui.adsabs.harvard.edu/abs/2024ApJ...963...67C}
}

@ARTICLE{2012RSPTA.370.2765A,
       author = {{Allard}, F. and {Homeier}, D. and {Freytag}, B.},
        title = "{Models of very-low-mass stars, brown dwarfs and exoplanets}",
      journal = {Philosophical Transactions of the Royal Society of London Series A},
         year = 2012,
        month = jun,
       volume = {370},
       number = {1968},
        pages = {2765-2777},
          doi = {10.1098/rsta.2011.0269},
archivePrefix = {arXiv},
       eprint = {1112.3591},
 primaryClass = {astro-ph.SR},
       adsurl = {https://ui.adsabs.harvard.edu/abs/2012RSPTA.370.2765A}
}

@ARTICLE{2015A&A...577A..42B,
       author = {{Baraffe}, Isabelle and {Homeier}, Derek and {Allard}, France and {Chabrier}, Gilles},
        title = "{New evolutionary models for pre-main sequence and main sequence low-mass stars down to the hydrogen-burning limit}",
      journal = {\aap},
         year = 2015,
        month = may,
       volume = {577},
          eid = {A42},
        pages = {A42},
          doi = {10.1051/0004-6361/201425481},
archivePrefix = {arXiv},
       eprint = {1503.04107},
 primaryClass = {astro-ph.SR},
       adsurl = {https://ui.adsabs.harvard.edu/abs/2015A&A...577A..42B}
}

@ARTICLE{2001ApJ...556..872A,
       author = {{Ackerman}, Andrew S. and {Marley}, Mark S.},
        title = "{Precipitating Condensation Clouds in Substellar Atmospheres}",
      journal = {\apj},
         year = 2001,
        month = aug,
       volume = {556},
       number = {2},
        pages = {872-884},
          doi = {10.1086/321540},
archivePrefix = {arXiv},
       eprint = {astro-ph/0103423},
 primaryClass = {astro-ph},
       adsurl = {https://ui.adsabs.harvard.edu/abs/2001ApJ...556..872A}
}

@ARTICLE{2013Sci...341.1492D,
       author = {{Dupuy}, Trent J. and {Kraus}, Adam L.},
        title = "{Distances, Luminosities, and Temperatures of the Coldest Known Substellar Objects}",
      journal = {Science},
         year = 2013,
        month = sep,
       volume = {341},
       number = {6153},
        pages = {1492-1495},
          doi = {10.1126/science.1241917},
archivePrefix = {arXiv},
       eprint = {1309.1422},
 primaryClass = {astro-ph.SR},
       adsurl = {https://ui.adsabs.harvard.edu/abs/2013Sci...341.1492D}
}

@ARTICLE{2018ApJS..234....1B,
       author = {{Best}, William M.~J. and {Magnier}, Eugene A. and {Liu}, Michael C. and {Aller}, Kimberly M. and {Zhang}, Zhoujian and {Burgett}, W.~S. and {Chambers}, K.~C. and {Draper}, P. and {Flewelling}, H. and {Kaiser}, N. and {Kudritzki}, R. -P. and {Metcalfe}, N. and {Tonry}, J.~L. and {Wainscoat}, R.~J. and {Waters}, C.},
        title = "{Photometry and Proper Motions of M, L, and T Dwarfs from the Pan-STARRS1 3{\ensuremath{\pi}} Survey}",
      journal = {\apjs},
         year = 2018,
        month = jan,
       volume = {234},
       number = {1},
          eid = {1},
        pages = {1},
          doi = {10.3847/1538-4365/aa9982},
archivePrefix = {arXiv},
       eprint = {1701.00490},
 primaryClass = {astro-ph.SR},
       adsurl = {https://ui.adsabs.harvard.edu/abs/2018ApJS..234....1B}
}

@ARTICLE{2020ApJ...905...46G,
       author = {{Gonzales}, Eileen C. and {Burningham}, Ben and {Faherty}, Jacqueline K. and {Cleary}, Colleen and {Visscher}, Channon and {Marley}, Mark S. and {Lupu}, Roxana and {Freedman}, Richard},
        title = "{Retrieval of the d/sdL7+T7.5p Binary SDSS J1416+1348AB}",
      journal = {\apj},
         year = 2020,
        month = dec,
       volume = {905},
       number = {1},
          eid = {46},
        pages = {46},
          doi = {10.3847/1538-4357/abbee2},
archivePrefix = {arXiv},
       eprint = {2010.01224},
 primaryClass = {astro-ph.SR},
       adsurl = {https://ui.adsabs.harvard.edu/abs/2020ApJ...905...46G},
}

@ARTICLE{2013ApJ...768..121A,
       author = {{Apai}, D{\'a}niel and {Radigan}, Jacqueline and {Buenzli}, Esther and {Burrows}, Adam and {Reid}, Iain Neill and {Jayawardhana}, Ray},
        title = "{HST Spectral Mapping of L/T Transition Brown Dwarfs Reveals Cloud Thickness Variations}",
      journal = {\apj},
         year = 2013,
        month = may,
       volume = {768},
       number = {2},
          eid = {121},
        pages = {121},
          doi = {10.1088/0004-637X/768/2/121},
archivePrefix = {arXiv},
       eprint = {1303.4151},
 primaryClass = {astro-ph.EP},
       adsurl = {https://ui.adsabs.harvard.edu/abs/2013ApJ...768..121A}
}

@ARTICLE{2006ApJ...648..614C,
       author = {{Cushing}, Michael C. and {Roellig}, Thomas L. and {Marley}, Mark S. and {Saumon}, D. and {Leggett}, S.~K. and {Kirkpatrick}, J. Davy and {Wilson}, John C. and {Sloan}, G.~C. and {Mainzer}, Amy K. and {Van Cleve}, Jeff E. and {Houck}, James R.},
        title = "{A Spitzer Infrared Spectrograph Spectral Sequence of M, L, and T Dwarfs}",
      journal = {\apj},
         year = 2006,
        month = sep,
       volume = {648},
       number = {1},
        pages = {614-628},
          doi = {10.1086/505637},
archivePrefix = {arXiv},
       eprint = {astro-ph/0605639},
 primaryClass = {astro-ph},
       adsurl = {https://ui.adsabs.harvard.edu/abs/2006ApJ...648..614C}
}

@ARTICLE{2023MNRAS.523.4739S,
       author = {{Su{\'a}rez}, Genaro and {Metchev}, Stanimir},
        title = "{Ultracool dwarfs observed with the Spitzer Infrared Spectrograph - III. Dust grains in young L dwarf atmospheres are heavier}",
      journal = {\mnras},
         year = 2023,
        month = aug,
       volume = {523},
       number = {3},
        pages = {4739-4747},
          doi = {10.1093/mnras/stad1711},
archivePrefix = {arXiv},
       eprint = {2306.01119},
 primaryClass = {astro-ph.SR},
       adsurl = {https://ui.adsabs.harvard.edu/abs/2023MNRAS.523.4739S}
}

@ARTICLE{2007MNRAS.377L..74L,
       author = {{Liddle}, Andrew R.},
        title = "{Information criteria for astrophysical model selection}",
      journal = {\mnras},
         year = 2007,
        month = may,
       volume = {377},
       number = {1},
        pages = {L74-L78},
          doi = {10.1111/j.1745-3933.2007.00306.x},
archivePrefix = {arXiv},
       eprint = {astro-ph/0701113},
 primaryClass = {astro-ph},
       adsurl = {https://ui.adsabs.harvard.edu/abs/2007MNRAS.377L..74L}
}

@ARTICLE{1978AnSta...6..461S,
       author = {{Schwarz}, Gideon},
        title = "{Estimating the Dimension of a Model}",
      journal = {Annals of Statistics},
         year = 1978,
        month = jul,
       volume = {6},
       number = {2},
        pages = {461-464},
       adsurl = {https://ui.adsabs.harvard.edu/abs/1978AnSta...6..461S}
}

@ARTICLE{Kass1995,
author = {Robert E. Kass and Adrian E. Raftery},
title = {Bayes Factors},
journal = {Journal of the American Statistical Association},
volume = {90},
number = {430},
pages = {773--795},
year = {1995},
publisher = {ASA Website},
doi = {10.1080/01621459.1995.10476572},
URL = {https://www.tandfonline.com/doi/abs/10.1080/01621459.1995.10476572},
eprint = {https://www.tandfonline.com/doi/pdf/10.1080/01621459.1995.10476572}
}

@ARTICLE{2002ApJ...568..335M,
       author = {{Marley}, Mark S. and {Seager}, S. and {Saumon}, D. and {Lodders}, Katharina and {Ackerman}, Andrew S. and {Freedman}, Richard S. and {Fan}, Xiaohui},
        title = "{Clouds and Chemistry: Ultracool Dwarf Atmospheric Properties from Optical and Infrared Colors}",
      journal = {\apj},
         year = 2002,
        month = mar,
       volume = {568},
       number = {1},
        pages = {335-342},
          doi = {10.1086/338800},
archivePrefix = {arXiv},
       eprint = {astro-ph/0105438},
 primaryClass = {astro-ph},
       adsurl = {https://ui.adsabs.harvard.edu/abs/2002ApJ...568..335M}
}

@ARTICLE{2002ApJ...575..264T,
       author = {{Tsuji}, Takashi},
        title = "{Dust in the Photospheric Environment: Unified Cloudy Models of M, L, and T Dwarfs}",
      journal = {\apj},
         year = 2002,
        month = aug,
       volume = {575},
       number = {1},
        pages = {264-290},
          doi = {10.1086/341262},
archivePrefix = {arXiv},
       eprint = {astro-ph/0204401},
 primaryClass = {astro-ph},
       adsurl = {https://ui.adsabs.harvard.edu/abs/2002ApJ...575..264T}
}

@ARTICLE{2002ApJ...571L.151B,
       author = {{Burgasser}, Adam J. and {Marley}, Mark S. and {Ackerman}, Andrew S. and {Saumon}, Didier and {Lodders}, Katharina and {Dahn}, Conard C. and {Harris}, Hugh C. and {Kirkpatrick}, J. Davy},
        title = "{Evidence of Cloud Disruption in the L/T Dwarf Transition}",
      journal = {\apjl},
         year = 2002,
        month = jun,
       volume = {571},
       number = {2},
        pages = {L151-L154},
          doi = {10.1086/341343},
archivePrefix = {arXiv},
       eprint = {astro-ph/0205051},
 primaryClass = {astro-ph},
       adsurl = {https://ui.adsabs.harvard.edu/abs/2002ApJ...571L.151B}
}

@ARTICLE{2010ApJ...723L.117M,
       author = {{Marley}, Mark S. and {Saumon}, Didier and {Goldblatt}, Colin},
        title = "{A Patchy Cloud Model for the L to T Dwarf Transition}",
      journal = {\apjl},
         year = 2010,
        month = nov,
       volume = {723},
       number = {1},
        pages = {L117-L121},
          doi = {10.1088/2041-8205/723/1/L117},
archivePrefix = {arXiv},
       eprint = {1009.6217},
 primaryClass = {astro-ph.SR},
       adsurl = {https://ui.adsabs.harvard.edu/abs/2010ApJ...723L.117M}
}

@ARTICLE{2025arXiv250500978T,
       author = {{Turner}, Savanah K. and {Stephens}, Denise C. and {Scoresby}, Conner B. and {Miller}, Josh A.},
        title = "{A Survey Of Model Fits to Brown Dwarf Spectra Through the L-T Sequence}",
      journal = {arXiv e-prints},
         year = 2025,
        month = may,
          eid = {arXiv:2505.00978},
        pages = {arXiv:2505.00978},
          doi = {10.48550/arXiv.2505.00978},
archivePrefix = {arXiv},
       eprint = {2505.00978},
 primaryClass = {astro-ph.SR},
       adsurl = {https://ui.adsabs.harvard.edu/abs/2025arXiv250500978T}
}

@ARTICLE{2018ApJ...858...41Z,
       author = {{Zhang}, Zhoujian and {Liu}, Michael C. and {Best}, William M.~J. and {Magnier}, Eugene A. and {Aller}, Kimberly M. and {Chambers}, K.~C. and {Draper}, P.~W. and {Flewelling}, H. and {Hodapp}, K.~W. and {Kaiser}, N. and {Kudritzki}, R. -P. and {Metcalfe}, N. and {Wainscoat}, R.~J. and {Waters}, C.},
        title = "{The Pan-STARRS1 Proper-motion Survey for Young Brown Dwarfs in Nearby Star-forming Regions. I. Taurus Discoveries and a Reddening-free Classification Method for Ultracool Dwarfs}",
      journal = {\apj},
         year = 2018,
        month = may,
       volume = {858},
       number = {1},
          eid = {41},
        pages = {41},
          doi = {10.3847/1538-4357/aab269},
archivePrefix = {arXiv},
       eprint = {1804.01533},
 primaryClass = {astro-ph.SR},
       adsurl = {https://ui.adsabs.harvard.edu/abs/2018ApJ...858...41Z}
}

@ARTICLE{2017ApJ...836..200G,
       author = {{Gully-Santiago}, Michael A. and {Herczeg}, Gregory J. and {Czekala}, Ian and {Somers}, Garrett and {Grankin}, Konstantin and {Covey}, Kevin R. and {Donati}, J.~F. and {Alencar}, Silvia H.~P. and {Hussain}, Gaitee A.~J. and {Shappee}, Benjamin J. and {Mace}, Gregory N. and {Lee}, Jae-Joon and {Holoien}, T.~W. -S. and {Jose}, Jessy and {Liu}, Chun-Fan},
        title = "{Placing the Spotted T Tauri Star LkCa 4 on an HR Diagram}",
      journal = {\apj},
         year = 2017,
        month = feb,
       volume = {836},
       number = {2},
          eid = {200},
        pages = {200},
          doi = {10.3847/1538-4357/836/2/200},
archivePrefix = {arXiv},
       eprint = {1701.06703},
 primaryClass = {astro-ph.SR},
       adsurl = {https://ui.adsabs.harvard.edu/abs/2017ApJ...836..200G}
}

@ARTICLE{2025AJ....170...64Z,
       author = {{Zhang}, Zhoujian and {Molli{\`e}re}, Paul and {Fortney}, Jonathan J. and {Marley}, Mark S.},
        title = "{ELemental Abundances of Planets and Brown Dwarfs Imaged around Stars (ELPIS). II. The Jupiter-like Inhomogeneous Atmosphere of the First Directly Imaged Planetary-mass Companion 2MASS 1207 b}",
      journal = {\aj},
         year = 2025,
        month = aug,
       volume = {170},
       number = {2},
          eid = {64},
        pages = {64},
          doi = {10.3847/1538-3881/addfcb},
archivePrefix = {arXiv},
       eprint = {2502.18559},
 primaryClass = {astro-ph.EP},
       adsurl = {https://ui.adsabs.harvard.edu/abs/2025AJ....170...64Z}
}

@ARTICLE{2009ApJ...692..729D,
       author = {{Dupuy}, Trent J. and {Liu}, Michael C. and {Ireland}, Michael J.},
        title = "{Dynamical Mass of the Substellar Benchmark Binary HD 130948BC}",
      journal = {\apj},
         year = 2009,
        month = feb,
       volume = {692},
       number = {1},
        pages = {729-752},
          doi = {10.1088/0004-637X/692/1/729},
archivePrefix = {arXiv},
       eprint = {0807.2450},
 primaryClass = {astro-ph},
       adsurl = {https://ui.adsabs.harvard.edu/abs/2009ApJ...692..729D}
}

@ARTICLE{2020AJ....160..196B,
       author = {{Brandt}, Timothy D. and {Dupuy}, Trent J. and {Bowler}, Brendan P. and {Bardalez Gagliuffi}, Daniella C. and {Faherty}, Jacqueline and {Brandt}, G. Mirek and {Michalik}, Daniel},
        title = "{A Dynamical Mass of 70 {\ensuremath{\pm}} 5 M$_{Jup}$ for Gliese 229B, the First T Dwarf}",
      journal = {\aj},
         year = 2020,
        month = oct,
       volume = {160},
       number = {4},
          eid = {196},
        pages = {196},
          doi = {10.3847/1538-3881/abb45e},
archivePrefix = {arXiv},
       eprint = {1910.01652},
 primaryClass = {astro-ph.SR},
       adsurl = {https://ui.adsabs.harvard.edu/abs/2020AJ....160..196B}
}

@ARTICLE{2013ApJ...777L..20L,
       author = {{Liu}, Michael C. and {Magnier}, Eugene A. and {Deacon}, Niall R. and {Allers}, Katelyn N. and {Dupuy}, Trent J. and {Kotson}, Michael C. and {Aller}, Kimberly M. and {Burgett}, W.~S. and {Chambers}, K.~C. and {Draper}, P.~W. and {Hodapp}, K.~W. and {Jedicke}, R. and {Kaiser}, N. and {Kudritzki}, R. -P. and {Metcalfe}, N. and {Morgan}, J.~S. and {Price}, P.~A. and {Tonry}, J.~L. and {Wainscoat}, R.~J.},
        title = "{The Extremely Red, Young L Dwarf PSO J318.5338-22.8603: A Free-floating Planetary-mass Analog to Directly Imaged Young Gas-giant Planets}",
      journal = {\apjl},
         year = 2013,
        month = nov,
       volume = {777},
       number = {2},
          eid = {L20},
        pages = {L20},
          doi = {10.1088/2041-8205/777/2/L20},
archivePrefix = {arXiv},
       eprint = {1310.0457},
 primaryClass = {astro-ph.EP},
       adsurl = {https://ui.adsabs.harvard.edu/abs/2013ApJ...777L..20L}
}

@ARTICLE{2015ApJ...805...56D,
       author = {{Dupuy}, Trent J. and {Liu}, Michael C. and {Leggett}, S.~K. and {Ireland}, Michael J. and {Chiu}, Kuenley and {Golimowski}, David A.},
        title = "{The Mass-Luminosity Relation in the L/T Transition: Individual Dynamical Masses for the New J-band Flux Reversal Binary SDSSJ105213.51+442255.7AB}",
      journal = {\apj},
         year = 2015,
        month = may,
       volume = {805},
       number = {1},
          eid = {56},
        pages = {56},
          doi = {10.1088/0004-637X/805/1/56},
archivePrefix = {arXiv},
       eprint = {1503.06212},
 primaryClass = {astro-ph.SR},
       adsurl = {https://ui.adsabs.harvard.edu/abs/2015ApJ...805...56D}
}

@ARTICLE{1999ApJ...519..793L,
       author = {{Lodders}, Katharina},
        title = "{Alkali Element Chemistry in Cool Dwarf Atmospheres}",
      journal = {\apj},
         year = 1999,
        month = jul,
       volume = {519},
       number = {2},
        pages = {793-801},
          doi = {10.1086/307387},
       adsurl = {https://ui.adsabs.harvard.edu/abs/1999ApJ...519..793L}
}

@ARTICLE{1999ApJ...512..843B,
       author = {{Burrows}, Adam and {Sharp}, C.~M.},
        title = "{Chemical Equilibrium Abundances in Brown Dwarf and Extrasolar Giant Planet Atmospheres}",
      journal = {\apj},
         year = 1999,
        month = feb,
       volume = {512},
       number = {2},
        pages = {843-863},
          doi = {10.1086/306811},
archivePrefix = {arXiv},
       eprint = {astro-ph/9807055},
 primaryClass = {astro-ph},
       adsurl = {https://ui.adsabs.harvard.edu/abs/1999ApJ...512..843B}
}

@ARTICLE{2001ApJ...556..357A,
       author = {{Allard}, France and {Hauschildt}, Peter H. and {Alexander}, David R. and {Tamanai}, Akemi and {Schweitzer}, Andreas},
        title = "{The Limiting Effects of Dust in Brown Dwarf Model Atmospheres}",
      journal = {\apj},
         year = 2001,
        month = jul,
       volume = {556},
       number = {1},
        pages = {357-372},
          doi = {10.1086/321547},
archivePrefix = {arXiv},
       eprint = {astro-ph/0104256},
 primaryClass = {astro-ph},
       adsurl = {https://ui.adsabs.harvard.edu/abs/2001ApJ...556..357A}
}

@ARTICLE{2016ApJ...830...96H,
       author = {{Hiranaka}, Kay and {Cruz}, Kelle L. and {Douglas}, Stephanie T. and {Marley}, Mark S. and {Baldassare}, Vivienne F.},
        title = "{Exploring the Role of Sub-micron-sized Dust Grains in the Atmospheres of Red L0-L6 Dwarfs}",
      journal = {\apj},
         year = 2016,
        month = oct,
       volume = {830},
       number = {2},
          eid = {96},
        pages = {96},
          doi = {10.3847/0004-637X/830/2/96},
archivePrefix = {arXiv},
       eprint = {1606.09485},
 primaryClass = {astro-ph.SR},
       adsurl = {https://ui.adsabs.harvard.edu/abs/2016ApJ...830...96H}
}

@ARTICLE{2007ApJ...658..617B,
       author = {{Burgasser}, Adam J.},
        title = "{The Physical Properties of HD 3651B: An Extrasolar Nemesis?}",
      journal = {\apj},
         year = 2007,
        month = mar,
       volume = {658},
       number = {1},
        pages = {617-621},
          doi = {10.1086/511176},
archivePrefix = {arXiv},
       eprint = {astro-ph/0611542},
 primaryClass = {astro-ph},
       adsurl = {https://ui.adsabs.harvard.edu/abs/2007ApJ...658..617B}
}

@ARTICLE{2008ApJ...681..579B,
       author = {{Burgasser}, Adam J. and {Liu}, Michael C. and {Ireland}, Michael J. and {Cruz}, Kelle L. and {Dupuy}, Trent J.},
        title = "{Subtle Signatures of Multiplicity in Late-type Dwarf Spectra: The Unresolved M8.5 + T5 Binary 2MASS J03202839-0446358}",
      journal = {\apj},
         year = 2008,
        month = jul,
       volume = {681},
       number = {1},
        pages = {579-593},
          doi = {10.1086/588379},
archivePrefix = {arXiv},
       eprint = {0803.0295},
 primaryClass = {astro-ph},
       adsurl = {https://ui.adsabs.harvard.edu/abs/2008ApJ...681..579B}
}

@ARTICLE{2011AJ....141...70B,
       author = {{Burgasser}, Adam J. and {Bardalez-Gagliuffi}, Daniella C. and {Gizis}, John E.},
        title = "{Hubble Space Telescope Imaging and Spectral Analysis of Two Brown Dwarf Binaries at the L Dwarf/T Dwarf Transition}",
      journal = {\aj},
         year = 2011,
        month = mar,
       volume = {141},
       number = {3},
          eid = {70},
        pages = {70},
          doi = {10.1088/0004-6256/141/3/70},
archivePrefix = {arXiv},
       eprint = {1011.0798},
 primaryClass = {astro-ph.SR},
       adsurl = {https://ui.adsabs.harvard.edu/abs/2011AJ....141...70B}
}

@ARTICLE{2013ApJ...772..129B,
       author = {{Burgasser}, Adam J. and {Sheppard}, Scott S. and {Luhman}, K.~L.},
        title = "{Resolved Near-infrared Spectroscopy of WISE J104915.57-531906.1AB: A Flux-reversal Binary at the L dwarf/T Dwarf Transition}",
      journal = {\apj},
         year = 2013,
        month = aug,
       volume = {772},
       number = {2},
          eid = {129},
        pages = {129},
          doi = {10.1088/0004-637X/772/2/129},
archivePrefix = {arXiv},
       eprint = {1303.7283},
 primaryClass = {astro-ph.SR},
       adsurl = {https://ui.adsabs.harvard.edu/abs/2013ApJ...772..129B}
}

@INPROCEEDINGS{2017ASInC..14....7B,
       author = {{Burgasser}, A.~J. and {Splat Development Team}},
        title = "{The SpeX Prism Library Analysis Toolkit (SPLAT): A Data Curation Model}",
    booktitle = {Astronomical Society of India Conference Series},
         year = 2017,
       series = {Astronomical Society of India Conference Series},
       volume = {14},
        month = jan,
        pages = {7-12},
          doi = {10.48550/arXiv.1707.00062},
archivePrefix = {arXiv},
       eprint = {1707.00062},
 primaryClass = {astro-ph.SR},
       adsurl = {https://ui.adsabs.harvard.edu/abs/2017ASInC..14....7B}
}

@ARTICLE{2011AJ....142...77D,
       author = {{Deacon}, Niall R. and {Liu}, Michael C. and {Magnier}, Eugene A. and {Bowler}, Brendan P. and {Goldman}, Bertrand and {Redstone}, Joshua A. and {Burgett}, W.~S. and {Chambers}, K.~C. and {Flewelling}, H. and {Kaiser}, N. and {Lupton}, R.~H. and {Morgan}, J.~S. and {Price}, P.~A. and {Sweeney}, W.~E. and {Tonry}, J.~L. and {Wainscoat}, R.~J. and {Waters}, C.},
        title = "{Four New T Dwarfs Identified in Pan-STARRS 1 Commissioning Data}",
      journal = {\aj},
         year = 2011,
        month = sep,
       volume = {142},
       number = {3},
          eid = {77},
        pages = {77},
          doi = {10.1088/0004-6256/142/3/77},
archivePrefix = {arXiv},
       eprint = {1106.3105},
 primaryClass = {astro-ph.SR},
       adsurl = {https://ui.adsabs.harvard.edu/abs/2011AJ....142...77D}
}

@ARTICLE{2011AJ....141...71F,
       author = {{Faherty}, Jacqueline K. and {Burgasser}, Adam J. and {Bochanski}, John J. and {Looper}, Dagny L. and {West}, Andrew A. and {van der Bliek}, Nicole S.},
        title = "{Identification of a Wide, Low-Mass Multiple System Containing the Brown Dwarf 2MASS J0850359+105716}",
      journal = {\aj},
         year = 2011,
        month = mar,
       volume = {141},
       number = {3},
          eid = {71},
        pages = {71},
          doi = {10.1088/0004-6256/141/3/71},
archivePrefix = {arXiv},
       eprint = {1012.4232},
 primaryClass = {astro-ph.SR},
       adsurl = {https://ui.adsabs.harvard.edu/abs/2011AJ....141...71F}
}

@ARTICLE{2011ApJ...732...56G,
       author = {{Gei{\ss}ler}, Kerstin and {Metchev}, Stanimir and {Kirkpatrick}, J. Davy and {Berriman}, G. Bruce and {Looper}, Dagny},
        title = "{A Cross-match of 2MASS and SDSS. II. Peculiar L Dwarfs, Unresolved Binaries, and the Space Density of T Dwarf Secondaries}",
      journal = {\apj},
         year = 2011,
        month = may,
       volume = {732},
       number = {1},
          eid = {56},
        pages = {56},
          doi = {10.1088/0004-637X/732/1/56},
archivePrefix = {arXiv},
       eprint = {1103.1160},
 primaryClass = {astro-ph.SR},
       adsurl = {https://ui.adsabs.harvard.edu/abs/2011ApJ...732...56G}
}

@ARTICLE{2006ApJ...639.1120K,
       author = {{Kirkpatrick}, J. Davy and {Barman}, Travis S. and {Burgasser}, Adam J. and {McGovern}, Mark R. and {McLean}, Ian S. and {Tinney}, Christopher G. and {Lowrance}, Patrick J.},
        title = "{Discovery of a Very Young Field L Dwarf, 2MASS J01415823-4633574}",
      journal = {\apj},
         year = 2006,
        month = mar,
       volume = {639},
       number = {2},
        pages = {1120-1128},
          doi = {10.1086/499622},
archivePrefix = {arXiv},
       eprint = {astro-ph/0511462},
 primaryClass = {astro-ph},
       adsurl = {https://ui.adsabs.harvard.edu/abs/2006ApJ...639.1120K}
}

@ARTICLE{2007AJ....134.1162L,
       author = {{Looper}, Dagny L. and {Kirkpatrick}, J. Davy and {Burgasser}, Adam J.},
        title = "{Discovery of 11 New T Dwarfs in the Two Micron All Sky Survey, Including a Possible L/T Transition Binary}",
      journal = {\aj},
         year = 2007,
        month = sep,
       volume = {134},
       number = {3},
        pages = {1162-1182},
          doi = {10.1086/520645},
archivePrefix = {arXiv},
       eprint = {0706.1601},
 primaryClass = {astro-ph},
       adsurl = {https://ui.adsabs.harvard.edu/abs/2007AJ....134.1162L}
}

@ARTICLE{2010AJ....139.1045S,
       author = {{Schmidt}, Sarah J. and {West}, Andrew A. and {Burgasser}, Adam J. and {Bochanski}, John J. and {Hawley}, Suzanne L.},
        title = "{Discovery of an Unusually Blue L Dwarf Within 10 pc of the Sun}",
      journal = {\aj},
         year = 2010,
        month = mar,
       volume = {139},
       number = {3},
        pages = {1045-1050},
          doi = {10.1088/0004-6256/139/3/1045},
archivePrefix = {arXiv},
       eprint = {0912.3565},
 primaryClass = {astro-ph.SR},
       adsurl = {https://ui.adsabs.harvard.edu/abs/2010AJ....139.1045S}
}

@ARTICLE{2007AJ....133.2320S,
       author = {{Siegler}, Nick and {Close}, Laird M. and {Burgasser}, Adam J. and {Cruz}, Kelle L. and {Marois}, Christian and {Macintosh}, Bruce and {Barman}, Travis},
        title = "{Discovery of a 66 mas Ultracool Binary with Laser Guide Star Adaptive Optics}",
      journal = {\aj},
         year = 2007,
        month = may,
       volume = {133},
       number = {5},
        pages = {2320-2326},
          doi = {10.1086/513273},
archivePrefix = {arXiv},
       eprint = {astro-ph/0702013},
 primaryClass = {astro-ph},
       adsurl = {https://ui.adsabs.harvard.edu/abs/2007AJ....133.2320S}
}

@ARTICLE{2020MNRAS.491.5925M,
       author = {{Manjavacas}, E. and {Lodieu}, N. and {B{\'e}jar}, V.~J.~S. and {Zapatero-Osorio}, M.~R. and {Boudreault}, S. and {Bonnefoy}, M.},
        title = "{Spectral library of age-benchmark low-mass stars and brown dwmarfs}",
      journal = {\mnras},
         year = 2020,
        month = feb,
       volume = {491},
       number = {4},
        pages = {5925-5950},
          doi = {10.1093/mnras/stz3441},
archivePrefix = {arXiv},
       eprint = {1912.02806},
 primaryClass = {astro-ph.SR},
       adsurl = {https://ui.adsabs.harvard.edu/abs/2020MNRAS.491.5925M}
}

@ARTICLE{2025A&A...701A.208P,
       author = {{Petrus}, S. and {Chauvin}, G. and {Bonnefoy}, M. and {Tremblin}, P. and {Morley}, C. and {Charnay}, B. and {Suarez}, G. and {Gagn{\'e}}, J. and {Palma-Bifani}, P. and {Denis}, A. and {Ravet}, M. and {Bayo}, A. and {B{\'e}zard}, B. and {Biller}, B. and {Delorme}, P. and {Faherty}, J. and {Goyal}, J.-M. and {Hoch}, K. and {Hoy}, K. and {Jenkins}, J.-S. and {Lagrange}, A.-M. and {Lavie}, B. and {Liu}, M.~C. and {Manjavacas}, E. and {Marleau}, G.-D. and {McElwain}, M. and {Molli{\`e}re}, P. and {Mordasini}, C. and {Phillips}, M. and {Rojo}, P. and {Zhang}, Z. and {Zurlo}, A.},
        title = "{X-SHYNE: X-Shooter spectra of young exoplanet analogs: II. Presentation and analysis of the full library}",
      journal = {\aap},
         year = 2025,
        month = sep,
       volume = {701},
          eid = {A208},
        pages = {A208},
          doi = {10.1051/0004-6361/202554986},
archivePrefix = {arXiv},
       eprint = {2507.16899},
 primaryClass = {astro-ph.EP},
       adsurl = {https://ui.adsabs.harvard.edu/abs/2025A&A...701A.208P}
}

@ARTICLE{2007ApJ...658.1217M,
       author = {{McLean}, Ian S. and {Prato}, L. and {McGovern}, Mark R. and {Burgasser}, Adam J. and {Kirkpatrick}, J. Davy and {Rice}, Emily L. and {Kim}, Sungsoo S.},
        title = "{The NIRSPEC Brown Dwarf Spectroscopic Survey. II. High-Resolution J-Band Spectra of M, L, and T Dwarfs}",
      journal = {\apj},
         year = 2007,
        month = apr,
       volume = {658},
       number = {2},
        pages = {1217-1235},
          doi = {10.1086/511740},
archivePrefix = {arXiv},
       eprint = {astro-ph/0612668},
 primaryClass = {astro-ph},
       adsurl = {https://ui.adsabs.harvard.edu/abs/2007ApJ...658.1217M}
}

@ARTICLE{2024RNAAS...8..114Z,
       author = {{Zhang}, Zhoujian},
        title = "{Initial Entropy and Potential Delayed Formation of the Directly Imaged Exoplanet AF Lep b}",
      journal = {Research Notes of the American Astronomical Society},
         year = 2024,
        month = apr,
       volume = {8},
       number = {4},
          eid = {114},
        pages = {114},
          doi = {10.3847/2515-5172/ad4481},
       adsurl = {https://ui.adsabs.harvard.edu/abs/2024RNAAS...8..114Z}
}

\end{CJK*}

\end{document}